\documentclass[12pt,preprint]{aastex}
\usepackage{amssymb,amsmath,hyperref,rotating,natbib,morefloats,multirow}
\shorttitle{Triggered Star Formation in Six H II Regions}
\shortauthors{Dirienzo et al.}

\begin{document}

\title{Testing Triggered Star Formation in Six H {\small II} Regions}

\author{William J. Dirienzo\altaffilmark{1}}
\email{dirienzo@virginia.edu}
\author{R\'emy Indebetouw\altaffilmark{1,2}}
\author{Crystal Brogan\altaffilmark{2}}
\author{Claudia J. Cyganowski\altaffilmark{3}}
\author{Edward B. Churchwell\altaffilmark{4}}
\author{Rachel K. Friesen\altaffilmark{2,5}}

\altaffiltext{1}{Department of Astronomy, University of Virginia, P.O. Box 3818, Charlottesville, VA 22903, USA}
\altaffiltext{2}{National Radio Astronomy Observatory, 520 Edgemont Road, Charlottesville, VA 22903, USA}
\altaffiltext{3}{Harvard-Smithsonian Center for Astrophysics, Cambridge, MA 02138, USA}
\altaffiltext{4}{Department of Astronomy, University of Wisconsin-Madison, 475 N Charter Street, Madison, WI 53706, USA}
\altaffiltext{5}{Dunlap Institute for Astronomy and Astrophysics, University of Toronto, 50 St. George Street, Toronto M5S 3H4, Ontario, Canada}

\begin{abstract}

We investigated six H {\footnotesize II} regions with infrared, bright rimmed bubble or cometary morphology, in search of quantitative evidence for triggered star formation, both collect and collapse and radiatively driven implosion. We identified and classified 458 Young Stellar Objects (YSOs) in and around the H {\footnotesize II} regions. YSOs were determined by fitting a collection of radiative transfer model spectral energy distributions (SEDs) to infrared photometry for a large sample of point sources. We determined areas where there exist enhanced populations of relatively unevolved YSOs on the bright rims of these regions, suggesting that star formation has been triggered there. We further investigated the physical properties of the regions by using radio continuum emission as a proxy for ionizing flux powering the H {\footnotesize II} regions, and ${}^{13}$CO (1-0) observations to measure masses and gravitational stability of molecular clumps. We used an analytical model of collect and collapse triggered star formation, as well as a simulation of radiatively driven implosion, and thus we compare the observed properties of the molecular gas with those predicted in the triggering scenarios. Notably, those regions in our sample that show evidence of cometary, or ``blister,'' morphology are more likely to show evidence of triggering.


\end{abstract}

\section{Introduction\label{sec-intro}}

While many of the details of isolated, low mass star formation are now understood, the precise process of massive star formation remains uncertain, mostly because of the additional difficulty of studying massive star forming regions. Most such regions are over 1 kpc away and their protostars are often observed through high extinction \citep{2007ARA&A..45..481Z}, which makes it difficult to identify and study these regions. Among the most important theoretical obstacles is the need for very high accretion rates to form a massive star in less time than it takes for radiation pressure and other feedback to halt formation.

High accretion rates may arise in the high-pressure environment hypothesized in triggered star formation scenarios. Originally termed ``sequential star formation'' by \cite{1977ApJ...214..725E}, this theory posits that star formation, and massive star formation in particular, is self-propagating through molecular gas. If at least one massive star can be formed initially, then this star produces ionizing radiation that advances into the surrounding gas, creating an H {\footnotesize II} region. As the star continues to energize the region, the ionized gas expands and displaces the molecular gas, thus causing overdensities along the advancing boundary. If the expansion of the ionization front is faster that the sound speed in the neutral gas, then the increase in pressure in these overdensities cannot be redistributed outward and the material continues to collect. Eventually this gas becomes so dense that it begins to fragment. These fragments will be compelled to collapse under self-gravity, and may form stars more quickly, and at preferentially higher mass, than quiescent, isolated star formation \citep[see for example][]{2003MNRAS.344..461M}.

This triggering mechanism is typically termed ``collect and collapse'' \citep[CnC; see][]{1994MNRAS.268..291W,2007MNRAS.375.1291D}, in contrast to another possible triggering process, like ``radiatively driven implosion'' \citep[RDI: see for example][]{1982ApJ...260..183S}. In the latter process, pre-existing overdensities in the molecular gas are enhanced when an ionization front sweeps away the less dense gas and begins to compress the overdensities from all sides, inducing collapse \citep{2009MNRAS.398..157H}. While this can enhance the local density of Young Stellar Objects (YSOs), it may not necessarily lead to more massive stars, depending on the properties of the pre-existing overdensities. \cite{2011ApJ...736..142B} modeled radiatively driven implosion and determined a range of values of the ionizing flux for which star formation is triggered, as well as a power law relationship between the ionizing flux and the timescale for collapse.

The theory of collect and collapse makes quantitative predictions that can be tested observationally. The ages of triggered YSOs, as well as the masses, sizes, and densities of molecular cloud fragments in the spherically expanding shell, can be predicted from the flux of ionizing radiation powering the H {\footnotesize II} region and the initial density and sound speed of the molecular gas \citep{1994MNRAS.268..291W}. Observationally, the initial density may be estimated from observations of ${}^{13}$CO (1-0) in these clouds as a tracer of the total molecular gas, and the ionizing flux can be determined from the properties of the existing massive stars or from radio observations that trace the amount of ionized gas and thus the ionizing flux. It is important to test these predictions because the presence of YSOs and molecular gas clumps around a bubble, while suggestive, is not enough evidence alone to show that collect and collapse triggering is taking place. For instance, simulations by \cite{2011arXiv1109.3478W} show that this morphology may be replicated by the expansion of an ionization front into fractal molecular clouds, even when no stable, self-gravitating shell fragments have formed. Star formation may additionally be triggered by radiatively driven implosion in this scenario.

The available predictions that are readily applied to observations assume a simple spherically symmetric geometry. This type of study is best performed in relatively isolated, simple H {\footnotesize II} regions with dense rims, and the predictions will be best applied to regions that are round, closed bubbles. However, it is important to cover a range of morphologies and apparent evolutionary states to keep the sample unbiased, as few H {\footnotesize II} regions exhibit this ideal morphology.

\cite{2006ApJ...649..759C} cataloged 322 visually identified partial and complete mid-infrared (MIR) rings in the Galactic Legacy Infrared Mid-Plane Survey Extraordinaire (GLIMPSE). They found that these structures were ubiquitous (about 1.5 per square degree), 88\% of them were less than 4' across, about 25\% of them were coincident with H {\footnotesize II} regions known at the time, and 13\% enclosed known star clusters. They proposed that these structures were in fact three-dimensional bubbles containing gas ionized by OB stars and surrounded by a photodissociation region (PDR). \cite{2007ApJ...670..428C} found an additional 269 bubbles, and more recent studies such as \cite{2012MNRAS.424.2442S} have confirmed that these structures are common across the galactic plane. Since this type of structure is consistent with theoretical models of triggered star formation, recent studies of triggering have frequently drawn samples from this catalog.

Previous observational searches for evidence of triggered star formation around \cite{2006ApJ...649..759C} bubbles have been conducted with varied results. \cite{2008ApJ...681.1341W} studied three apparently wind-blown, parsec-sized mid-infrared bubbles, including N49, studied in this work. They identified central ionizing sources, as well as YSO populations around the rims, and determined all 3 regions to be possible sites of triggering. \cite{2010ApJ...716.1478W} looked for YSOs around 46 infrared bubbles, but reported that only 20\% of their sample showed a significant population of associated YSOs; however they did not use photometry at wavelengths longward of 8 $\mu$m, which is useful to identify and classify YSOs. \cite{2010A&A...523A...6D} investigated 102 bubbles, extending to the submillimeter wavelengths using the ATLASGAL survey at 870 $\mu$m to probe the cold dust, while also analyzing radio continuum and the YSO populations. They found that 86\% of the bubbles enclosed H {\footnotesize II} regions, and 20\% showed evidence of massive star formation on their rims. \cite{2012MNRAS.421..408T} analyzed the distribution of massive YSOs (MYSOs) from the RMS survey compared to the locations of all 322 \cite{2006ApJ...649..759C} bubbles. They reported a statistically significant overdensity of MYSOs coincident with the bubbles, and the rims in particular, which was not explained by intrinsic clustering of MYSOs. They estimated that 14-30\% of MYSOs in the Milky Way may be formed by triggering in bubbles, though they did not find any evidence that MYSOs associated with bubbles had higher luminosity (mass) than field MYSOs.

While the aforementioned studies have concentrated on the \cite{2006ApJ...649..759C} bubbles, several studies have investigated regions not in that catalog as well, often with results consistent with triggering scenarios. \cite{2011A&A...525A.132P} studied the single H {\footnotesize II} region G35.673-00.847, a region with ``semi-ring'' mid-infrared morphology and two distinct but neighboring PDRs. They identified YSOs in the immediate vicinity of the region using infrared colors and then classified them using spectral energy distribution (SED) fitting. Using the same methods of testing collect and collapse and most of the same datasets as this work, they rejected it as a plausible scenario for that region. \cite{2009ApJ...700..506S} identified YSOs in NGC 2467 using infrared colors (then confirmed by SED fitting) and found that they were largely located where the ionization front had compressed the molecular gas. They estimated that 25-50\% of the YSOs in that region were triggered, though they ruled out radiatively driven implosion as the mechanism. \cite{2009A&A...494..987P} found several YSOs on the boundary of RCW 82, but determined that the region was too young to have triggered star formation.

\cite{2008ApJ...688.1142K} analyzed the W5 H {\footnotesize II} region, which has two cometary regions in the same complex. YSOs within W5 were identified and classified using infrared colors and multiple clusters were seen. They found that both radiatively driven implosion and collect and collapse were plausible scenarios in this region. \cite{2006A&A...446..171Z} studied RCW 79, a fairly isolated H {\footnotesize II} region with somewhat cometary morphology, and found several massive fragments identified by millimeter continuum in a shell around the ionized gas. Additionally, the presence of several Class I YSOs identified by infrared color selection coincident with these fragments was consistent with triggering by collect and collapse. \cite{2008A&A...482..585D} studied Sh2-212, a round, isolated H {\footnotesize II} region. They found fragments of molecular gas arranged in a shell around the region, with strong evidence for a massive YSO coincident with the most massive fragment. Studies of other individual H {\footnotesize II} regions with similar promising results have appeared in \cite{2007A&A...472..835Z,2010A&A...518L.101Z,2010A&A...518L..81Z}. The existing literature suggests that collect and collapse is a viable star formation mechanism, but its relative importance and under which physical conditions it operates are still undetermined.

Several other studies searched for evidence of radiatively driven implosion in similar regions, and we summarize only a few here. \cite{2010RAA....10..777C} analyzed the Cepheus B molecular cloud and claimed it was a good RDI candidate because of its morphology, the presence of an age gradient in young stars leading back to the ionizing source, and the temperature, density, and velocity structure of the molecular gas around its bright rim. \cite{2007A&A...467.1125U} conducted a detailed study of the region BRC SFO 75 in millimeter continuum, ${}^{13}$CO, and NH${}_{3}$ emission. They identified two dense cores; one was being influenced by ionizing radiation while the other was still beyond the ionization front. They reported three YSOs near the core under the influence of the ionizing radiation, while the other core appears nearly spherical and devoid of stars. \cite{2010MNRAS.408..157M} observed the NH${}_{3}$ (1,1), (2,2), (3,3), and (4,4) transitions towards 42 bright-rimmed regions under the influence of an ionizing source. Using previously published submillimeter continuum and CO data, as well as locations of known outflows and masers, they identified many of the regions with active star formation as likely sites of triggering. The NH${}_{3}$ data showed that these regions have higher velocity dispersions than the counterparts that were not triggering candidates. They proposed that the higher velocity dispersions are an indication either that shock fronts have induced star formation in these regions, or that they are a result of increased star formation activity. These studies have shown that radiatively driven implosion is also a viable mechanism for triggering star formation, though again its global importance is not known.

\cite{2007MNRAS.377..535D} performed SPH simulations of a molecular cloud with and without a central ionizing source. They compared the cores that formed in each scenario and found that the star formation efficiency was approximately 30\% higher when including the ionizing source. This increase in efficiency was due to both an acceleration in the formation time of cores that would have formed in the simulation without an ionizing source, as well as the formation of additional, apparently triggered cores. However, they did not see a significant change in the masses of the cores, nor an age gradient with position. Furthermore, the velocity of the cores primarily reflected the initial turbulent conditions rather than the velocity of the expanding shell. These simulations are consistent with an increase in star formation due to triggering, but show that it can be quite difficult to gather convincing observational evidence of this process.

The aim of this work is to study multiple isolated H {\footnotesize II} regions with varied morphologies in a homogeneous way to analyze triggered star formation in H {\footnotesize II} regions and determine whether the H {\footnotesize II} region morphology has any effect. Additionally, we use SED fitting to identify and classify YSOs not only in the immediate vicinity of the infrared bubbles and rims, but also in the surrounding field to quantify any enhancement in the YSO surface density. The benefit of SED fitting over infrared color selection is the improved ability to estimate the mass and other physical parameters of the YSOs.

We have adopted a sample of six previously identified H {\footnotesize II} regions that are relatively isolated and have simple morphologies, but range from round, closed bubbles to rims of so-called cometary, or ``blister,'' H {\footnotesize II} regions. \cite{1978A&A....70..769I} developed the term ``blister model'' to describe cometary H {\footnotesize II} regions and asserted that most optically visible H {\footnotesize II} regions were in fact cometary. The sample is comprised of G028.83-0.25, G041.10-0.15, G041.91-0.12, G041.92+0.04, G044.28+0.11, and G044.34-0.82. Mid-infrared images of these regions are presented in Figure \ref{f1}, and coordinates are given in Table \ref{t1}. \cite{2006ApJ...649..759C} previously identified four of these regions, G028.83-0.25, G041.92+0.04, G044.28+0.11, and G044.34-0.82, as N49, N80, N91, and N92, respectively. They argue that nearly all of the bubbles of this type that they identified were formed by hot, young stars.

The determination of the distances to the regions in our sample is presented in \S\ref{sec-dist}. An overview of the infrared data and the YSO selection and categorization process is given in \S\ref{sec-irdata}. The radio continuum images and its relationship to the ionizing sources powering these regions is discussed in \S\ref{sec-radiocont}. The analysis of molecular gas data is in \S\ref{sec-molecular}. Tests of triggered star formation are discussed in \S\ref{sec-trigger}. Results for each region are given in \S\ref{sec-HII}. Finally, a discussion of the evidence for triggering is presented in \S \ref{sec-discussion}.

\section{Methodology and Analysis \label{sec-method}}

\subsection{Distance Determination \label{sec-dist}}

Many of our quantitative results depend on the distance to the H {\footnotesize II} regions. We calculated the kinematic distances using the galactic rotation curve of \cite{2009ApJ...700..137R}. They adopted a galactocentric radius $R_{\circ}=8.4 \pm 0.6$ kpc and a circular rotation speed $\Theta_{\circ}=254 \pm 16$ km s${}^{-1}$ kpc${}^{-1}$, based on the results of their measured trigonometric parallaxes of massive star-forming regions. Radio recombination line velocities are known for four of our regions from \cite{2009ApJ...690..706A}: G028.83-0.25 at 90.6 km s${}^{-1}$, G041.10-0.15 at 59.4 km s${}^{-1}$, G041.91-0.12 at 18.1 km s${}^{-1}$, and G044.34-0.82 at 59.6 km s${}^{-1}$. All four of these regions have significant molecular gas emission at similar velocities. CO (3-2) velocities for G041.92+0.04 and G044.34-0.82 are known to be 17.7 km s${}^{-1}$ and 62.0 km s${}^{-1}$, respectively, from \cite{2010ApJ...709..791B}.

All of the regions in our sample lie in the $|\ell| < 90^{\circ}$ regime, so there is naturally a near-far distance ambiguity. \cite{2009ApJ...690..706A} attempted to resolve this ambiguity for 291 H {\footnotesize II} regions, including G028.83-0.25, G041.10-0.15, G041.91-0.12, and G044.28+0.11 (named C28.82-0.23, C41.10-0.21, D41.91-0.12, and U44.26+0.10 in their work, respectively). They did this by analyzing H {\footnotesize I} spectra using two different methods.

The first method is based on the features in the H {\footnotesize I} spectra due to absorption against the H {\footnotesize II} radio continuum by foreground H {\footnotesize I} clouds. \cite{2009ApJ...690..706A} searched for evidence of this absorption process by looking at the difference in spectra along the line of sight towards H {\footnotesize II} regions and towards nearby off-source positions. It is expected that all regions will show a difference in the H {\footnotesize I} emission and absorption features between the on and off source positions at velocities less than the radio recombination line velocity, but regions at the far distance will also show differences in these features between the recombination line velocity and the tangent point velocity.

The second method relies on the cold H {\footnotesize I} gas within the molecular clouds associated with the H {\footnotesize II} regions to absorb some of the emission from the warmer background H {\footnotesize I}. Narrow H {\footnotesize I} absorption at a velocity coincident with the velocity of ${}^{13}$CO (1-0) emission associated with the H {\footnotesize II} region indicates a source is at the near distance, while the absence of narrow H{ {\footnotesize I} absorption at the ${}^{13}$CO (1-0) velocity indicates the source is at the far distance. 

\cite{2009ApJ...690..706A} used H {\footnotesize I} data from the Very Large Array (VLA) Galactic Plane Survey (VGPS) \citep{2006AJ....132.1158S} and the ${}^{13}$CO (1-0) Boston University Galactic Ring Survey (BU-GRS) data \citep{2006ApJS..163..145J}. The VGPS was a survey of the 21 cm H {\footnotesize I} line and 21 cm continuum, combining interferometric data from the VLA with single dish data from the Robert C. Byrd Green Bank Telescope (GBT). The data have angular resolution of 1 arcminute, velocity resolution of 1.56 km s${}^{-1}$, and 2 K rms sensitivity. The BU-GRS is a large scale survey of the 110.2 GHz ${}^{13}$CO (1-0) transition in the disk of the Milky Way using the Five College Radio Astronomy Observatory (FCRAO) 14 meter single dish telescope. The publicly available data cubes have velocity resolution of 0.2 km s${}^{-1}$, angular resolution of 46'', and typical antenna temperature rms sensitivity of 0.13 K \citep{2006ApJS..163..145J}.

The results of \cite{2009ApJ...690..706A} are summarized for the regions in our sample as follows. They find that G028.83-0.25 is likely at the near distance, G041.10-0.15 is likely at the far distance, G041.91-0.12 may be at the far distance, though with low confidence, and G044.28+0.11 is likely at the far distance. We also analyzed all six of our regions using the same methods and data. A detailed analysis of the spectra shows that the data remain at least consistent with the near distance for our sources. In particular, we note that the molecular gas associated with G041.10-0.15 shows a velocity gradient that should be considered when applying the H {\footnotesize I} self-absorption method. We thus assume the near kinematic distance for all of our sources for the remainder of this paper. The resulting distances are presented in Table \ref{t1}.

In the event that any of the regions lie at the far kinematic distance, the biggest effect will be that our bolometric luminosity estimates for the YSOs will be too low. For four of our sources, this effect would cause us to underestimate the bolometric luminosities by approximately a factor of four, since the far distances are about twice the near distances. The other two regions, G041.91-0.12 and G041.92+0.04, have a difference of about a factor of 8 between the near and far distances, and it is unlikely that the YSOs in these regions would be 64 times as bright as our current estimates. Our tests of triggered star formation (see \S\ref{sec-trigger}) depend on the distance as well, though fairly insensitively. Our selection of YSOs is relatively insensitive to the distance, since the shapes of the SEDs will not be significantly changed. The longest wavelengths are the most important in the SED for identifying YSOs, and are also the least sensitive to a change in extinction associated with a change in distance.

\subsection{Infrared Data and YSO Identification \label{sec-irdata}}

The \emph{Spitzer Space Telescope} has been revolutionary in collecting data of use for star formation studies. The Infrared Array Camera (IRAC) \citep{2004ApJS..154...10F} and Multiband Imaging Photometer for \emph{Spitzer} (MIPS) \citep{2004ApJS..154...25R} instruments, together with the ground-based Two Micron All-Sky Survey (2MASS) \citep{2006AJ....131.1163S}, provide the wide wavelength coverage important for reliable identification and classification of YSOs. The 2MASS Survey provides images at the near-infrared \emph{J} (1.25 $\mu$m), \emph{H} (1.65 $\mu$m), and \emph{K${}_{s}$} (2.16 $\mu$m) bands covering the entire sky. The Galactic Legacy Infrared Mid-Plane Survey Extraordinaire (GLIMPSE) \citep{2003PASP..115..953B, 2009PASP..121..213C} covers the region $|b| \le 1^{\circ}$ and $ 10^{\circ} \le |\ell| \le 65^{\circ}$ in all the IRAC bands (3.6, 4.5, 5.8, and 8.0 $\mu$m) with 1''.5 to 1''.9 resolution, while the MIPS Galactic Plane Survey (MIPSGAL) \citep{2009PASP..121...76C} is a complimentary survey in the 24 and 70 $\mu$m MIPS wavebands with 6'' and 18'' resolution, respectively. The GLIMPSE Point Source Catalog (GPSC) \citep{2003PASP..115..953B} is a publicly available, highly reliable catalog of automatedly identified point sources from the GLIMPSE survey. The catalog itself provides coordinates and flux density measurements from point spread function (PSF) fitting in each of the 2MASS and IRAC wavebands. Coordinates bounding the regions of the sky for which we used the GPSC and searched for YSOs are given in Table \ref{t2} and shown in Figure \ref{f2}. We chose the regions to contain the entirety of the infrared bubbles and rims, as well as a significant area surrounding them for use as a control for comparisons of YSO spatial density.

In addition to the sources from the GPSC, we also identified point sources that were seen in IRAC 8 $\mu$m and/or MIPS 24 $\mu$m images but were missing from the catalog. The GPSC was finely tuned to have very high reliability in regions of complex diffuse emission, at the cost of the inevitable loss of some completeness. To improve the completeness of our final YSO list, we included these manually identified sources. An additional 179 sources were added to the sample in this manner, compared to 15,798 from the GPSC. We refer to these additional 179 sources as ``MI'' (manually identified) sources.

We employed a custom written Interactive Data Language (IDL) code to perform aperture photometry on all of the point sources in our sample. This was done to obtain photometry of the MI sources and MIPS photometry for all sources, as well as to obtain upper limits on nondetections from the GPSC. We used 3'' radius circular apertures for the 2MASS and IRAC images, and 6'' and 12'' radius circular apertures for the MIPS 24 $\mu$m and 70 $\mu$m images, respectively, owing to the poorer resolution at longer wavelengths. In all cases the background emission was estimated from the mean value in an annulus extending 1.75 to 2.5 times the radius of the aperture. The criterion for detection was that the background-subtracted flux was at least one standard deviation of the background variation above the mean background level. We used this relatively low threshold above the background in individual bands because we later require detections in at least 4 photometric bands for a source to be considered part of our sample. In cases of a nondetection, we adopted the value of the background plus one standard deviation as an upper limit. We corrected for the fraction of missing flux from the comparison of aperture size to the PSF following \cite{2007MNRAS.374..979C}. In the case of 70 $\mu$m, the images often suffered from their lower resolution as well as much more extended diffuse emission that made it nearly impossible to obtain reliable flux density measurements, except for the very brightest sources. We therefore only used the 70 $\mu$m images to determine upper limits for each source. Uncertainties were computed using Poisson counting statistics, with a minimum uncertainty of 10\% of the photometric value used for all sources and wavebands.

When evaluating the aperture photometry values for sources in the GPSC within our sample, they generally agree with the GPSC photometry value to within ten percent. The biggest disagreements are at the very highest ($\gtrsim 0.5$ Jy) and very lowest ($\lesssim 2$ mJy) flux densities. Discrepancy at the highest flux densities is due to saturation in the images, which is better handled by PSF fitting than aperture photometry. In some cases, the aperture photometry allows us to obtain a measurement where the GPSC does not supply one, and so we adopted such values. In all other cases we adopt an uncertainty-weighted average of the values from the two methods. Since the two methods generally agree well, averaging the two does not significantly change the values, but we do get a larger, more realistic estimate of the uncertainty in cases where the two methods do disagree.

We required that only sources that are detected in at least four wavebands are analyzed, which helped to verify that sources are not spurious and that there was enough photometric information for each source to be studied reliably. Sources with fewer than four photometric data points were removed from further consideration. This requirement reduced our sample to 15,685 GPSC sources and 78 MI sources. The remaining sample sizes in each region are presented in Table \ref{t3}.

With our sample of infrared point sources with flux density measurements or upper limits in nine wavebands, we begin to classify the sources using the SED fitter described in \cite{2007ApJS..169..328R}. The fitter calculates a $\chi^{2}$ value for each point source paired with each SED model from a selection of radiative transfer models. Because our point source sample contains sources with different numbers of photometric measurements, $n_{\textrm{data}}$ (not counting upper limits), we use the $\chi^{2}$ divided by the number of data points, $\chi^{2}/n_{\textrm{data}}$, as a measure of goodness of fit.

To get a reliable list of YSO candidates, we took steps to remove sources that could plausibly be main sequence or giant stars. We began by first fitting a grid of 7853 stellar atmosphere radiative transfer models from \cite{2005ESASP.576..565B} to our sample, using the SED fitter from \cite{2007ApJS..169..328R}. The SED models spanned a range of effective temperatures, metallicities, and gravities ($ 2.7 \times 10^{3} \textrm{ K} \le T_{\textrm{eff}} \le 10^{4} \textrm{ K} $, $ -0.4 \le \left[Z/H\right] \le 0.5 $, and $ -0.5 \le \log (g) \le 5.5 $, respectively). The extinction, $A_{V}$, along the line of sight to the source was a free parameter of the fitting process that we restricted to be between 0 and 20 magnitudes. The choice of this range was informed by \cite{2005ApJ...619..931I} who found that $A_{V}$ in the galactic plane is approximately 0.5-2 magnitudes kpc${}^{-1}$. Since the $A_{V}$ determination for each source is independent of every other source, this should account for variations in extinction across the field. The stellar atmosphere model fitting was distance independent, i.e. the absolute flux density scale is arbitrary at this stage, and only the SED shape is considered. Any source for which the best-fit SED met the criteria $\chi^{2}_{\textrm{best}}/n_{\textrm{data}} < 3$ was classified as ``stellar,'' while the remaining sources were used as the sample for fitting YSO SEDs. The majority of GPSC point sources in our sample were well fit by the stellar atmosphere models (see Table \ref{t3}). Removing these sources from further consideration, we were left with 1794 GPSC sources and 73 MI sources.

With sources consistent with stellar atmospheres identified and removed, we then performed SED fitting of YSO radiative transfer models from \cite{2006ApJS..167..256R} on the remaining sample. When performing the SED fitting, the line of sight extinction (to the ``source,'' where source is defined as the outermost boundary of the radiative transfer model, not all the way to the surface of the central object) was again a free parameter between 0 and 20 magnitudes. The fitted distance was allowed to be within the ranges listed in Table \ref{t2}, which were chosen to be consistent with the near kinematic distances following \S\ref{sec-dist}. The distance range for G041.91-0.12 and G041.92+0.04 was between 0.5 and 2.5 kpc, and 3.5 to 5.5 kpc for all the other regions. Again, any source for which the best-fit SED met the criteria $\chi^{2}_{\textrm{best}}/n_{\textrm{data}} < 3$ was classified as a good fit, while the remaining sources were excluded from further consideration. Examples of a good and marginally acceptable fit for each YSO stage (see \S\ref{sec-classification}) are presented in Figure \ref{f3}. A total of 598 YSO candidates (538 GPSC and 60 MI) were identified in this manner.

Once a source was identified as a YSO candidate, we also identified other SED models for which $\chi^{2}/n_{\textrm{data}}$ was within 6 of the best fit, $\chi^{2}_{\textrm{best}}/n_{\textrm{data}}$. We did this so that once a source was reliably identified as a YSO candidate, we could investigate the full range of physical parameters that fit the data. We knew the central mass, accretion rate, disk mass, inclination angle, etc. of each model, so we calculated an average and uncertainty of several key physical parameters for each source based on the distribution ``good fit'' models, weighted by the probability, $\exp \left( -\chi^{2}/2 \right)$, of each model. Most importantly, we estimated the mass and evolutionary stage of each of these YSO candidates. Using $\left( \chi^{2} - \chi^{2}_{\textrm{best}} \right) /n_{\textrm{data}} < 6$ as a threshold allowed for a more realistic estimate of the uncertainties in the physical parameters.

To check how much our YSO candidate sample depends on MIPS photometry, we repeated this same fitting process without 24 $\mu$m data. We find that 329 YSO candidates are recovered (55\% of the YSO candidate sample using the MIPS photometry). This quantitatively illustrates the importance of long wavelength data for the identification of YSOs. We proceed with further analysis using the YSO candidate sample identified using the MIPS photometry. We note that of the 542 MIPS 24 $\mu$m point sources we report, 358 (66\%) are detected at the 3-$\sigma$ level or greater.

A likely residual contaminant in this sample of YSO candidates is AGB stars, which tend to have SEDs similar to certain YSO models in this wavelength range. In the absence of additional data to discern the true YSOs, such as spectroscopic observations, we evaluate our candidates in color-magnitude space. \cite{2008AJ....136.2413R} have analyzed AGB stars (both sAGB and xAGB) and clustered (likely YSO) sources to see where they each fall in [8.0]-[24.0] versus [4.5] space. They have determined areas in this space that are predominantly occupied by each population, however the populations do overlap somewhat in this space, and there will likely be contamination in both samples. The results of \cite{2008AJ....136.2413R} are derived for the entire galactic plane, so shifts in criteria to optimize the cut for each region are expected. We have determined our own selection criteria for removing AGB stars for each H {\footnotesize II} region using the \cite{2008AJ....136.2413R} results as a guide and being as conservative as possible in the removal of sources so as not to lower the completeness of our final YSO sample drastically. The decision to be conservative in this removal process is justified by our completeness estimates in \S\ref{sec-complete}. Color-magnitude diagrams of all YSO candidates are presented in Figure \ref{f4}, showing our criteria for discrimination between AGB and YSO sources. Objects with limits on [4.5] or [8.0]-[24.0] are only removed if they lie entirely on the AGB side of our cuts.

One can see that each sample of YSO candidates roughly separates into two populations that are better separated in [4.5] than in [8.0]-[24.0]. The coordinates and properties of YSOs remaining after this color-magnitude cut are presented in Table \ref{t4}. A comparison of the sample sizes to the numbers of stellar sources, YSOs, and AGBs are presented in Table \ref{t3}. A total of 458 YSOs remained after 140 AGBs were removed. We refer to this final sample of 458 as ``YSOs,'' while we refer to the previously combined sample of YSOs and AGBs as ``YSO candidates.'' All additional analysis in this work only makes use of the YSO sample.

\subsubsection{YSO Classification \label{sec-classification}}

Historically, low-mass YSOs have been identified by their spectral indices and infrared colors. The reddest sources are classified as Class I objects, slightly bluer objects as Class II objects, and yet bluer objects as Class III objects. Class I corresponds to sources in a relatively early evolutionary phase, with significant accretion from a surrounding envelope. Class II objects have optically thick disks and potentially the remains of an envelope. Class III objects are the most evolved with only an optically thin disk remaining \citep{1987ApJ...312..788A}. We investigated both GPSC and MI sources using the color selection method of \cite{2004ApJS..154..363A} and found minimal differences between the YSOs identified by color and those identified by SED fitting. Most discrepancies were in confused regions of high background emission. This result is in good agreement with a similar study in M16 (the Eagle Nebula) by \cite{2007ApJ...666..321I}. The major advantage of using the SED fitting method to identify YSOs, though it is more labor intensive than other methods, is that it allows an estimate of the physical properties of the source from the parameters of the radiative transfer models that best fit the data.

Throughout the remainder of this paper, we adopt the YSO ``Stage'' classification scheme of \cite{2006ApJS..167..256R}. This is physically similar to the common ``Class'' system corresponding to the relative evolutionary state of low-mass YSOs, described above, but can be simply determined in our case from the accretion rate, disk mass, and central source mass as determined by SED fitting and the \cite{2006ApJS..167..256R} models. As noted by \cite{2006ApJS..167..256R}, the use of spectral index classification can lead to confusion as it is motivated more by observation than physical state and does not properly account for changes in viewing angle between individual sources. Furthermore, \cite{2004ApJ...617.1177W} note that for high mass sources, both $T_{\textrm{eff}}$ and the evolutionary state affect the mid-infrared spectral index, so the model-derived ``Stage'' is less ambiguous than a simple spectral index. Stage I objects are defined as those that have $\dot{M_{\textrm{env}}}/M_{\star} > 10^{-6} \textrm{ yr}^{-1}$, where $\dot{M_{\textrm{env}}}$ is the envelope accretion rate and $M_{\star}$ is the mass of the central source. Stage II objects are defined by $\dot{M_{\textrm{env}}}/M_{\star} < 10^{-6} \textrm{ yr}^{-1}$ and $M_{\textrm{disk}}/M_{\star} > 10^{-6}$, where $M_{\textrm{disk}}$ is the disk mass. Finally, Stage III objects are defined by $\dot{M_{\textrm{env}}}/M_{\star} < 10^{-6} \textrm{ yr}^{-1}$ and $M_{\textrm{disk}}/M_{\star} < 10^{-6}$.

\subsubsection{Completeness \label{sec-complete}}

To estimate the completeness of our YSO sample, we determined which of the YSO SED models from \cite{2006ApJS..167..256R} would be both detectable by \emph{Spitzer} and 2MASS and identifiable as YSOs by our selection method if the models represented real YSOs within the bubble or cometary regions. To start, we calculated the flux density of each model SED in each waveband at the distance of each H {\footnotesize II} region and applied an extinction of $A_{V} = 1$ mag kpc${}^{-1}$. For each H {\footnotesize II} region, we independently adopted values of the limiting flux density to qualify as a detection at each wavelength. These limiting values were determined by plotting the source counts from the GPSC in each region as a function of magnitude to identify the sensitivity limit in the bubble or cometary structures, which are known to have high backgrounds. We adopted a single value for each region and wavelength, though the actual background can vary by as much as a factor of five at the longer wavelengths. Furthermore, we applied the 2MASS and GLIMPSE saturation limits presented in \cite{2006AJ....131.1163S} and \cite{2003PASP..115..953B}. We were thus able to determine the wavebands in which each YSO model would be detectable in each region, and so generate a set of simulated photometric data points for each YSO model. We then determined which of these model sources would be identified as YSOs following our selection method in \S\ref{sec-irdata}.

We account for the fact that the distribution of physical parameters in the model grid does not necessarily represent the distribution of the true YSO population by using a simulated sample. \cite{2006ApJS..167..256R} simulated a large ``virtual'' cluster of YSOs drawing from a uniform age distribution (implying a constant rate of star formation) in the range from $10^{3}$ years to 2 Myr, and a \cite{2001MNRAS.322..231K} IMF ranging from 0.1 to 30 solar masses. \cite{2006ApJS..167..256R} chose this mass range to approximate a real cluster, though the SED models have masses as high as 50 solar masses. They then calculated ``weights'' that scale with the likelihood of each model to appear in the simulated cluster. We used the list of models that would be detected and identified as YSOs within each region, weighted by these probabilities, to represent our simulated sample. We then plotted the mass distribution of this simulated population and compared it to the distribution of our observed YSO sample normalized by solid angle. The plots for each region are shown in Figure \ref{f5}, using an arbitrary scaling of the simulated sample. For comparison, we also plotted the distribution of the virtual cluster without any observational or methodological effects considered, i.e. a simulated sample that is 100\% complete across the entire mass range. The ratio of these two simulated distributions with and without observational effects, seen in Figure \ref{f6}, provides a completeness estimate as a function of mass, while the correspondence between the simulated samples and the real samples seen in Figure \ref{f5} provides evidence that the completeness estimates are valid. We have not made an attempt to account for the effects of sampling small populations, so our simulated distributions are unable to account for stochasticity. For example, we do not detect any YSOs in G041.91-0.12 \& G041.92+0.04 in multiple bins around 2-4 $M_{\odot}$ despite the fact that we estimate our completeness at over 50\% in this mass range.

The loss of completeness at lower mass is dominated by photometric sensitivity limits, whereas the loss of completeness at high mass is dominated by AGB color-magnitude cuts that predominantly remove the sources among the brightest at 4.5 $\mu$m. The benefit of additional completeness at higher mass by altering or emitting the AGB color-magnitude cuts is outweighed by the likely contamination of AGBs falsely identified as high mass YSOs. The small number of massive YSO candidates makes it difficult to further investigate the uncertainty introduced by applying a population-based cut. Clearly G041.91-0.12 and G041.92+0.04 are the most complete samples, as is expected for the closest regions in this work, being over 50\% complete between approximately 1.5 and 5 $M_{\odot}$. The remaining regions show a dramatic loss in completeness below 3-4 $M_{\odot}$.

We performed this completeness analysis only as a guide to interpreting our results, especially in considering our estimated completeness of high mass YSOs. We do not apply any kind of completeness correction to our sample because of the relatively small number of YSOs in each region and the inherent uncertainty in applying such a correction. The incompleteness in our sample, particularly at the high mass end, precludes us from realistically analyzing whether possibly triggered YSOs have systematically higher masses than the field population.

We also note that the youngest, least evolved YSOs will be missed by our selection method regardless of mass because, while they may be bright in the far-infrared or in molecular tracers, they are not detectable in IRAC images shortward of 5 $\mu$m. Indeed, there is a small population of infrared sources in these regions that are identified in 24 $\mu$m and sometimes in 8 $\mu$m as well, but are not detectable in shorter wavelengths. However, these sources were excluded from our sample because they are detectable in fewer than the minimum four wavebands.

\subsection{Ionizing Sources \label{sec-radiocont}}

Our analysis of the H {\footnotesize II} regions in the context of triggered star formation requires knowledge of the ionizing luminosity from stars that power the regions. To estimate this, we make use of the Very Large Array Galactic Plane survey (VGPS) 21 cm continuum data (approximately 1' resolution) described by \cite{2006AJ....132.1158S}. Images of our sample regions with radio continuum contours are in Figures \ref{f7} to \ref{f12}. We use custom apertures around the continuum emission associated with each H {\footnotesize II} region to measure the flux density. Carefully drawn source and background apertures are necessary because of the varying shapes of the radio continuum, as well as the varying background emission. For measurements of G041.10-0.15, we are particularly careful to avoid emission from the supernova remnant SNR G041.1-00.3 (3C 397) \citep{2010ApJ...712.1147J}.

Assuming the emission is optically thin free-free thermal continuum, we calculate the ionizing luminosity following \cite{1992ARA&A..30..575C}:
\begin{equation} \label{eq:Qlim}
Q_{\textrm{Ly}} \gtrsim 7.54 \times 10^{46} \left( \dfrac{T_{e}}{10^{4} \textrm{ K}}\right)^{-0.45} \left( \dfrac{\nu}{\textrm{GHz}}\right)^{0.1} \left( \dfrac{S_{\nu}}{\textrm{Jy}} \right) \left( \dfrac{D}{\textrm{kpc}} \right)^{2} \textrm{s}^{-1} \textrm{,}
\end{equation}
where $T_{e}$ is the electron temperature, $\nu$ is the frequency of the observation, $S_{\nu}$ is the observed specific flux density, and $D$ is the distance to the H {\footnotesize II} region. This quantity is a lower limit because the fraction of photons absorbed by dust or leaking out of the region is unknown. The regions in our sample that have observed recombination lines in \cite{1989ApJS...71..469L} or \cite{1996ApJ...472..173L} are G028.83-0.25, G041.10-0.15, G041.91-0.12, and G044.34-0.82. The line widths are 19.9 $\pm$ 1.7 km s${}^{-1}$, 26.7 $\pm$ 1.9 km s${}^{-1}$, 36.7 $\pm$ 7.0 km s${}^{-1}$, and 30.4 $\pm$ 4.7 km s${}^{-1}$, respectively. These line widths imply temperatures of $0.87 \pm 0.15 \times 10^{4}$ K, $1.6 \pm 0.2 \times 10^{4}$ K, $2.9 \pm 1.1 \times 10^{4}$ K, and $2.0 \pm 0.6 \times 10^{4}$ K, respectively. \cite{1989ApJS...71..469L} notes that relatively large line widths, such as the one reported for G041.91-0.12, may be the result of multiple nebulae along the same line of sight that are not well separated in velocity. The resulting temperature should then be interpreted as an upper limit. Assuming a uniform value of $10^{4}$ K for the electron temperature, we calculate the ionizing luminosity necessary to power each H {\footnotesize II} region. We estimate the uncertainty in the radio flux at 30\%, the uncertainty in the electron temperature at a factor of 2 (100 \%), and uncertainty in the distance at 50\%, which yields an estimate of a factor of 1.75 uncertainty in the ionizing luminosity.

From $Q_{\textrm{Ly}}$, we determined the spectral type of a single ionizing star using \cite{1996ApJ...460..914V} and \cite{2002MNRAS.337.1309S}. We also determined the spectral type of the most massive star in a cluster with a \cite{1955ApJ...121..161S} IMF that would provide the same ionizing luminosity. Because of the steep relationship between mass and ionizing luminosity, the ionizing luminosity is dominated by the most massive member of the cluster, and therefore the star with the earliest spectral type. Thus, the results for a cluster were only approximately one spectral subtype later than when using a single star. We considered the effect of dust, by estimating that half of the ionizing luminosity was absorbed by dust grains \citep[consistent with][]{1989ApJS...69..831W}. The result was about one spectral subtype earlier. The factor of two uncertainty in the electron temperature also introduces an uncertainty of approximately one subtype. The $Q_{\textrm{Ly}}$ values and the equivalent spectral types of a single ionizing source are presented in Table \ref{t5}.

To verify whether the radio continuum emission is thermal, we calculated the spectral index from 11 cm to 21 cm, incorporating the Bonn 11 cm survey \citep{1984A&AS...58..197R} from the Effelsburg 100 meter telescope. These single dish observations have angular resolution of about 4'.3, and 50 mK rms sensitivity. To get a consistent measurement, we smoothed the VGPS images to the resolution of the Bonn images and measured the photometry using the same apertures on each. Using $S_{\nu} \propto \nu^{\alpha_{\textrm{cm}}}$, we calculated the spectral index, $\alpha_{\textrm{cm}}$, for the regions and present the values in Table \ref{t5}. We found that all of the regions have spectral indices of $\alpha_{\textrm{cm}} \approx -0.1$ within uncertainties, consistent with optically thin free-free emission.

\subsection{Molecular Gas \label{sec-molecular}}

To fully understand whether star formation in these H {\footnotesize II} regions is triggered, we need to understand the molecular gas environment. We make use of the Boston University Galactic Ring Survey (BU-GRS) introduced in \S\ref{sec-dist}. The publicly available data cubes have velocity resolution 0.2 km s${}^{-1}$, angular resolution of 46'', and typical rms sensitivity of 0.13 K \citep{2006ApJS..163..145J}. G028.83-0.25 has two velocity components seen in ${}^{13}$CO (1-0) at 88.0 and 95.6 km s${}^{-1}$ (see Figure \ref{f13}), and either one or both may be associated with the H {\footnotesize II} region. We consider both, and they are evaluated separately for their triggering analysis (\S\ref{sec-G28}).

To calculate the column density, we assume that the gas is in local thermodynamic equilibrium (LTE) and is optically thin, and thus use the standard equation:
\begin{equation}
N = \dfrac{3k}{8 \pi^{3} \nu S \mu^{2}} \dfrac{Q_{\textrm{rot}}}{g_{u} g_{K} g_{\textrm{nuclear}}} \exp \left( \dfrac{E_{u}}{k T_{\textrm{ex}}} \right) \int T_{B}~dv \textrm{,}
\end{equation}
where $\nu$ = 110.201353 GHz, $S = J_{u}/(2J_{u} + 1)$, $J_{u} = 1$, $\mu$ = 0.112 debye, $g_{u} = 2J_{u} + 1$, $g_{K} = 1$, $g_{\textrm{nuclear}} = 1$, and $E_{u}/k = 5.29$ K. We assume a partition function of the form $Q_{\textrm{rot}} \approx 0.38(T_{\textrm{ex}}/\textrm{K} + 0.88)$ \citep{2009tra..book.....W}. We note the intrinsic assumption that the level populations are determined by a single parameter, the excitation temperature, which is not necessarily the same as the kinetic temperature.

For the excitation temperature, we adopt a value consistent with similar environments. \cite{2002ApJ...566..931S} reported ammonia temperatures of approximately 20 K for massive cores without strong centimeter continuum emission. \cite{2011ApJ...739L..16B} observed massive YSOs with the VLA and found ammonia temperatures in the 20-30 K range in kinematically simple cores. \cite{2008ApJS..175..509R} observed ammonia in dense cores in Perseus with the GBT, including sources both with and without submillimeter continuum, and found temperatures as low as 11 K in the cold gas. Admittedly Perseus is more quiescent than our regions. Furthermore, we do not know that the ammonia and ${}^{13}$CO have the same $T_{\textrm{ex}}$ or trace the same volume. \cite{2011A&A...525A.132P} assumed $T_{\textrm{ex}}$ = 20 K for similar analysis around the H {\footnotesize II} region G35.673-00.847. \cite{2008A&A...482..585D} estimated the kinetic temperature of molecular gas to be between 14 K and 30 K in Sh2-212 based on ${}^{12}$CO, ${}^{13}$CO, and C${}^{18}$O. We assume a value of $T_{\textrm{ex}}$ = 20 K with an uncertainty of 10 K, and thus we obtain
\begin{equation}
N({}^{13}\textrm{CO (1-0)}) = 1.25 \times 10^{15} \dfrac{\int T_{B}~dv}{\textrm{K km s}^{-1}} \textrm{ cm}^{-2} \textrm{.}
\end{equation}
We adopt a conversion factor $N(\textrm{H}_{2})/N({}^{13}\textrm{CO}) = 5 \times 10^{5}$ from \cite{2001ApJ...551..747S}, and thus the column density of the total molecular gas is
\begin{equation}
N(\textrm{H}_{2}) = 6.24 \times 10^{20} \dfrac{\int T_{B}~dv}{\textrm{K km s}^{-1}} \textrm{ cm}^{-2} \textrm{.}
\end{equation}
We note that a 50\% uncertainty in $T_{\textrm{ex}}$ (20 $\pm$ 10 K) introduces an uncertainty of about 35\% in each of the column density and mass of molecular gas. Both column density and mass also scale linearly with our choice of $N(\textrm{H}_{2})/N({}^{13}\textrm{CO} (1\textrm{-}0))$, and rely heavily on our assumption of LTE. The mass additionally depends our adopted distance. Contours of column density are shown in Figures \ref{f7}, \ref{f8}, \ref{f9}, \ref{f10}, \ref{f11}, and \ref{f12}. To calculate the mass of the gas in each region, we use
\begin{equation}
M = \mu m_{p} D^{2} \Omega \sum N(\textrm{H}_{2}) = 0.16 \left( \dfrac{D}{\textrm{kpc}} \right)^{2} \left( \dfrac{\Omega}{484\textrm{ ss}} \right) \dfrac{\int T_{B}~dv}{\textrm{K km s}^{-1}}~M_{\odot} \textrm{,}
\end{equation}
where $\mu$ is the mean molecular weight in multiples of the proton mass (assumed to be 2.8), D is the distance to the region, and $\Omega$ is the solid angle occupied by the gas (484 square arcseconds is the solid angle of one pixel in the BU-GRS data cubes).

Next, we identify individual clumps within the molecular cloud structure using the Clumpfind code \citep{1994ApJ...428..693W} that identifies local maxima in the data cubes and grows the clumps outward, down to lower evenly spaced signal levels until the noise floor is reached. We use the recommended value of twice the rms for setting both the noise floor and the interval between adjacent contour levels. The result is a catalog of ${}^{13}$CO (1-0) clumps in position-position-velocity space, with size (full width at half maximum) measurements in galactic longitude ($\Delta \ell_{\textrm{FWHM,cl}}$), galactic latitude ($\Delta b_{\textrm{FWHM,cl}}$), and velocity ($\sigma_{\textrm{cl}}$); we use the subscript ``cl'' to refer to clumps identified by Clumpfind. Clumpfind also calculates an effective radius, $R_{\textrm{cl}}$, which is the radius of a circle that has the same solid angle on the sky as the clump, though the clump may itself be irregularly shaped.

There are a large number of unresolved or barely resolved clumps in this catalog that are near the noise floor, which are probably not real clumps. Clumpfind may provide several false positives in complex regions, so we apply additional ``quality control'' cuts to the list of clumps as follows. We first merge the clumps that have antenna temperature peaks in the data cube within 22'' (1 pixel) of each other in $\ell$ or $b$ and within $(\sigma_{\textrm{cl},i}+\sigma_{\textrm{cl},j})/2$ of each other, where the indices $i$ and $j$ correspond to two clumps. Then, we remove the clumps with $\Delta \ell_{\textrm{FWHM,cl}}$ or $\Delta b_{\textrm{FWHM,cl}}$ less than 66'' (3 pixels), or $\sigma_{\textrm{cl}}$ less than 0.6 km s${}^{-1}$ (3 channels), and all of the clumps that have an average antenna temperature below 3 times the rms of the data. These criteria were determined to (1) produce a final catalog of high confidence clumps and (2) to balance the effects of Clumpfind's tendency to identify extraneous clumps in complicated data sets with the unintended consequence of merging clumps that are truly separate structures. Some of our H {\footnotesize II} regions still show a large number of clumps, but we consider them to be plausibly distinct structures in the data cubes. For the remaining reliable catalog, we computed the mass, peak column density, average number density, and nearest neighbor (peak-to-peak) separation in the plane of the sky.

A plot of the masses and effective radii of all the clumps of molecular gas identified in our sample is presented in Figure \ref{f14}. The distribution we find is consistent with the mass-size relation found empirically in 7 molecular clouds by \cite{2010ApJ...716..433K}, though our clumps do not reach the smallest scales as they do for the closer regions presented in that study. \cite{2003AJ....126..286R} observed ${}^{13}$CO in 30 young stellar clusters within 1 kpc. The molecular clumps we identify in G041.91-0.12 and G041.92+0.04 are consistent with the range of cloud masses and radii seen in that study (marked in Figure \ref{f14}), however we likely cannot resolve the fragments associated with individual clusters in the more distant regions. We interpret the agreement with \cite{2010ApJ...716..433K} as an indication that our quality control cuts are sufficient to remove most spurious Clumpfind detections. However, our inability to resolve parsec-scale clumps must be considered when comparing the observed properties of these clumps to those predicted during collect and collapse.

As noted by \cite{2005A&A...433..565D}, dense molecular gas forming part or all of a shell can be an indication of collect and collapse triggered star formation, particularly if the shell is composed of dense fragments of gas. We calculated the column densities and masses of each individual clump as outlined above. To get the number density of the gas, $n(\textrm{H}_{2})$, we treat each clump as a sphere with a radius that is equal to the effective radius, $R_{\textrm{cl}}$. We also calculate the virial parameter,
\begin{equation}
\alpha_{\textrm{vir}} = \dfrac{M_{\textrm{vir}}}{M_{\textrm{cl}}} = \dfrac{5 \sigma_{\textrm{cl}}^2 R_{\textrm{cl}}}{G M_{\textrm{cl}}} \textrm{,}
\end{equation}
for each clump. A virial parameter less than 1 indicates that the clump is likely to collapse under self-gravity. This assumes that the clumps are spherically symmetric and isothermal, which is clearly not the case, so these values should be viewed with caution. The clumps typically have $\alpha_{\textrm{vir}} \approx 1$, with 23\% of all clumps prone to collapse.

All of the measured and calculated parameters for individual molecular clumps are presented in Table \ref{t6}, with a summary of median values and region-wide parameters in Table \ref{t7}.

\section{Results \label{sec-results}}

\subsection{Assessment of Triggered Star Formation \label{sec-trigger}}

There are two primary criteria we have checked to see if star formation is plausibly triggered by the collect and collapse process. The first is that we expect an enhanced population of Stage I YSOs on or within the infrared bright bubbles and rims surrounding the H {\footnotesize II} regions (a similar enhancement is expected from radiatively driven implosion as well). Second, \cite{1994MNRAS.268..291W} predicted the fragmentation timescale and size, column density, mass, and separation of the typical fragments of the molecular gas comprising the expanding spherical shell of the H {\footnotesize II} region. Assuming a single (or compact) ionizing source and an initially uniform number density of gas, these values are given by:
\begin{equation} \label{eq:Whit94t}
t_{\textrm{frag}} \approx 1.56 \left( \dfrac{a_{s}}{0.2 \textrm{ km s}^{-1}} \right)^{7/11} \left( \dfrac{Q_{\textrm{Ly}}}{10^{49} \textrm{ s}^{-1}} \right)^{-1/11} \left( \dfrac{n_{i}}{10^{3} \textrm{ cm}^{-3}} \right)^{-5/11} \textrm{Myr,}
\end{equation}
\begin{equation} \label{eq:Whit94R}
R_{\textrm{frag}} \approx 5.8 \left( \dfrac{a_{s}}{0.2 \textrm{ km s}^{-1}} \right)^{4/11} \left( \dfrac{Q_{\textrm{Ly}}}{10^{49} \textrm{ s}^{-1}} \right)^{1/11} \left( \dfrac{n_{i}}{10^{3} \textrm{ cm}^{-3}} \right)^{-6/11} \textrm{pc,}
\end{equation}
\begin{equation} \label{eq:Whit94N}
N_{\textrm{frag}} \approx 6.0 \times 10^{21} \left( \dfrac{a_{s}}{0.2 \textrm{ km s}^{-1}} \right)^{4/11} \left( \dfrac{Q_{\textrm{Ly}}}{10^{49} \textrm{ s}^{-1}} \right)^{1/11} \left( \dfrac{n_{i}}{10^{3} \textrm{ cm}^{-3}} \right)^{5/11} \textrm{cm}^{-2}\textrm{,}
\end{equation}
\begin{equation} \label{eq:Whit94M}
M_{\textrm{frag}} \approx 23 \left( \dfrac{a_{s}}{0.2 \textrm{ km s}^{-1}} \right)^{40/11} \left( \dfrac{Q_{\textrm{Ly}}}{10^{49} \textrm{ s}^{-1}} \right)^{-1/11} \left( \dfrac{n_{i}}{10^{3} \textrm{ cm}^{-3}} \right)^{-5/11} M_{\odot} \textrm{, and}
\end{equation}
\begin{equation} \label{eq:Whit94s}
d_{\textrm{frag}} \approx 0.83 \left( \dfrac{a_{s}}{0.2 \textrm{ km s}^{-1}} \right)^{18/11} \left( \dfrac{Q_{\textrm{Ly}}}{10^{49} \textrm{ s}^{-1}} \right)^{-1/11} \left( \dfrac{n_{i}}{10^{3} \textrm{ cm}^{-3}} \right)^{-5/11} \textrm{pc,}
\end{equation}
where $a_{s}$ is the sound speed in the neutral gas, $Q_{\textrm{Ly}}$ is the Lyman continuum (ionizing) luminosity in photons s${}^{-1}$, $n_{i}$ is the initial density of the molecular gas before H {\footnotesize II} region expansion, $t_{\textrm{frag}}$ is the timescale for fragmentation to begin, $R_{\textrm{frag}}$ is the radius of the fragments, $N_{\textrm{frag}}$ is the column density of the fragments, $M_{\textrm{frag}}$ is the mass of the fragments, and $d_{\textrm{frag}}$ is the separation of fragments. We use the term ``clump'' to refer to collections of gas identified in the data by Clumpfind, and ``fragment'' to refer to theoretical collections of molecular gas predicted by \cite{1994MNRAS.268..291W}. Observing molecular cloud clumps consistent with the quantities predicted in Equations \ref{eq:Whit94t}, \ref{eq:Whit94R}, \ref{eq:Whit94N}, \ref{eq:Whit94M}, and \ref{eq:Whit94s} indicates that collect and collapse is at least plausible in a particular region.

\cite{2007MNRAS.375.1291D} performed SPH simulations of expanding H {\footnotesize II} regions to test the validity of this analytical model. They found that fragmentation in the expanding shell did occur and that the time for fragmentation agreed with the prediction of \cite{1994MNRAS.268..291W} to within 20\%. Furthermore, they found that the fragment masses were approximately half of the values predicted by the analytical model.

The predicted values of \cite{1994MNRAS.268..291W} all depend on three parameters: the sound speed, the ionizing luminosity, and the initial density. The sound speed of neutral gas is expected to vary in the range 0.2 km s${}^{-1}$ - 0.6 km s${}^{-1}$, and so without a method of measuring this parameter we assume a value of 0.2 km s${}^{-1}$ \citep{1994MNRAS.268..291W,2011ApJ...741..110D}. The predicted fragment masses can vary by an order of magnitude because of a factor of 2 uncertainty in the sound speed, whereas the other predicted molecular fragment properties are fairly insensitive to this uncertainty. We have the ionizing luminosity measurements with uncertainties from analyzing the 21 cm radio continuum. To estimate the initial density of the gas before H {\footnotesize II} region expansion, we calculate the average number density in the bubble region using the analysis from \S\ref{sec-molecular} to calculate the mass integrated over the gas apparently associated with the bubble or rim (i.e. the mass in the shell exterior to the ionization front). We integrate over the full velocity range of the associated emission and within an irregular aperture determined by eye, though using the distribution of positions in the molecular clump catalog as a guide. We then follow the method of \cite{2011A&A...525A.132P} to estimate the volume by assuming the thickness along the line of sight is equal to the radius of the region, $R_{\textrm{HII}}$. If the gas is being collecting by an expanding shell, then the \emph{average} density in this volume now should still be equal to the average density before expansion began (i.e. the same amount of gas in the same volume, but distributed differently). Assuming 25\% uncertainty in the angular size of the regions and using the uncertainties stated in \S\ref{sec-molecular}, the initial densities of molecular gas are uncertain to within a factor of 3.6 larger or smaller than our quoted values. We assume uncertainty of a factor of 2 in the sound speed, thus the uncertainties in our collect and collapse predictions for the formation time, size, column density, mass, and separation of molecular clumps are approximately factors of 2.4, 2.5, 2.2, 5.4, and 3.43, respectively.

We estimate the ages of the H {\footnotesize II} regions using a dynamical age from \cite{1980pim..book.....D}, assuming spherical expansion:
\begin{equation}
t_{\textrm{HII}} = 7.2 \times 10^{4} \left( \dfrac{R_{\textrm{HII}}}{\textrm{pc}} \right)^{4/3} \left( \dfrac{Q_{\textrm{Ly}}}{10^{49} \textrm{ s}^{-1}} \right)^{-1/4} \left( \dfrac{n_{i}}{10^{3} \textrm{ cm}^{-3}} \right)^{-1/2} \textrm{yr,}
\end{equation}
where $R_{\textrm{HII}}$ is the radius of the region, $Q_{\textrm{Ly}}$ is the ionizing luminosity, and $n_{i}$ is the initial number density of the gas. We compare these ages with the fragmentation timescales for collect and collapse from \cite{1994MNRAS.268..291W} to see if they are consistent with the collect and collapse scenario, $t_{\textrm{HII}}/t_{\textrm{frag}} \ge 1$. Following the uncertainties above, these ages may be a factor of 3 larger or smaller than our quoted values. Quantitative results are in \S\ref{sec-G28}, \S\ref{sec-G41.1}, \S\ref{sec-G41.91}, \S\ref{sec-G41.92}, \S\ref{sec-G44.28}, and \S\ref{sec-G44.34}.

\cite{2011ApJ...736..142B} modeled radiatively driven implosion through simulations of ionizing flux permeating into a molecular cloud. They find that for a 5 $M_{\odot}$ Bonnor-Ebert sphere, star formation is triggered by the radiation field if the ionizing flux, $\Phi_{\textrm{Ly}}$, meets the criterion $10^{9} \lesssim \Phi_{\textrm{Ly}} \lesssim 3 \times 10^{11}$ cm${}^{-2}$ s${}^{-1}$. The first stars then form when the age of the H {\footnotesize II} region is approximately
\begin{equation}
t_{\star} \approx 0.19 \left( \dfrac{\Phi_{\textrm{Ly}}}{10^{9} \textrm{~cm}^{-2} \textrm{~s}^{-1}} \right)^{-1/3} \textrm{ Myr.}
\end{equation}
We estimate the ionizing flux $\Phi_{\textrm{Ly}}$ from the ionizing luminosity, $Q_{\textrm{Ly}}$, and the size of the regions. We are then able to predict whether YSOs may have been formed from radiatively driven implosion by taking the ratio of the dynamical age of the H {\footnotesize II} region and the time for RDI to begin: $t_{\textrm{HII}}/t_{\star} \ge 1$. We note, however, that we used the current sizes of the regions to calculate $\Phi_{\textrm{Ly}}$, while the size would have been smaller at any time in the past. Furthermore, sites of triggering may be closer to the ionizing sources than the rims, so our $t_{\star}$ values will be upper limits, whereas the values of $\Phi_{\textrm{Ly}}$ and $t_{\textrm{HII}}/t_{\star}$ will only be lower limits.

Due to the nature of radiatively driven implosion, quantitative predictions of the outcome of this process require detailed knowledge of the molecular gas before the expansion of the ionization front. This fact, combined with our ability to only place limits on the timescale for YSO formation, make it impossible to make strong statements about the contribution from radiatively driven implosion in our sample. We can only say that all of our regions are at least consistent with this scenario.

\subsection{Results of Individual H {\footnotesize II} Regions \label{sec-HII}}

\subsubsection{G028.83-0.25 (N49) \label{sec-G28}}

As seen in Figure \ref{f13} and previously mentioned, G028.83-0.25 (coincident with IRAS 18421-0348) has two velocity components in ${}^{13}$CO (1-0) possibly associated with the region, both part of larger structures. The components, located at about 87 and 95 km s${}^{-1}$, are on either side of the recombination-line velocity, 90.6 km s${}^{-1}$, from \cite{2009ApJ...690..706A}. It may be that these two components are the front and back of an expanding shell of molecular gas, or they may be two unrelated clouds and either one could be association with the infrared bubble. We favor the 87 km s${}^{-1}$ component as the morphology in the integrated map better matches the infrared rim, but we present results below from analyzing the two components separately.

The infrared bubble is nearly circular and shows no indication of cometary morphology. The region is 2.'8 across, corresponding to 2 pc at the near kinematic distance. The radio continuum seen by VGPS is highly peaked at the center of the infrared bubble, with a flux density of $\sim$ 1 Jy. We calculate the ionizing luminosity necessary to power the region, $Q_{\textrm{Ly}}$, to be $10^{48.3}$ photons s${}^{-1}$, which corresponds to a spectral type O8-O9, following \cite{2002MNRAS.337.1309S} (assuming solar metallicity and luminosity class V).

\cite{2010A&A...523A...6D} analyzed this region at 870 $\mu$m and reported a ``half ring'' of material with massive clumps coincident with the infrared rim. They concluded that this region is a good candidate for triggered star formation, specifically collect and collapse. Furthermore, they determined the mass in the dense shell to be 4200 $M_{\odot}$, with clumps of 2300, 350, 240, and 190 $M_{\odot}$. \cite{2010A&A...523A...6D} also reported an ionizing luminosity of $10^{48.48}$ photons s${}^{-1}$ (corresponding to an O7 V - O7.5 V star) based on MAGPIS 20 cm radio continuum data. MAGPIS data have 6'' resolution images made from VLA B, C, and D array and Effelsburg 100 m observations \citep{2006AJ....131.2525H}.

\cite{2008AJ....136.2391C} found an Extended Green Object (EGO), G028.83-0.25, located on the southern portion of the bright rim ($(\alpha,\delta)_{\textrm{J}2000} = (18^{h}44^{m}51.3^{s}, -03^{\circ}45'48'')$) (in this paper, we use the terms north, south, east, and west defined so that ``north'' describes the direction of increasing galactic latitude, and ``west'' describes the direction of increasing galactic longitude). EGOs are extended objects that are bright in the 4.5 $\mu$m IRAC band and are thought to be the result of jets or other outflow activity from a protostar \citep[e.g.][]{2010AJ....140..196D,2010ApJ...714..469Y}. The emission is expected to be from shocked H${}_{2}$ in the outflow. Additionally, this site is coincident with Class I and Class II methanol maser emission in the velocity range 79.5 - 92.67 km s${}^{-1}$ \citep{2009ApJ...702.1615C}, consistent with the lower velocity ${}^{13}$CO (1-0) component. The EGO does not correspond to a YSO because there is not a source detectable in enough wavebands to be fit via our method (it is either not seen or only appears as an extended object larger than our aperture in our wavebands). A second nearby EGO, G28.28-0.36, is identified by \cite{2009ApJ...702.1615C} who, along with \cite{1998MNRAS.301..640W}, report Class II methanol maser emission coincident with this EGO and with velocities consistent with the higher velocity ${}^{13}$CO (1-0) component. We describe these EGOs and masers here because they are evidence of massive star formation on the rim of this region. We do not identify YSOs at these locations, as no point source is visible in this portion of the rim longward of 4.5 $\mu$m. The absence of these two objects from the sample is not evidence that they do not contain protostars, but rather is consistent with heavily embedded MYSOs. We note that there is one Stage I YSO identified approximately 10'' to the west of the EGO.

Shown in Figure \ref{f7}, the concentration of Stage I (unevolved) YSOs peaks on the infrared bubble. The peak density is more than three times the density of Stage I YSOs surrounding the region. The overall distribution of YSOs follows features in the molecular gas distribution from each velocity component.

\cite{2008ApJ...681.1341W} analyze the structure and YSO population of G028.83-0.25 as part of their sample of 3 bubbles. They report an ionizing luminosity of $10^{48.89}$ photons s${}^{-1}$, almost four times as large as our value. This is in part due to their adoption of a slightly larger distance to the region (5.7 kpc), and likely also because they use the MAGPIS 20 cm survey to measure the radio continuum. The MAGPIS data have 6'' angular resolution \citep{2006AJ....131.2525H}, better than the VGPS images used in this work (we use the VGPS because it covers our entire sample). Due to the weak dependence on ionizing luminosity, the collect and collapse predictions change by less than 15\% from a factor of 4 difference in $Q_{\textrm{Ly}}$ (see Equations \ref{eq:Whit94t}, \ref{eq:Whit94R}, \ref{eq:Whit94N}, \ref{eq:Whit94M}, and \ref{eq:Whit94s}). \cite{2008ApJ...681.1341W} note that the local minimum in 24 $\mu$m emission at the center of the bubble suggests that there is a central wind from the ionizing star evacuating the dust in the central region. \cite{2008ApJ...681.1341W} also identify the likely ionizing source, an O5 V star coincident with the bubble center and the 24 $\mu$m minimum. They additionally report 7 YSOs in the immediate vicinity of this region using the SED fitting method presented here. Our sample of YSOs includes 5 of the YSOs reported in \cite{2008ApJ...681.1341W}, and the reported physical parameters are generally in good agreement. We find discrepant values for one source in particular, the YSO G28.8299-00.2532. \cite{2008ApJ...681.1341W} report a mass of 29 $M_{\odot}$ and accretion rate of $8.9 \times 10^{-4} M_{\odot}$ yr${}^{-1}$, compared to our values of 6 $M_{\odot}$ and accretion rate of $8.0 \times 10^{-5} M_{\odot}$ yr${}^{-1}$. The most likely reason for this discrepancy is that \cite{2008ApJ...681.1341W} specifically decided to use a lower limit on the 24 $\mu$m photometry for this source, whereas our method applied an upper limit at a different value. This YSO is in a region of high, nonuniform background, so 24 $\mu$m photometry is not straightforward.

\cite{2010ApJ...713..592E} modeled the dust distribution in this region with the Cloudy software package and simulated the 24 $\mu$m emission. They were able to match the observations using a model of a wind-blown bubble (WBB), providing further evidence that a stellar wind is at work in the central cavity. Their model is consistent with an age of 0.5-1 Myr. \cite{1994MNRAS.268..291W} give a different set of equations to predict the properties of the molecular gas in regions triggered by stellar winds than by expanding H {\footnotesize II} regions. It is difficult to assess the relative contribution of these two reasons for region expansion with currently available data, particularly without a way to measure the power in the stellar wind. We therefore proceed with the analysis for an expanding H {\footnotesize II} for all of our sources. We do note that for G028.83-0.25, the estimated dynamical age ignoring the effects of wind agrees with the age from the WBB model of \cite{2010ApJ...713..592E}, so our analysis is still viable when viewed with caution.

The identification of clumps of molecular gas and the characterization of this gas is more complicated for G028.83-0.25 than any other H {\footnotesize II} region in our sample. The two velocity components that lie along the line of sight to this region are marginally resolved in velocity in the BU-GRS data. In Table \ref{t6}, one can see that over half of the clumps in each component have virial parameters indicating likelihood to collapse under self-gravity. Coincidentally, the average density of the gas over the bubble region is nearly the same for both velocity components, so our estimates of the dynamical age and expected fragment parameters for collect and collapse are largely unaffected by our choice of component. If instead we included the total emission from both components but assumed the same volume occupied by the gas, the collect and collapse predictions and dynamical age of the region would change by less than a factor of 1.4; $N_{\textrm{frag}}$ and $t_{\textrm{HII}}$ would increase, while the other quantities would decrease.

The ages of the regions using \cite{1980pim..book.....D} are presented in Table \ref{t7}, and the predicted timescales and molecular fragment properties from \cite{1994MNRAS.268..291W} and \cite{2011ApJ...736..142B} are given in Table \ref{t8}. We find that G028.83-0.25 has a dynamical age of about 0.8 Myr and has a total mass in molecular gas of at least a few times $10^{4}~M_{\odot}$. The dynamical age depends on the physical size of the region, the initial density, and the Lyman continuum luminosity. The mass is dependent on the distance, the integrated ${}^{13}$CO intensity, and assumed values of the excitation temperature and the conversion factor $N(\textrm{H}_{2})/N({}^{13}\textrm{CO}) = 5 \times 10^{5}$ \citep{2001ApJ...551..747S}. The mass we report is several times that reported by \cite{2010A&A...523A...6D}, however they focused on a significantly smaller region immediately around the infrared rim (which we cannot probe with the resolution of the BU-GRS data).

For either velocity component, we calculate that the dynamical age of the region, 0.79 or 0.83 Myr for the low or high velocity component, respectively, is within uncertainty of being consistent with the timescale for collect and collapse to begin, 0.9 Myr for both components. As happens to be true of all of our regions, the limits on the formation timescales for YSOs triggered by radiatively driven implosion, less than 0.08 Myr in the case of either velocity component of G028.83-0.25, are consistent with the dynamical ages of the H {\footnotesize II} regions. For this region, we predict that molecular fragments experiencing collect and collapse should be approximately 2 pc in radius, have column densities of about $1 \times 10^{22}$ cm${}^{-2}$, be 13 $M_{\odot}$, and be separated by 0.5 pc. The median values of the observed molecular clumps for each of the two velocity components are about 3 pc in radius (for both components) and have column densities of $2.8 \times 10^{21}$ cm${}^{-2}$ and $6.5 \times 10^{21}$ cm${}^{-2}$, 1200 and 1500 $M_{\odot}$, and have median separations of 1.8 pc and 1.7 pc for the high and low velocity components, respectively. For both velocity components the dynamical ages and median clump sizes are consistent with collect and collapse predictions. The median column density of clumps in the lower velocity component is also consistent with collect and collapse. The median separation of clumps in each component is approximately at the edge of the uncertainty range for being consistent with the collect and collapse prediction. These values are presented in Table \ref{t7}.

\cite{2010A&A...518L.101Z} presented a case study of G028.83-0.25 using \emph{Herschel}-PACS and -SPIRE data from the Hi-GAL survey, in addition to GLIMPSE, MIPSGAL, and ATLASGAL 870 $\mu$m data, to investigate this region as a candidate for triggering. They identified four condensations at 870 $\mu$m, and applied the same SED fitter and YSO models used here, though using apertures 40-100'' in size and only employing \emph{Herschel} photometry. Four of our YSOs (as well as the EGO identified by \cite{2008AJ....136.2391C}) are coincident with three of their condensations. We determine all four of these YSO to be Stage I, with masses 1.5-6.2 $M_{\odot}$. They further use \cite{1994MNRAS.268..291W} to estimate the parameters $t_{\textrm{frag}}$, $R_{\textrm{frag}}$, and $N_{\textrm{frag}}$ for collect and collapse as 0.5 Myr, 1.55 pc, and $1.6 \times 10^{22}$ cm${}^{-2}$, respectively. These values agree with ours, taking into account that \cite{2010A&A...518L.101Z} used a slightly larger distance (5.5 kpc). They also conclude that this region would be better evaluated using a model accounting for the dynamics of the apparent stellar wind. 

\cite{2010ApJ...709..791B} performed a study of several infrared bubbles, including G028.83-0.25, G041.92+0.04, and G044.34-0.82, using JCMT CO (3-2) and MAGPIS survey 20 cm emission. For G028.83-0.25, they analyze the molecular gas at 87.5 $\pm$ 3.1 km s${}^{-1}$. They list the size of the bubble as 1.77 $\pm$ 0.43 pc, the 20 cm flux as 0.985 Jy, and an ionizing luminosity of $10^{48.21}$ photons s${}^{-1}$. These quantities are all consistent with ours.

\subsubsection{G041.10-0.15 \label{sec-G41.1}}

G41.1-0.15 is a cometary region in 8 $\mu$m emission, and spans 6 pc at 4 kpc. The morphology of the molecular gas very closely follows the 8 $\mu$m rim, but also extends to the east. The radio continuum peaks very close to the infrared rim, and has a total flux density of 5.5 Jy. We calculate the ionizing luminosity necessary to power the region, $Q_{\textrm{Ly}}$, to be $10^{48.8}$ photons s${}^{-1}$, which corresponds to a spectral type O7-O7.5, following \cite{2002MNRAS.337.1309S}. The supernova remnant 3C 397 (also known as SNR G041.1-00.3) is seen in the 24 $\mu$m and radio continuum images to the south of the region. At a distance of 10.3 kpc \citep{2010ApJ...712.1147J}, it is unrelated to G041.10-0.15.

One might naively assume that the very bright point source present near the center of the bubble is the star powering this region; in fact this region shows a minimum in the 24 $\mu$m emission around this source, which may be evidence of a central stellar wind as in G028.83-0.25. However, this star is identified as V844 Aql, an M6 variable AGB star \citep{2000A&A...355L..27H}. We verified this classification with spectral observations performed with the Fan Observatory Bench Optical Spectrograph at the Fan Mountain Observatory operated by the University of Virginia. V844 Aql is thus most likely in the foreground and unrelated to the region of interest. G041.10-0.15 may contain a wind-blown bubble (WBB), though the true ionizing source may be obfuscated by V844 Aql. The morphology of the 24 $\mu$m emission may also be explained by the cometary nature of the region.

G041.10-0.15 has a very clearly enhanced population of Stage I YSOs around and in the bubble region (Figure \ref{f8}). The area of enhanced YSO density follows the molecular gas in general, however the Stage I YSOs show the greatest density within the bubble only and not as greatly in the extended molecular gas. The collection of YSOs in the eastern portion of the bubble shows preferentially less evolved YSOs compared to the field YSOs in this region. There are multiple 24 $\mu$m point sources located around the rim that were not detected at shorter wavelengths and thus not identified as YSOs by our SED fitting, but are candidate embedded protostars. In addition, there are two infrared dark clouds (IRDCs) seen against the emission of the rim; one in the east and one in the south-southwest. We cannot be certain whether or not these clouds are part of the same structure as the infrared rim.

Unlike G028.83-0.25, the molecular gas around the rim is easily separated into distinct clumps with Clumpfind, particularly the gas coincident with the infrared rim (see Table \ref{t6}). The total mass in molecular gas is about $2 \times 10^{4}~M_{\odot}$. There are over 60 distinct clumps, however only about 15\% of them are prone to collapse. The region has a dynamical age of about 2.1 Myr, and we calculate the expected timescale for collect and collapse to begin to be 1.4 Myr. The limit on the timescale for radiatively driven implosion is less than 0.10 Myr. The collect and collapse fragments are predicted to be 4.8 pc in radius and have $6.6 \times 10^{21}$ cm${}^{-2}$ column density, 20.8 $M_{\odot}$, and 0.7 pc separation. The observed clumps have typical radius 1.8 pc, $2.6 \times 10^{21}$ cm${}^{-2}$ column, 202 $M_{\odot}$, and 1 pc separation (see Table \ref{t8}). The dynamical age and separations of clumps are within uncertainties of the values necessary for collect and collapse, but the sizes, column densities, and masses are not.

\subsubsection{G041.91-0.12 \label{sec-G41.91}}

G041.91-0.12 is a great example of a so-called cometary, or blister, H {\footnotesize II} region. The morphology indicates that after some initial expansion it opened on one side, possibly due to ambient gas of lower density on the side of the opening. The opening is currently 0.8 pc across. Nevertheless, G041.91-0.12 is interesting as a potential location for triggered star formation because of its bright rim and very closely matching morphology in molecular gas. The H {\footnotesize II} region is seen immediately to the east of the infrared rim, with a flux density of 0.5 Jy. We calculate the ionizing luminosity necessary to power the region, $Q_{\textrm{Ly}}$, to be $10^{46.9}$ photons s${}^{-1}$, which corresponds to a spectral type B0.5-B1. The image of radio continuum emission (Figure \ref{f9}) confirms that the ionized gas is extended in the direction of the opening.

G041.91-0.12, seen in Figure \ref{f9}, is in close proximity on the sky and in radial velocity to G041.92+0.04 (Figure \ref{f10}), so we surveyed one continuous region from the GLIMPSE point source catalog to search for YSOs and to provide the field sample. We then performed the clump decomposition and evaluated the evidence for triggering in each region separately. Both regions are part of the same larger, diffuse structure of molecular gas that is continuous in position-position-velocity space, though the dense shells where distinct clumps were identified were well separated. The spatial density of Stage I YSOs seen in Figure \ref{f9} peaks on the infrared rim coincident with a molecular gas peak.

Despite being much closer than most of the other regions in our sample, it is still very difficult to resolve distinct clumps in the molecular gas. Only 6 distinct clumps are identified in Table \ref{t6}, with a total mass of less than 200 $M_{\odot}$, and none of them have virial parameters indicating likelihood to collapse. The region has a dynamical age of 0.3 Myr, which is about 1 Myr less than the predicted time for fragments to start experiencing collect and collapse. The limit on the timescale for radiatively driven implosion is less than 0.12 Myr. The fragments are predicted to be 2 pc in radius and have $6.5 \times 10^{21}$ cm${}^{-2}$ column density, 21.3 $M_{\odot}$, and 0.8 pc separation. The observed clumps have typical radius 0.58 pc, $3.4 \times 10^{21}$ cm${}^{-2}$ column, 18 $M_{\odot}$, and 0.7 pc separation (see Table \ref{t7}). The median column density, mass, and separation of the molecular clumps are consistent with collect and collapse.

\subsubsection{G041.92+0.04 (N80) \label{sec-G41.92}}

G041.92+0.04, seen in Figure \ref{f10}, has a round geometry, and is 0.6 pc across. The radio continuum peaks within the bubble, with a total flux density of 0.45 Jy at 21 cm. There is a local minimum in the 24 $\mu$m emission at the center of the bubble, possibly indicating this is a WBB. We calculate the ionizing luminosity necessary to power the region, $Q_{\textrm{Ly}}$, to be $10^{46.8}$ photons s${}^{-1}$, which corresponds to a spectral type B0.5-B1. The molecular gas emission, though relatively weak, shows two spatially separated components around the bubble, one to the northwest and one to the southeast. \cite{2010A&A...523A...6D} reported that the 870 $\mu$m emission shows several clumps located around the bubble, indicating this region is a good candidate for collect and collapse triggering.

A look at the YSO sample reveals that there is only slight evidence that there is a significant YSO population on the rim in Figure \ref{f10}. The enhancement of YSO density coincident with the bubble is weaker than the enhancement to the southeast that is not coincident with strong molecular gas or radio continuum emission (see \S\ref{sec-G41.91}). \cite{2010ApJ...716.1478W} reported that they also did not find a significant YSO population associated with this bubble. Of the four brightest point sources on the rim, three are classified as Stage II YSOs.

Like G041.91-0.12, the molecular gas is not readily identified as distinct clumps. A total of 16 clumps are found, presented in Table \ref{t6}, with a total mass of about 350 $M_{\odot}$, though again none are prone to collapse as determined by their virial parameters. The dynamical age for this region is 0.2 Myr, which is over 1 Myr less than the calculated time for collect and collapse to begin triggering. The limit on the timescale for radiatively driven implosion is less than 0.11 Myr. Collect and collapse fragments are predicted to be 1.9 pc in radius and have $6.5 \times 10^{21}$ cm${}^{-2}$ column density, 21.3 $M_{\odot}$, and 0.8 pc separation. The observed clumps have a median radius 0.67 pc, $2.3 \times 10^{21}$ cm${}^{-2}$ column, 23 $M_{\odot}$, and 0.3 pc separation (see Table \ref{t7}). The median column density, separation, and mass of the molecular gas clumps are within uncertainty of the collect and collapse predictions.

\cite{2010ApJ...709..791B} found that G041.92+0.04 has a size of 0.48 $\pm$ 0.11 pc, a 20 cm flux of 0.254 Jy, and an ionizing luminosity of $10^{46.21}$ photons s${}^{-1}$. We measured a radio continuum flux about a factor of two higher using the VGPS data, but the other quantities are consistent with ours.

\subsubsection{G044.28+0.11 (N91) \label{sec-G44.28}}

G044.28+0.11, seen in Figure \ref{f11}, is among the more interesting regions in the sample. It is the largest in both angular and physical extent. There is a portion of a round, bright rim in the east, but is a cometary H {\footnotesize II} region overall. The radio continuum peaks immediately to the west of the 8 $\mu$m rim, coincident with the 24 $\mu$m emission. The total flux density from VGPS is 1.3 Jy, indicating an ionizing luminosity of $10^{48.3}$ photons s${}^{-1}$, which corresponds to a spectral type O8-O9. The molecular gas is concentrated along the infrared rim, and the region is approximately 5.6 pc across. \cite{2010A&A...523A...6D} noted several 870 $\mu$m condensations, including one coincident with the PDR. They suggest this region is a good candidate for triggering through either collect and collapse or radiatively driven implosion.

Seen in Figure \ref{f11}, the YSO sample clearly shows locations of enhanced densities of Stage I YSOs coincident with the bright rim and the molecular gas. The greatest concentration is in the southeastern portion of the rim, which has 14 Stage I YSOs, though there are also concentrations on the northern portion of the rim and near the end of the southern portion of the rim. The estimated evolutionary stages of the YSOs show that these areas also have systematically less evolved YSOs than the surrounding regions. MSX6C G044.3103+00.0416 is a MYSO previously identified by \cite{2009A&A...501..539U} as part of the RMS survey. They further detected a 6.4 mJy 6 cm continuum source toward this source with the VLA. This source was classified as a Stage I YSO by our YSO fitting (as G044.3102+00.0410), with a mass of 7.5$\pm$1.79 $M_{\odot}$ and luminosity $10^{3.19 \pm 3.17}~L_{\odot}$. The SED fitting determines a distance of 4.48$\pm$0.8 kpc and $A_{V}$=14.05$\pm$5.73. \cite{2007ApJ...656..255P} also report a Class II methanol maser coincident with this location.

There is a lack of YSOs in the center of the infrared rim, neighboring the radio continuum peak and the 24 $\mu$m emission. There is a fairly sharp transition in 8 $\mu$m emission, and there is an ``elephant trunk,'' or pillar, feature in the rim with one faint, moderately extended source near the end of it. This feature is an indication of radiatively driven implosion \citep{1994A&A...289..559L}. The faint point source may be an embedded protostar missed by our SED fitting, and the trunk structure may be formed by the advancing ionization front clearing away gas surrounding the overdensity progenitor to this protostar.

Like G041.10-0.15, a large number of molecular clumps are easily identified by Clumpfind in this region, particularly along the infrared rim (see Table \ref{t6}). A total of 45 clumps are identified, with a total mass of $3 \times 10^{4}~M_{\odot}$, of which 20\% are prone to collapse. The dynamical age of G044.28+0.11 is 2.4 Myr, and the collect and collapse fragmentation timescale is only 1.7 Myr. The limit on the timescale for radiatively driven implosion is less than 0.15 Myr. Collect and collapse fragments are predicted to be 4.5 pc in radius and have $5.6 \times 10^{21}$ cm${}^{-2}$ column density, 24.7 $M_{\odot}$, and 0.9 pc separation. The observed clumps have a median radius of 2.5 pc, column density of $3.7 \times 10^{21}$ cm${}^{-2}$, 534 $M_{\odot}$, and 1.7 pc separation (see Table \ref{t7}). The dynamical age of the region and the median size, separation, and column density of the molecular clumps are consistent with the predicted values for collect and collapse.

\subsubsection{G044.34-0.82 (N92) \label{sec-G44.34}}

G044.34-0.82, seen in Figure \ref{f12}, is 3.4 pc across. It does appear to be a cometary H {\footnotesize II} region, though it is still fairly round and contained. We may be seeing the back wall of an open shell. The radio continuum and molecular gas emission are concentrated on the infrared rim. With a 21 cm flux density of 0.1 Jy, the region is expected to be powered by a B0-B0.5 star. The molecular gas observations show that it is part of the same larger structure as G044.28+0.11. G044.34-0.82 is far enough away in angular extent from G044.28+0.11 that we use separate YSO field samples. \cite{2010A&A...523A...6D} reported 870 $\mu$m emission coincident with the IRDC crossing the region.

Shown in Figure \ref{f12}, there is a greatly enhanced population of YSOs, mostly Stage I, near the center of the region, rather than on the rim. One of the YSOs is coincident with the IRDC. G044.34-0.82 shows the greatest contrast in YSO density between the center of the region and the surrounding field seen anywhere in this study. \cite{2010ApJ...716.1478W} also reported a significant YSO population in this region using the same SED fitting method used here, though without 24 $\mu$m photometry. They identified 7 YSOs within 3 bubble radii of the center; we identify 14 in the same region.

A total of 13 clumps, with 38\% prone to collapse, have a combined mass of $10^{4}~M_{\odot}$ (see Table \ref{t6}). The dynamical age of G044.34-0.82 is 2.1 Myr, while the collect and collapse fragmentation timescale is 1.7 Myr. The limit on the timescale for radiatively driven implosion is less than 0.23 Myr. Collect and collapse fragments are predicted to be 3 pc in radius and have $5.4 \times 10^{21}$ cm${}^{-2}$ column density, 25.5 $M_{\odot}$, and 0.9 pc separation. The observed clumps have a median radius 2.5 pc, $4 \times 10^{21}$ cm${}^{-2}$ column density, 656 $M_{\odot}$, and 1.6 pc separation (see Table \ref{t7}). The dynamical age of the region and the median size, separation, and column density of the molecular clumps are consistent with the predicted values for collect and collapse.

\cite{2010ApJ...709..791B} found that G044.34-0.82 has a size of 2.09 $\pm$ 0.57 pc, but did not measure the radio continuum.

\section{Discussion \label{sec-discussion}}

Every one of our H {\footnotesize II} regions shows at least some enhancement of the density of YSOs, particularly Stage I YSOs, on the infrared bubbles or rims. The YSO density enhancements also often follow the molecular gas. An overdensity of relatively unevolved YSOs on the rims is suggestive of triggering as opposed to spontaneous collapse throughout the cloud. The rims have the highest and most complex diffuse emission complicating the extraction of point source photometry, so it is even more remarkable that so many YSOs can be identified in such areas. Because the ages of individual YSOs have considerable uncertainty, the relative evolutionary states of YSOs are a more robust measure than the absolute ages. Images of the YSO density maps are in Figures \ref{f7}, \ref{f8}, \ref{f9}, \ref{f10}, \ref{f11}, and \ref{f12}. All of our regions are consistent with radiatively driven implosion, though we cannot be more precise given that we can only put limits on the expected ionizing fluxes and timescales. We do however note the morphology on the infrared rim/PDR boundary in G044.28+0.11 suggestive of radiatively driven implosion.

For collect and collapse models, the observed values that match the theoretical predictions are summarized in Table \ref{t9}. We first consider the dynamical ages, the time required for fragments to begin to form YSOs, and the molecular gas fragment sizes. We note that, when considering the uncertainties in the parameters, only G041.91-0.12 and G041.92+0.04 are too young to have collect and collapse triggered star formation. G041.91-0.12 is decidedly not a round H {\footnotesize II} region, so the quantitative predictions of fragment properties we perform should be viewed with caution. The opening would cause the intact portion of the region to cease or significantly slow its expansion, thus causing our estimated age to be a lower limit. A change in the expansion speed would also likely change the time for the shell to collect enough neutral gas to begin fragmentation and collapse. Thus we cannot be certain that this region is in actuality too young to experience collect and collapse without a method of age estimation that correctly accounts for the geometry. In regard to the sizes of the molecular clumps, G041.10-0.15, G041.91-0.12, and G041.92+0.04 do not have any clumps physically large enough to be consistent with collect and collapse in the simple model of \cite{1994MNRAS.268..291W}.

Again referring to Table \ref{t9}, only G028.83-0.25 has clump separations not consistent with the predicted separation within our estimated uncertainty, though only just. The peak column densities we observe are consistent with the predicted values within the uncertainties, with the exceptions of G041.10-0.15 and the higher velocity component of G028.83-0.25. The peak column densities in the lower velocity component, however, are consistent with the predictions. Finally, the predicted masses of the clumps for all the regions are in the range 13-26 $M_{\odot}$. Only G041.91-0.12 and G041.92+0.04 have observed clumps consistent with their predicted values; the other regions have clumps much more massive than the predicted values. This is the reverse of the situation with the clump sizes. We note that many of the clumps we identify are at the limit of what we are able to resolve and detect in ${}^{13}$CO (1-0). The 46'' resolution corresponds to 0.3 pc at 1.35 kpc and 1.0 pc at 4.5 kpc, roughly representative of the distances in our sample. It is possible that the physical clumps are actually smaller than what we have identified. This may account for this mass discrepancy, however it is difficult to reconcile this scenario with the clumps appearing too small to be collect and collapse fragments in some regions.

Given this evidence, we conclude the following. G041.10-0.15, G041.91-0.12, G044.28+0.11, and G044.34-0.82 are good candidates for triggered star formation. G044.28+0.11 is almost certainly experiencing ongoing radiatively driven implosion in the center of its infrared bright rim, if not also collect and collapse around the rest of the edges of the region. G041.10-0.15 and G044.34-0.82 show great enhancement of unevolved YSOs around the bubble region. Seen in Table \ref{t9}, G044.28+0.11 is consistent with all the predictions for collect and collapse fragments from \cite{1994MNRAS.268..291W} except the masses. G041.10-0.15 is only consistent with the timescales and fragment separations, which might indicate that radiatively driven implosion is more important in this region. The inconsistency with \cite{1994MNRAS.268..291W} may also be explained by deviations from the assumed, simple geometry. G041.91-0.12 has the least quantitative evidence for collect and collapse of these four, however the incredible match of Stage I YSOs to the infrared rim cannot be ignored, and the excellent example of cometary morphology can easily account for the discrepancy between the predicted and observed parameters. To illustrate this point, we note that \cite{1973ApJ...183..863Z} originally proposed the idea of a blister in the Orion Nebula to resolve an apparent discrepancy between the age of the Trapezium stars and the H {\footnotesize II} region.

G028.83-0.25 appears morphologically to be a good candidate for triggering given the enhanced density of YSOs. The presence of EGOs and several masers coincident with both the infrared rim and a peak in ${}^{13}$CO (1-0) emission is highly suggestive, however the quantitative analysis shows that the molecular gas clumps do not have properties consistent with this scenario. Invoking the two velocity components of molecular gas to increase the density estimate is not sufficient to improve the correspondence between predictions and observations. The resulting increase in $t_{\textrm{HII}}$ and decrease in $t_{\textrm{frag}}$ in that scenario makes the timescales consistent with collect and collapse, however the other predicted physical parameters of the molecular gas fragments would show larger disagreement with observed properties.

Finally, G041.92+0.04 is unlikely to be a good example of triggered star formation. The YSO density enhancement is moderate, and predicted collect and collapse parameters typically do not agree with the observed parameters in this region. Since there is no compelling geometric region to doubt the calculated parameters, it is likely true that this region is not yet old enough to experience collect and collapse. We note that it may in the future, and it is also possible that radiatively driven implosion is responsible for the slight enhancement of YSOs along the infrared rim.

\cite{2010A&A...523A...6D} report that most of the bubbles in their sample that were good candidates for triggering were large and apparently evolved regions. They note that G028.83-0.25 (N49) is a relatively smaller and less evolved region despite being a good candidate for triggering, though this can be explained by a relatively high, homogeneous ambient density of gas into which the region is expanding. This assessment is consistent with our findings. We see the greatest evidence for triggering in regions with cometary morphology, which we interpret as an age effect, consistent with \cite{2010A&A...523A...6D}. This is, however, at odds with \cite{2012MNRAS.421..408T}, who find that bubbles with overdensities of YSOs are systematically the smaller bubbles, which they interpret as the youngest bubbles. It may be that at least some of these bubbles appear small because they are expanding into relatively denser gas, or that they are cometary regions that have slowed their expansion, rather than being younger. Further study, particularly with large samples, is necessary to resolve this issue.

Molecular gas observations with better angular and spectral resolution, thus allowing for more reliable clump decomposition and analysis, would allow for stronger conclusions to be drawn. Also, better estimates of the age of the H {\footnotesize II} region that do not assume a spherical geometry could provide stronger evidence. The results of \cite{2007MNRAS.375.1291D} indicate that the problem is well enough understood that more complex geometries and sets of initial conditions may be explored in simulations.

\section{Conclusions \label{sec-conclusions}}

The importance of triggered star formation is a key open question in understanding star formation on Galactic scales. Many recent studies of triggering have focused on the sample of MIR-bright bubbles identified in the GLIMPSE survey of the Galactic Plane (see \S\ref{sec-intro}); the citizen-science Milky Way Project has recently increased the number of such bubbles cataloged in GLIMPSE by an order of magnitude \citep{2012MNRAS.424.2442S}. With this explosion in the number of candidate triggered regions, it is important to understand whether the presence or absence of triggering around any given H {\footnotesize II} region can be reliably evaluated based on existing survey data.

We have performed SED fitting on a large number of infrared point sources around several H {\footnotesize II} regions, using 2MASS and \emph{Spitzer} near to mid-infrared photometry. We identified 458 objects that are consistent with radiative transfer models of YSOs, but not with stellar atmosphere models or AGB colors. We report properties of the individual candidates, including mass, evolutionary stage, and accretion rate, based on the physical parameters of the best matching model SEDs. The distribution of the YSOs along the bright rims of infrared bubbles as compared to the field populations, as well as their relatively early evolutionary state, provides evidence that triggered star formation is at work. We find that the regions with cometary morphology are the strongest candidates for triggered star formation.

We searched for further evidence of triggered star formation by \emph{quantitatively} comparing the predictions of collect and collapse and radiatively driven implosion triggering models to observations for 6 H {\footnotesize II} regions spanning a range of morphologies. To evaluate the consistency of models and data from as many angles as possible, we combined publicly available MIR, cm continuum, and ${}^{13}$CO (1-0) surveys to constrain the properties of YSOs and ionized and molecular gas. While the results for many of our regions are suggestive of triggering, the data are insufficient to draw firm conclusions about the triggering mechanism(s). Our analysis suggests that to distinguish collect and collapse and radiatively driven implosion in an individual region, additional data and modeling would be required, including: (1) high-resolution molecular line data to resolve clumps; (2) additional long-wavelength data to identify younger and more deeply embedded YSOs and improve SED coverage; and/or (3) models that account for source geometry to better constrain H {\footnotesize II} region ages. While (1) would require dedicated observations for each source of interest, the necessary data for (2) will be provided by \emph{Herschel} survey catalogs, allowing the application of statistical techniques \citep[e.g.][]{2012MNRAS.421..408T,2012ApJ...755...71K} to younger YSOs.

\acknowledgments

This work is based (in part) on observations made with the \emph{Spitzer Space Telescope}, which is operated by the Jet Propulsion Laboratory, California Institute of Technology under a contract with NASA. Support for this work was provided by NASA through an award issued by JPL/Caltech.
Claudia J .Cyganowski is supported by an NSF Astronomy and Astrophysics Postdoctoral Fellowship under award AST-1003134.
This publication makes use of data products from the Two Micron All Sky Survey, which is a joint project of the University of Massachusetts and the Infrared Processing and Analysis Center/California Institute of Technology, funded by the National Aeronautics and Space Administration and the National Science Foundation.
This publication makes use of molecular line data from the Boston University-FCRAO Galactic Ring Survey (GRS). The GRS is a joint project of Boston University and Five College Radio Astronomy Observatory, funded by the National Science Foundation under grants AST-9800334, AST-0098562, \& AST-0100793.
This research has made use of NASA's Astrophysics Data System Bibliographic Services.
This research has made use of the SIMBAD database, operated at CDS, Strasbourg, France.
This research has made use of SAOImage DS9, developed by Smithsonian Astrophysical Observatory.
The National Radio Astronomy Observatory is a facility of the National Science Foundation operated under cooperative agreement by Associated Universities, Inc.
This work was funded in part by the Jefferson Scholars Foundation at the University of Virginia.

\bibliographystyle{apj}
\bibliography{bibtex}

\clearpage

\begin{figure}
\begin{center}
\includegraphics{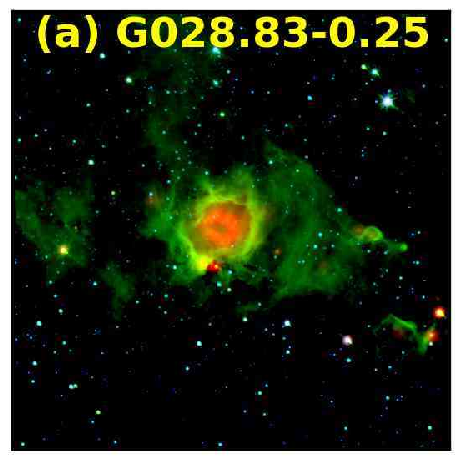}
\includegraphics{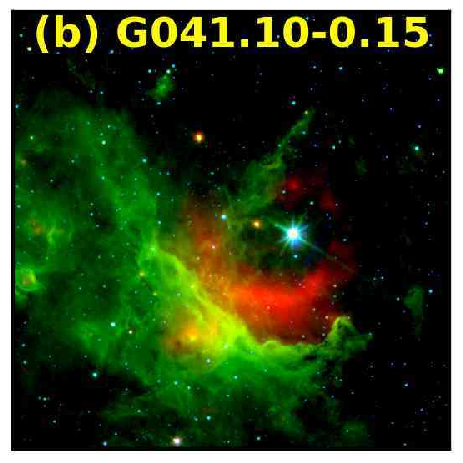}
\includegraphics{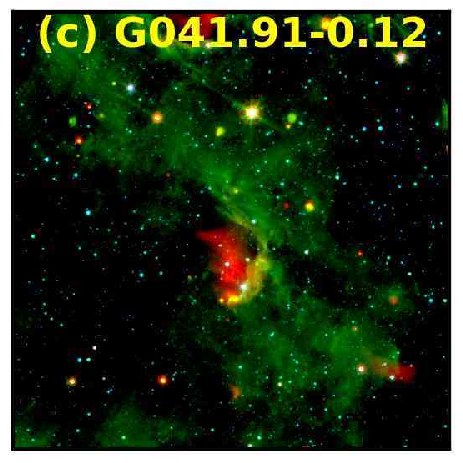}
\includegraphics{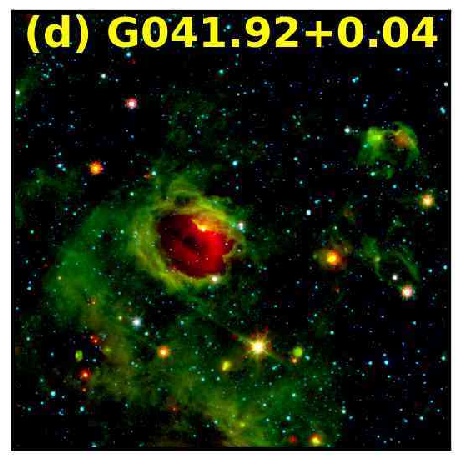}
\includegraphics{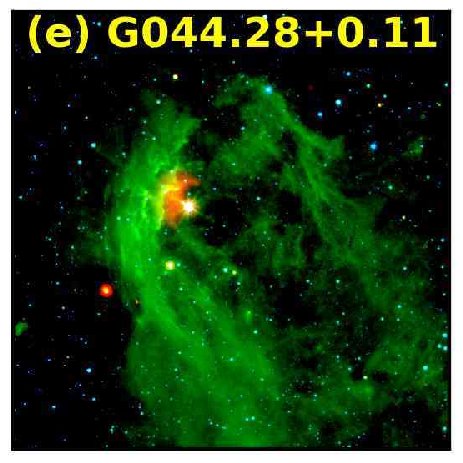}
\includegraphics{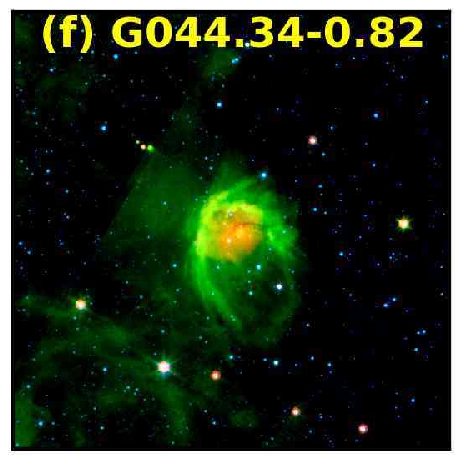}
\caption{H {\footnotesize II} regions in our sample as seen by \emph{Spitzer}, presented on a uniform angular scale 18' across. Red is 24 $\mu$m, green is 8 $\mu$m, and blue is 3.6 $\mu$m. The strong 24 $\mu$m emission located within the bubbles is likely from heated dust grains within the H {\footnotesize II} regions, while the 8 $\mu$m emission along the rims of the bubbles is likely from PAHs. PAHs are destroyed within the H {\footnotesize II} regions, but on the edges are excited by the radiation leaking out of the region. \label{f1}}
\end{center}
\end{figure}

\begin{figure}
\begin{center}
\includegraphics{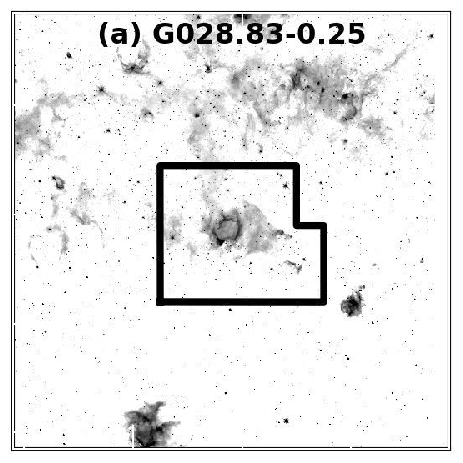}
\includegraphics{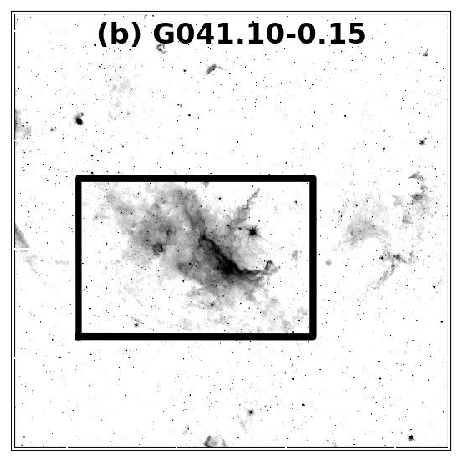}
\includegraphics{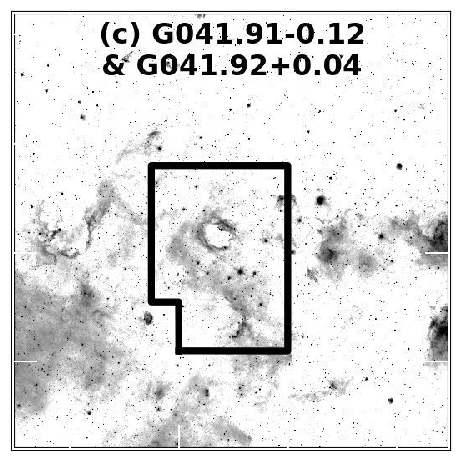}
\includegraphics{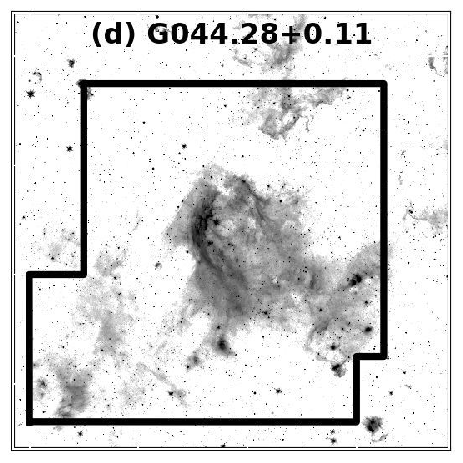}
\includegraphics{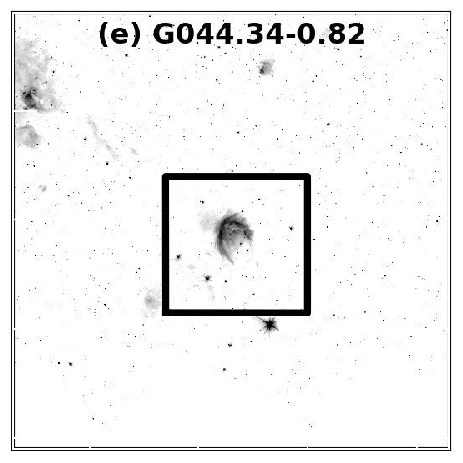}
\caption{H {\footnotesize II} regions in our sample presented at uniform angular scale 48' across. The grayscale images are IRAC 8 $\mu$m. The black boxes outline the areas of the sky over which we took point sources from the GPSC, identified additional point sources, and performed SED fitting to search for YSOs. The regions were chosen to include the entirety of the infrared bubble or rim, the associated molecular emission, and a significant field sample. Coordinates of the bounds of these boxes are given in Table \ref{t2}. \label{f2}}
\end{center}
\end{figure}

\begin{figure}
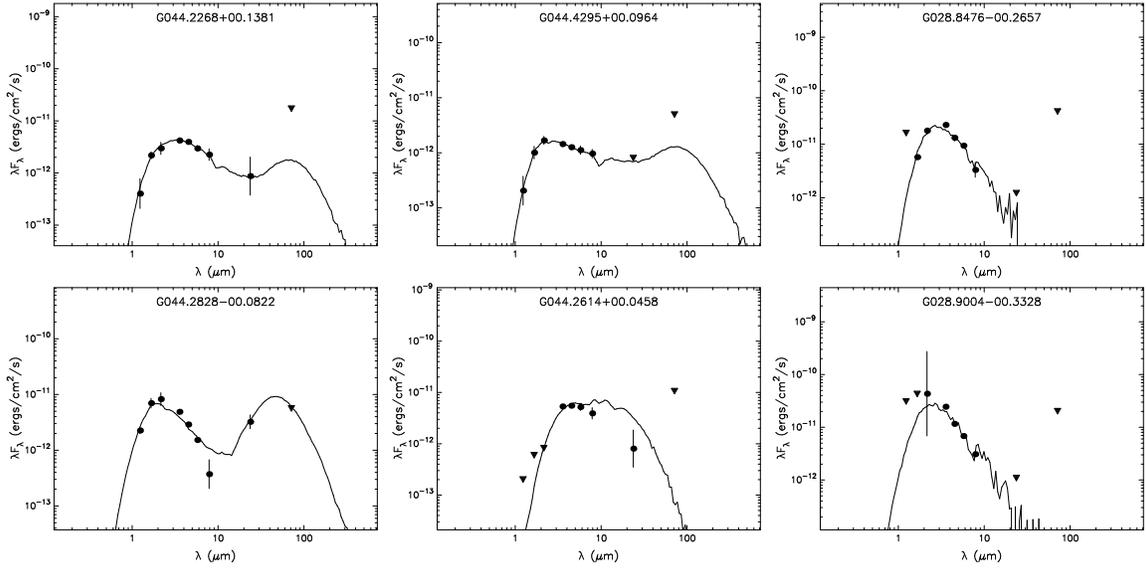

\begin{center}
\includegraphics[width=0.3\textwidth]{f3a.eps}
\includegraphics[width=0.3\textwidth]{f3b.eps}
\includegraphics[width=0.3\textwidth]{f3c.eps}
\includegraphics[width=0.3\textwidth]{f3d.eps}
\includegraphics[width=0.3\textwidth]{f3e.eps}
\includegraphics[width=0.3\textwidth]{f3f.eps}
\caption{Examples of point source SEDs with best fit YSO models. All six objects have been identified as YSOs by our fitting method and were not removed as AGBs following our color cut. Stage I sources are in the left column, Stage II are in the middle column, and Stage III are in the right column. Examples that are typical of the lowest $\chi^{2}/n_{\textrm{data}}$ are in the top row, while examples typical of highest $\chi^{2}/n_{\textrm{data}}$ (marginally acceptable fit) are in the bottom row. Circles with error bars (often too small to see) represent detections, while downward arrows represent upper limits. The solid line is the best fit SED model. \label{f3}}
\end{center}
\end{figure}

\begin{figure}
\begin{center}
\includegraphics[width=0.3\textwidth]{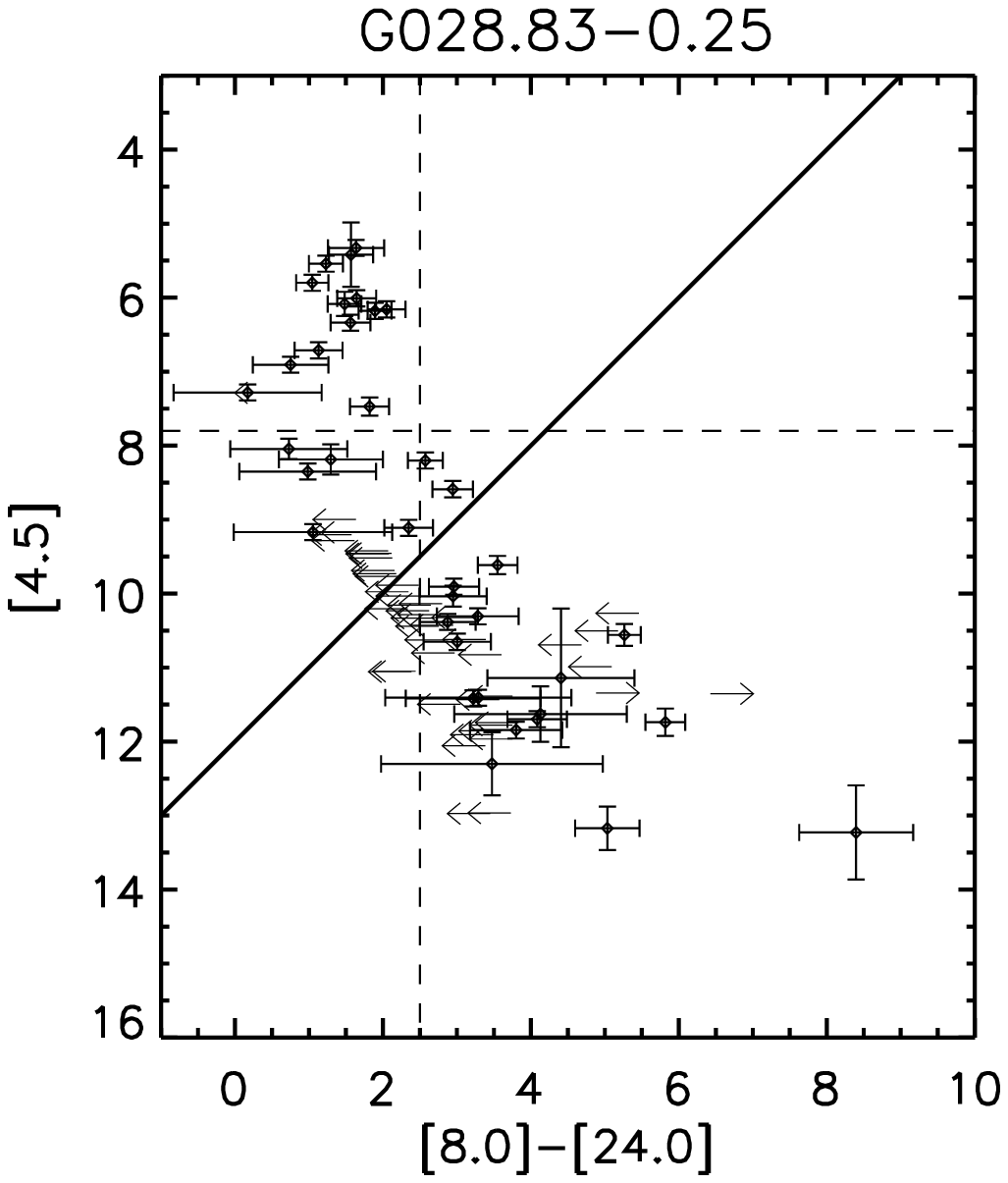}
\includegraphics[width=0.3\textwidth]{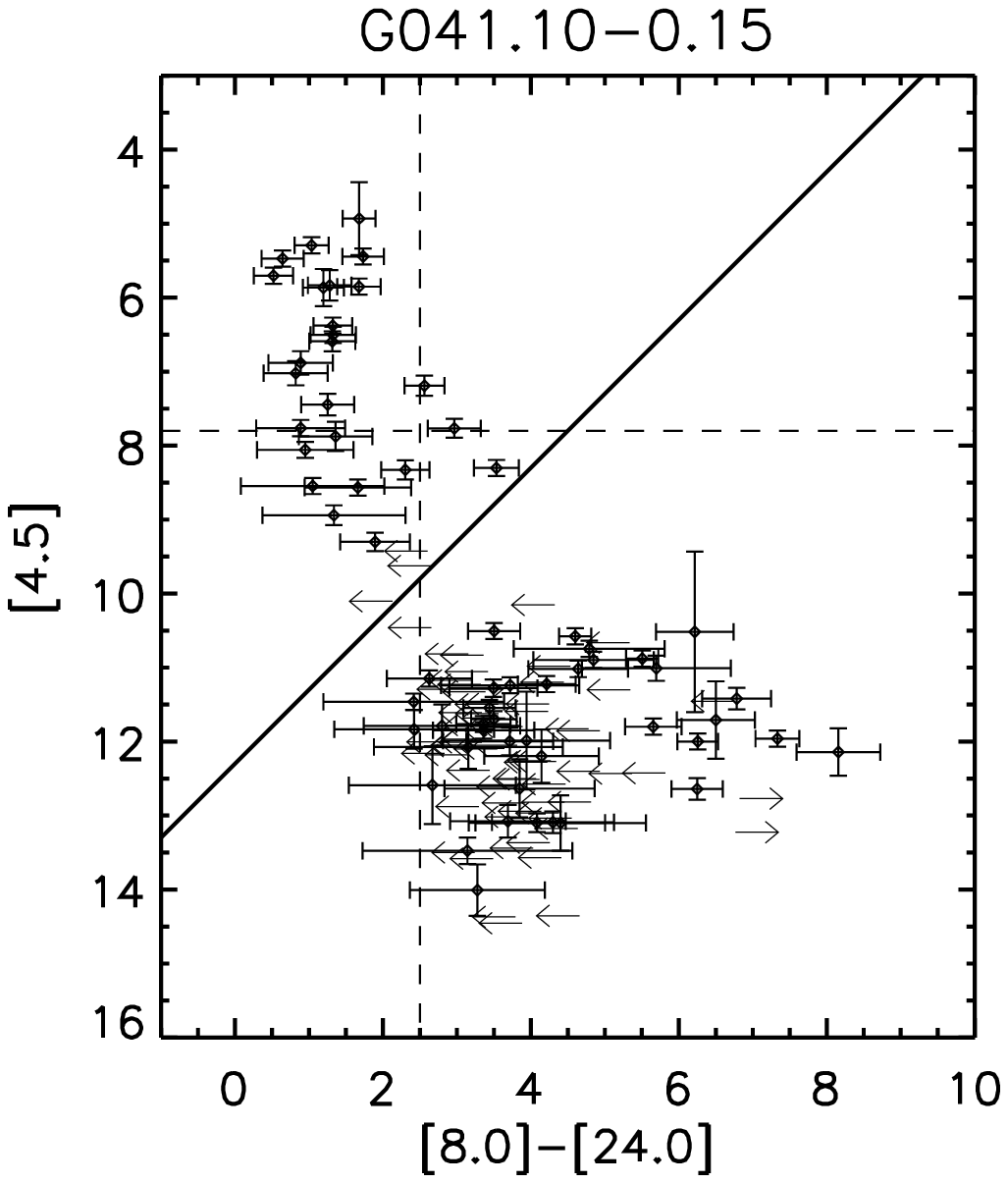}
\includegraphics[width=0.3\textwidth]{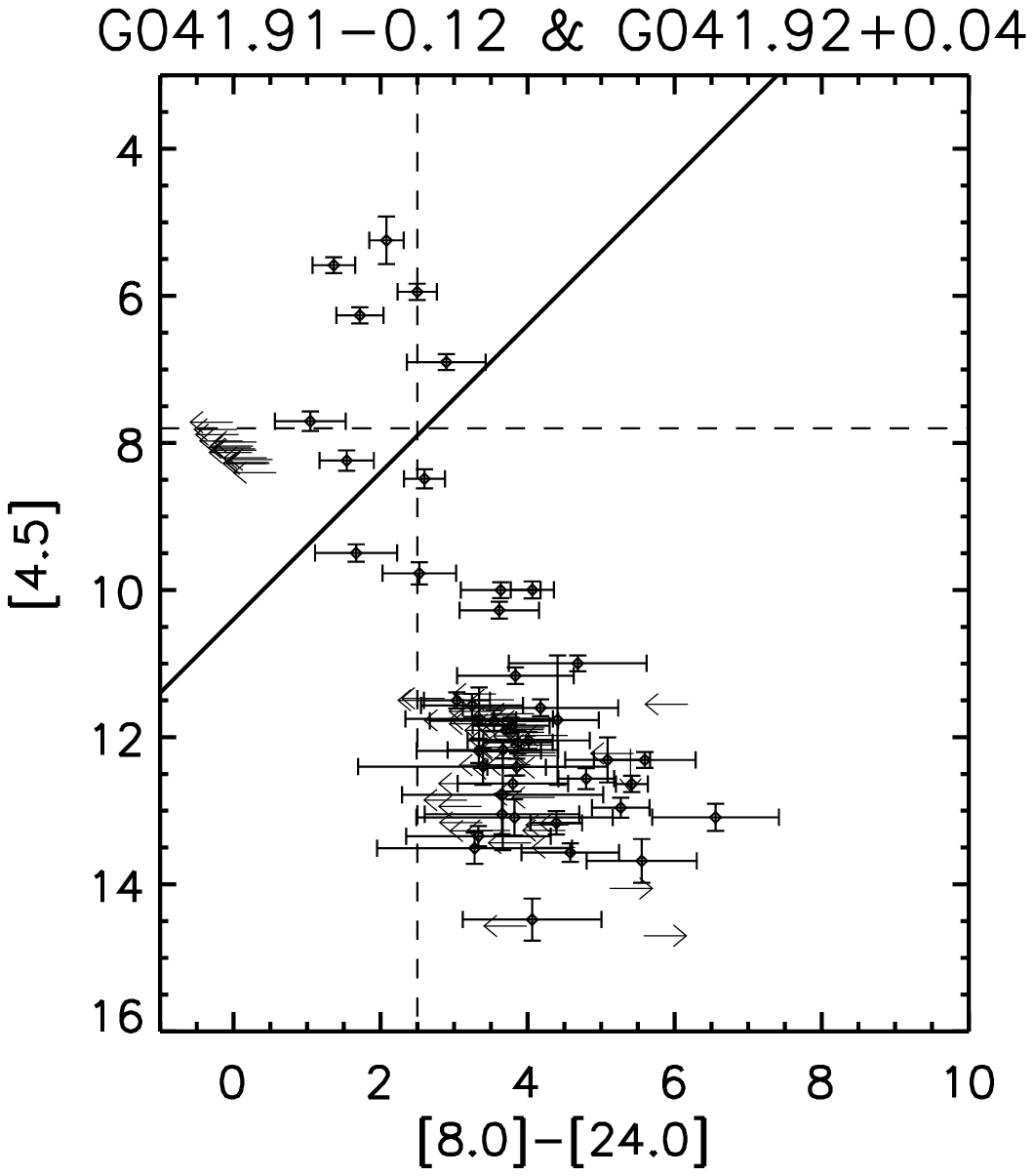}
\includegraphics[width=0.3\textwidth]{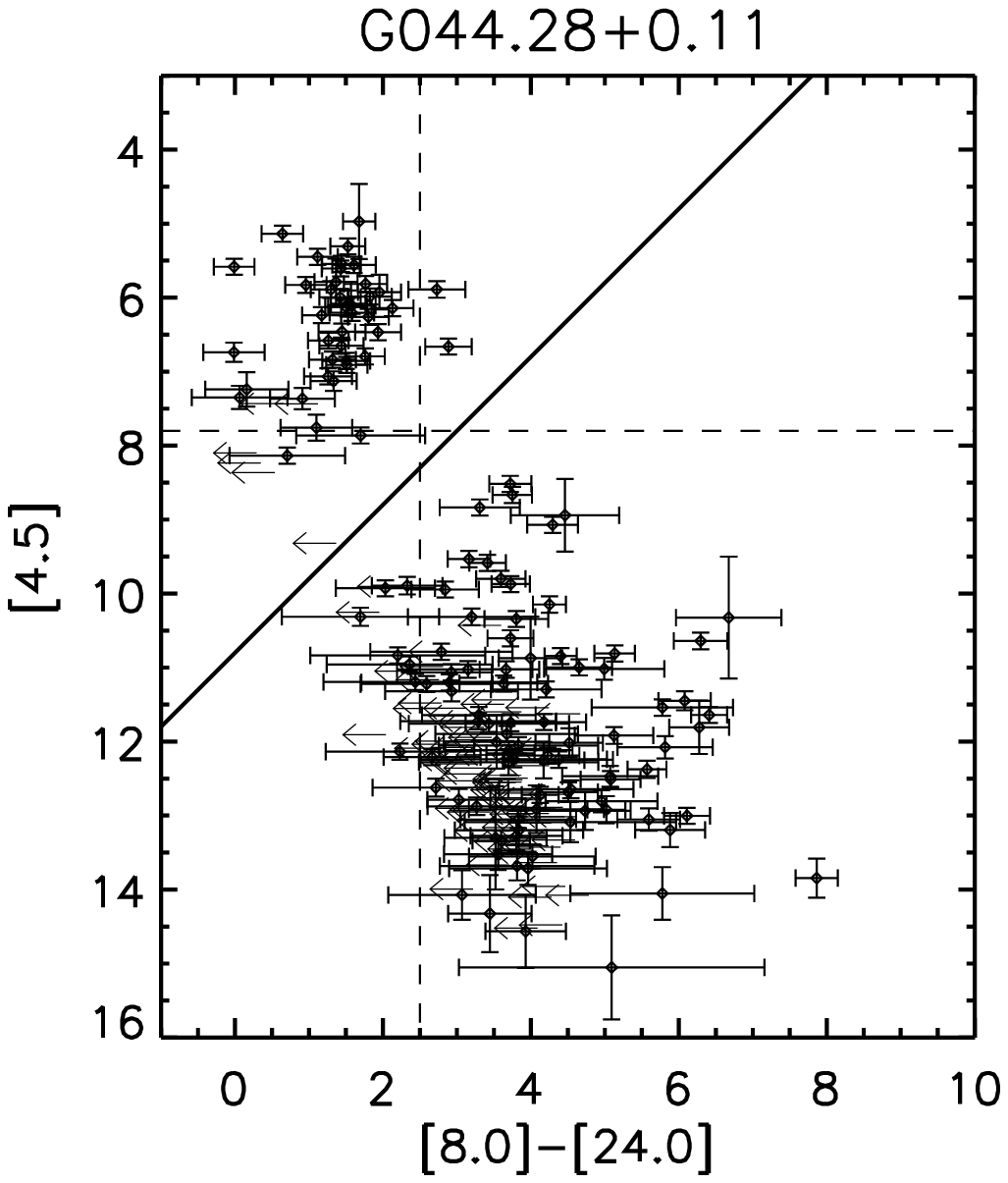}
\includegraphics[width=0.3\textwidth]{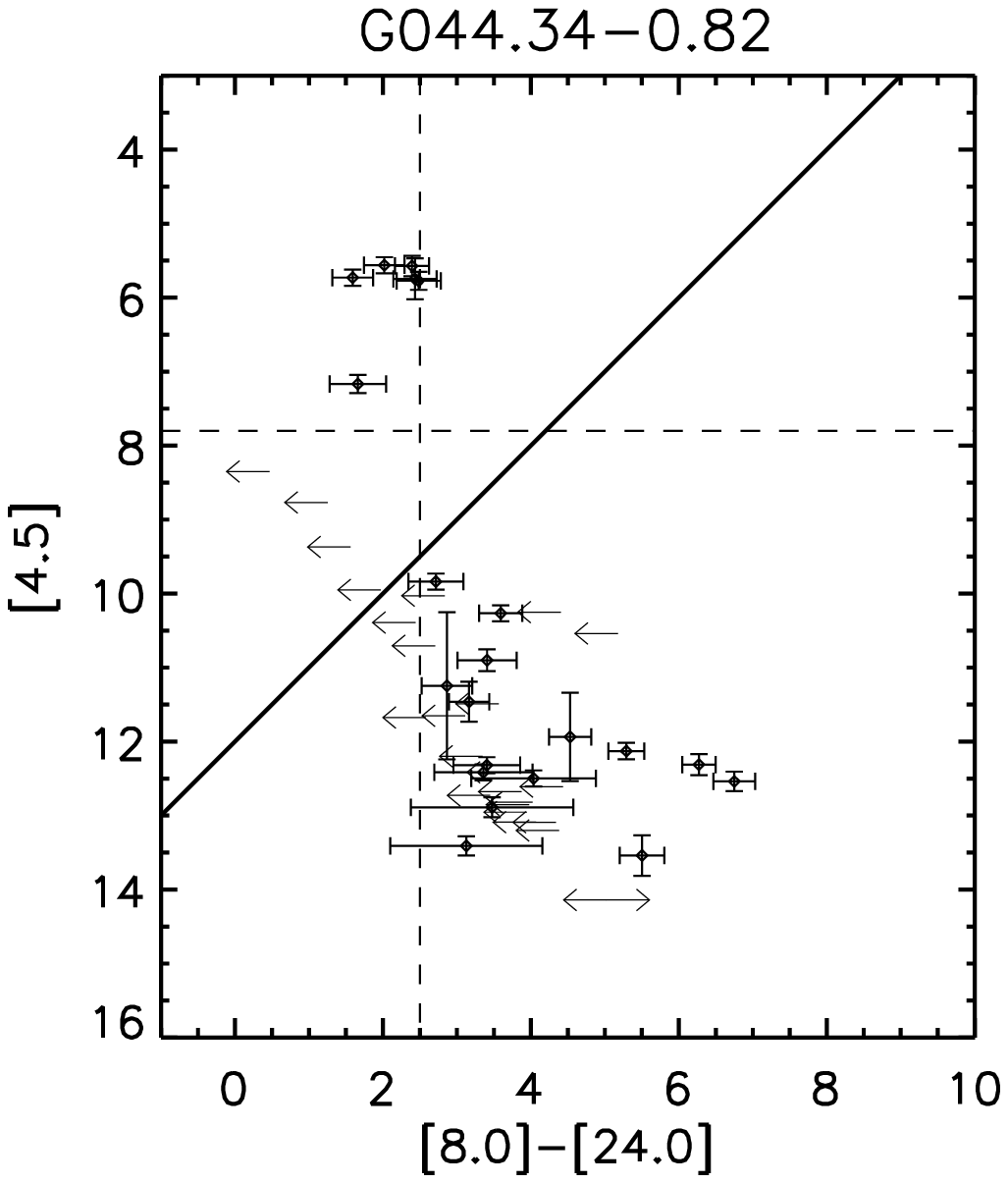}
\caption{Color-magnitude plots for our initial sample of YSO candidates used to remove contamination by AGB stars. Sources with upper or lower limits are plotted as arrows. In general the samples separate into two populations. Following \cite{2008AJ....136.2413R}, the brighter, bluer populations (upper left in this color space) are dominated by AGB stars over YSOs. We therefore exclude these objects from our final analysis. The dashed lines show the guidelines from \cite{2008AJ....136.2413R}, while our cuts determined for each region are the solid line. \label{f4}}
\end{center}
\end{figure}

\begin{figure}
\begin{center}
\includegraphics[width=0.3\textwidth]{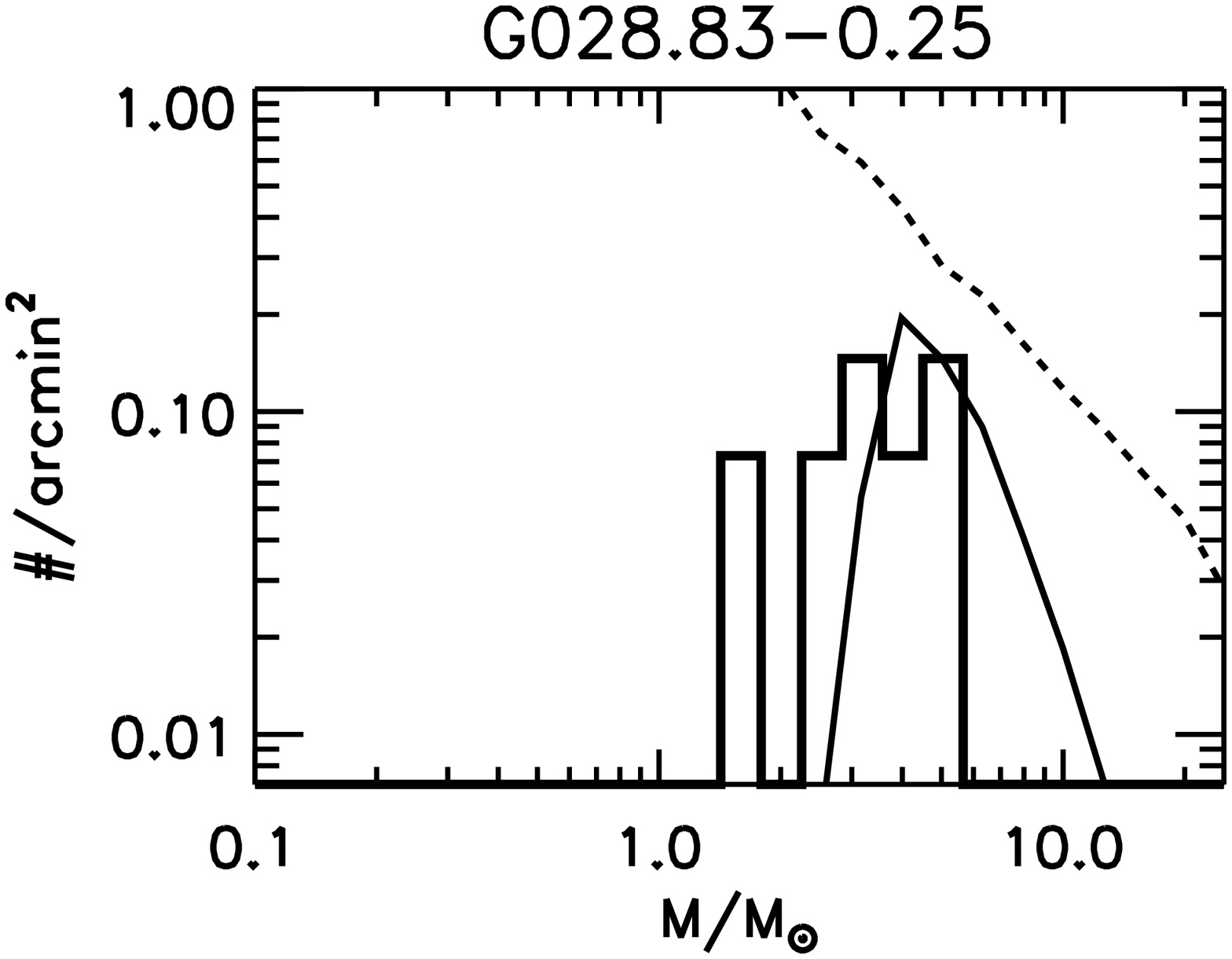}
\includegraphics[width=0.3\textwidth]{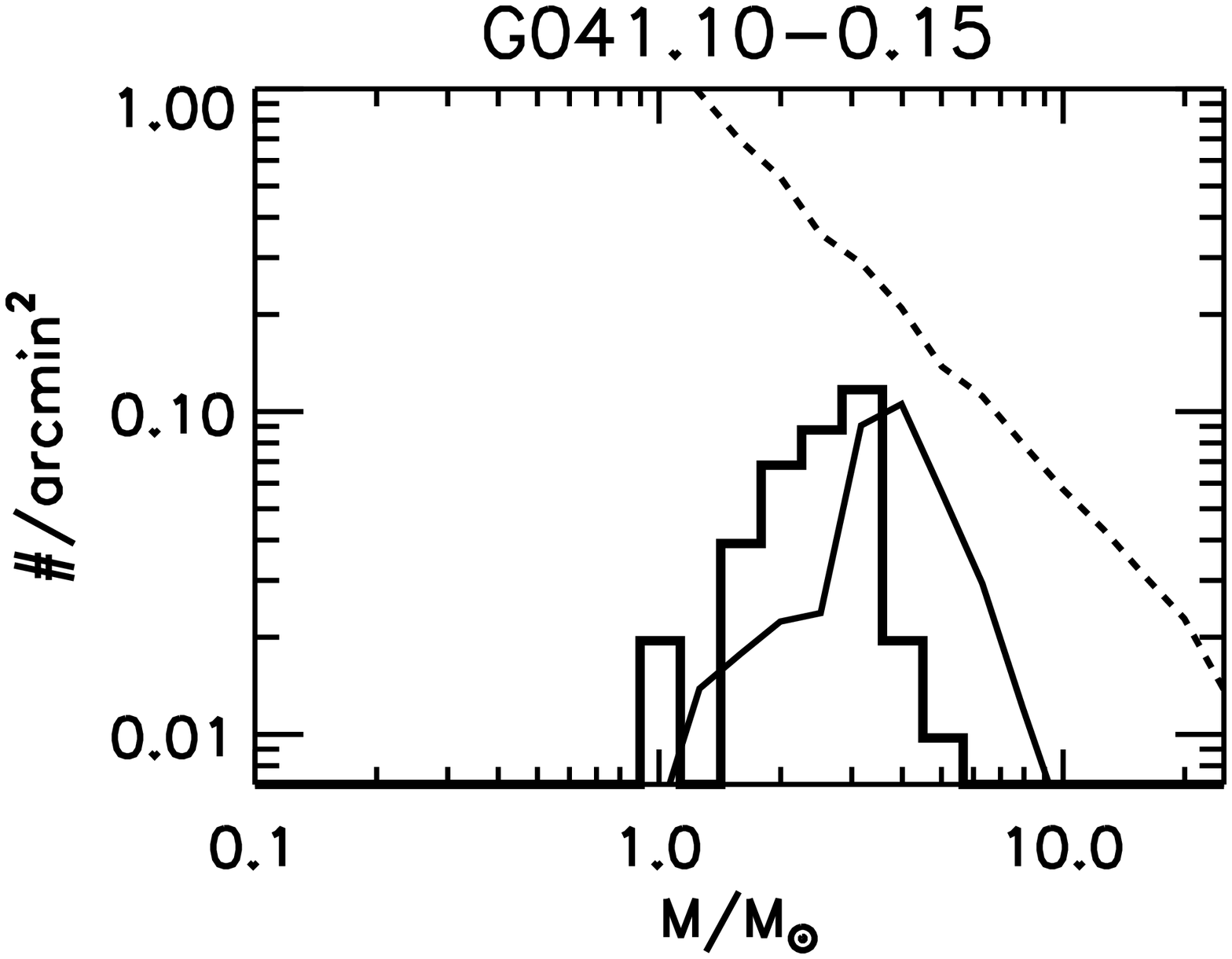}
\includegraphics[width=0.3\textwidth]{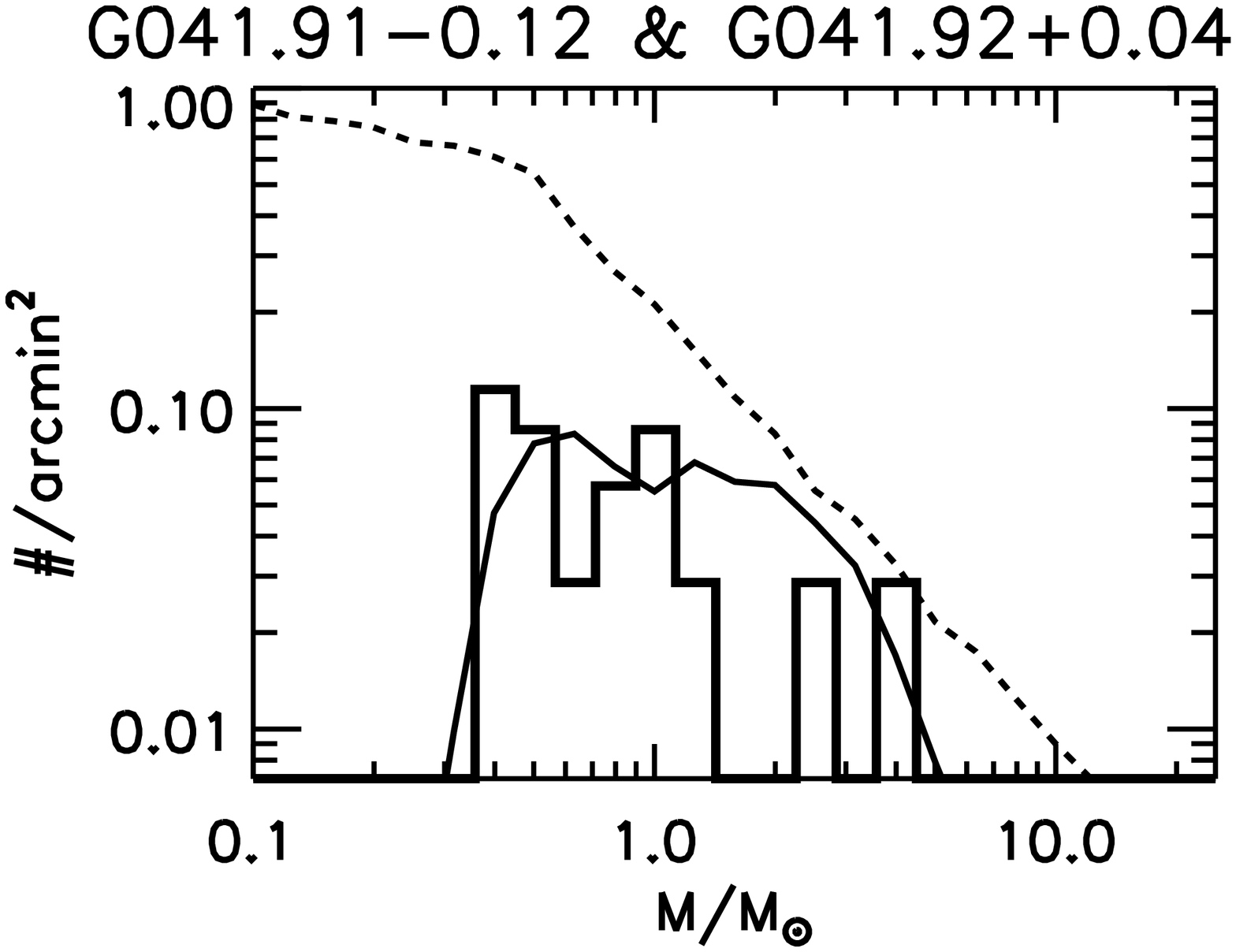}
\includegraphics[width=0.3\textwidth]{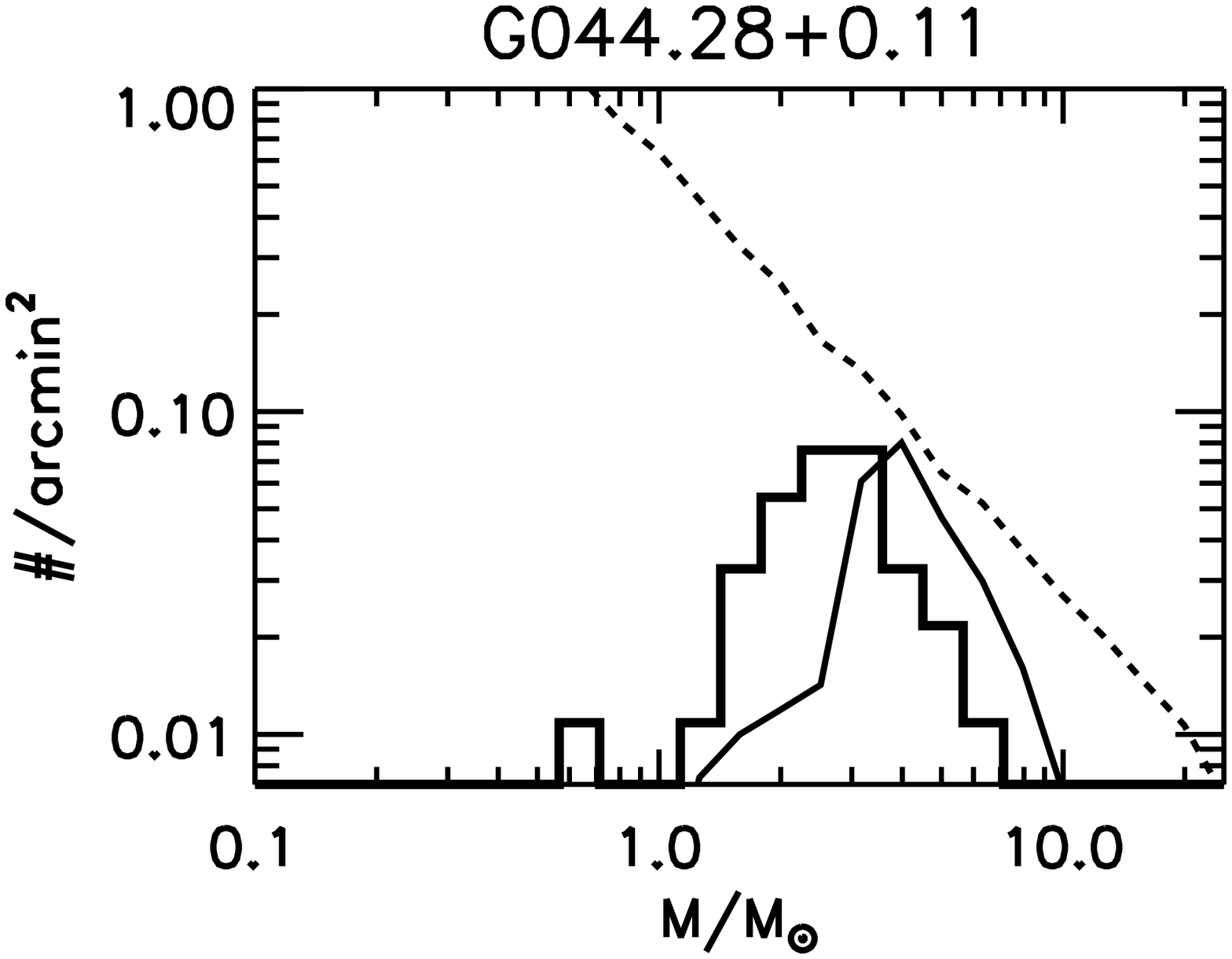}
\includegraphics[width=0.3\textwidth]{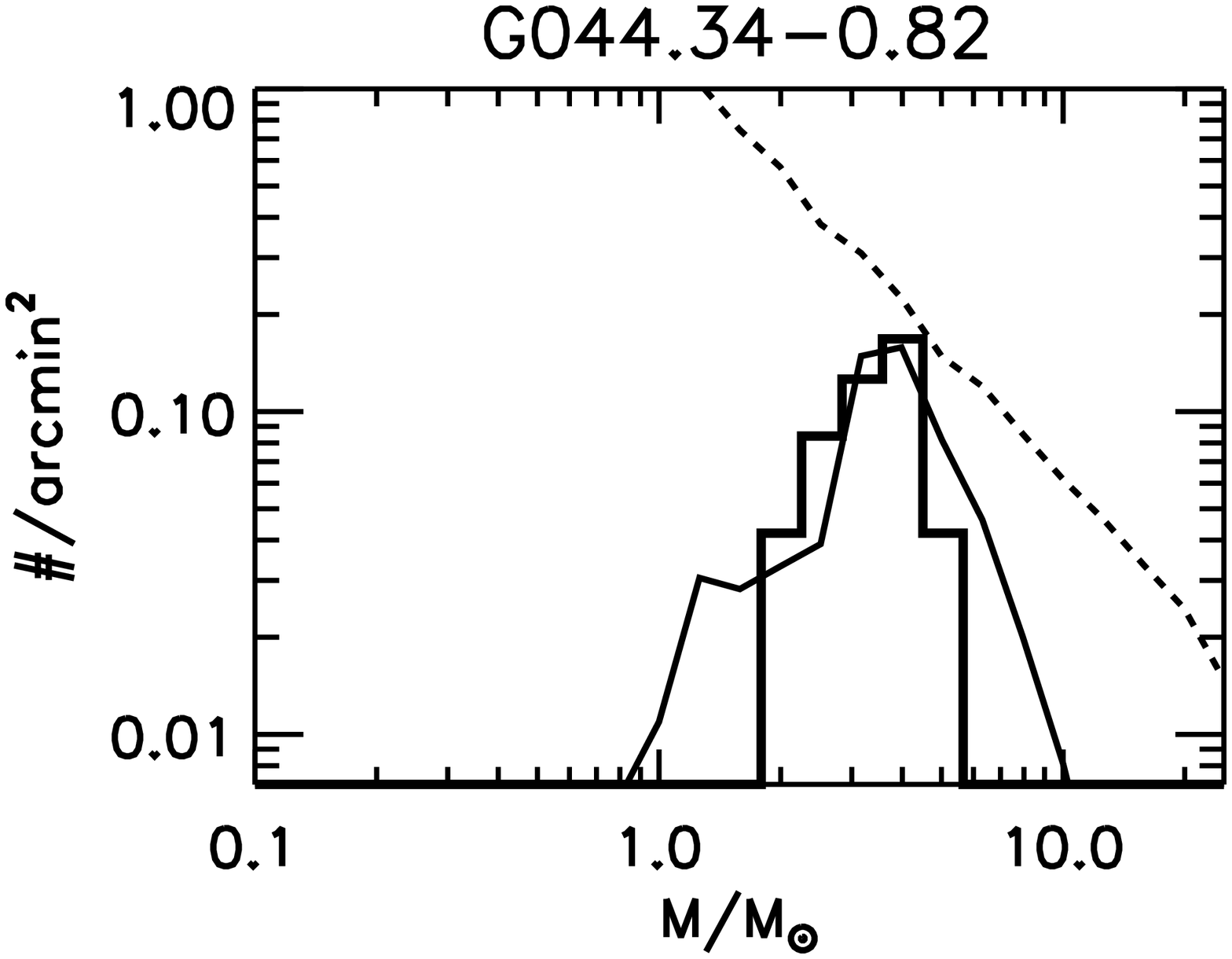}
\caption{Mass distribution of our observed YSO samples from this work identified by SED fitting, with likely AGBs removed using color-magnitude cuts, restricted to YSOs within or on the bright rim regions, plotted by the thick-lined histogram. Overplotted are the distributions of the virtual cluster YSO populations from \cite{2006ApJS..167..256R}. The dashed lines show the simulated populations ignoring any sources of completeness, while the thin solid lines show the distributions remaining after applying extinction corrections, considering sensitivity and saturation limits, fitting stellar atmosphere models to the SEDs, and applying color-magnitude cuts to remove AGB stars. Both simulated samples are presented with arbitrary scaling that is consistent within each region (see \S\ref{sec-complete}). The correspondence between the simulated sample with source incompleteness and the observed sample provides evidence that the completeness estimates are valid. \label{f5}}
\end{center}
\end{figure}

\begin{figure}
\begin{center}
\includegraphics[width=0.3\textwidth]{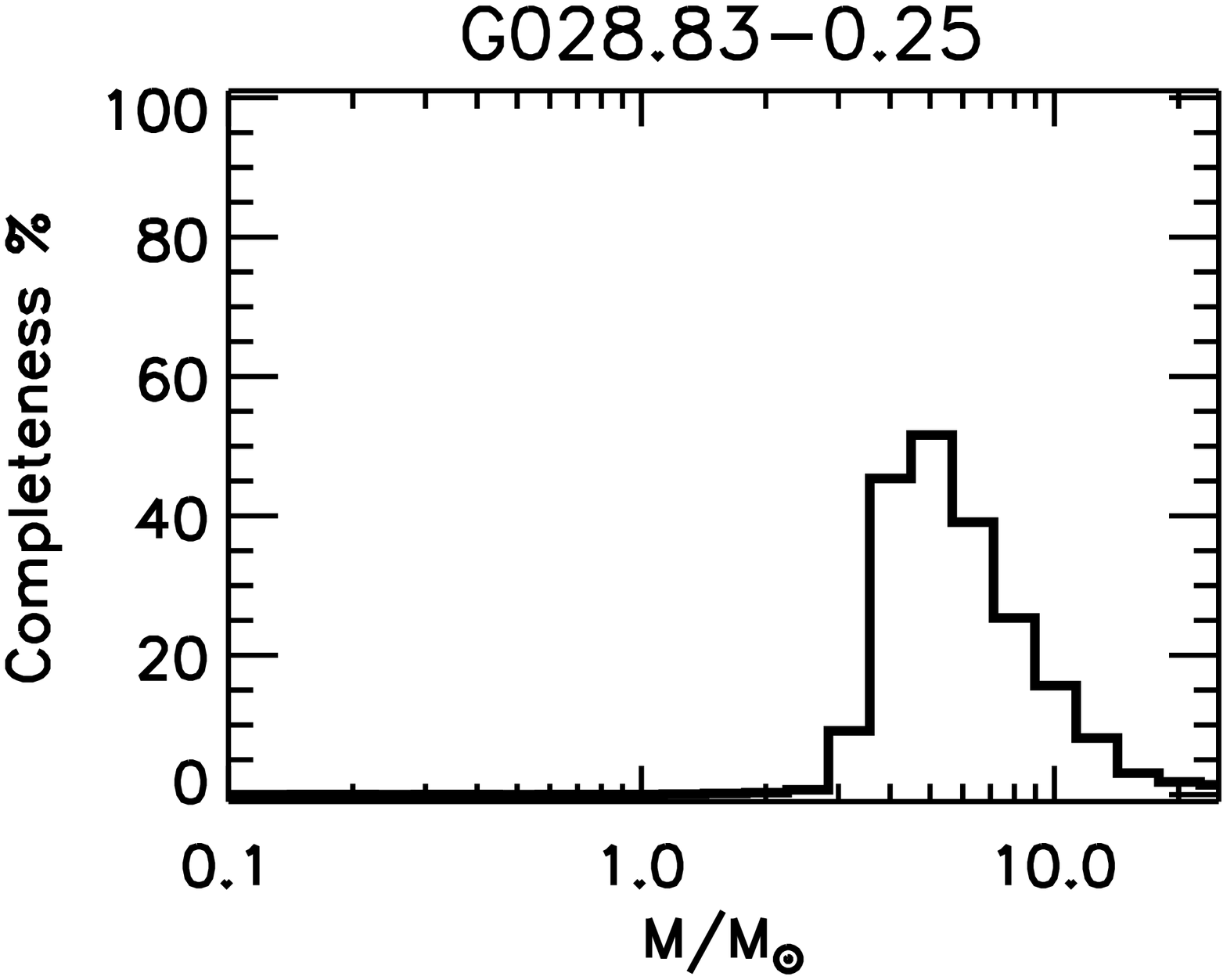}
\includegraphics[width=0.3\textwidth]{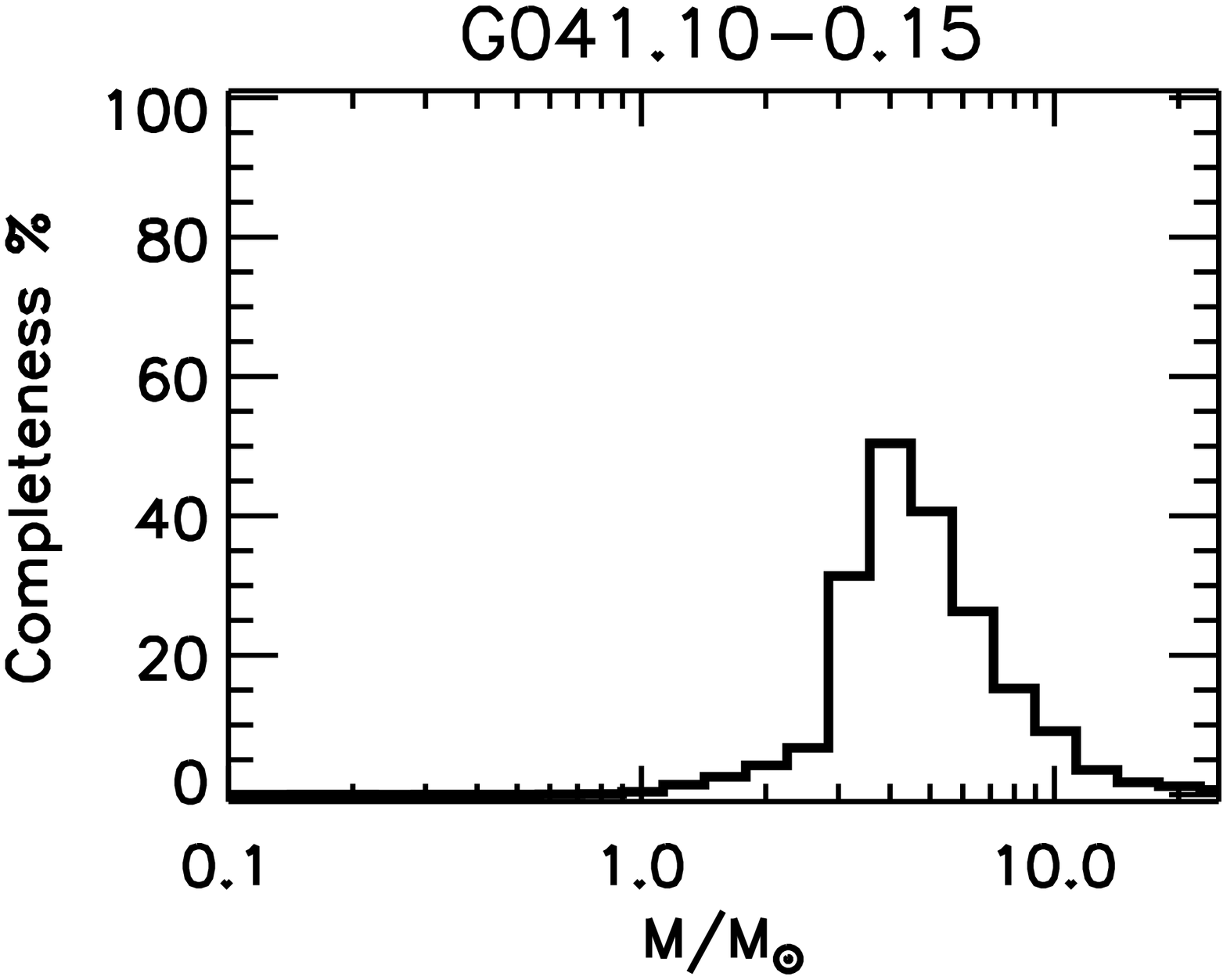}
\includegraphics[width=0.3\textwidth]{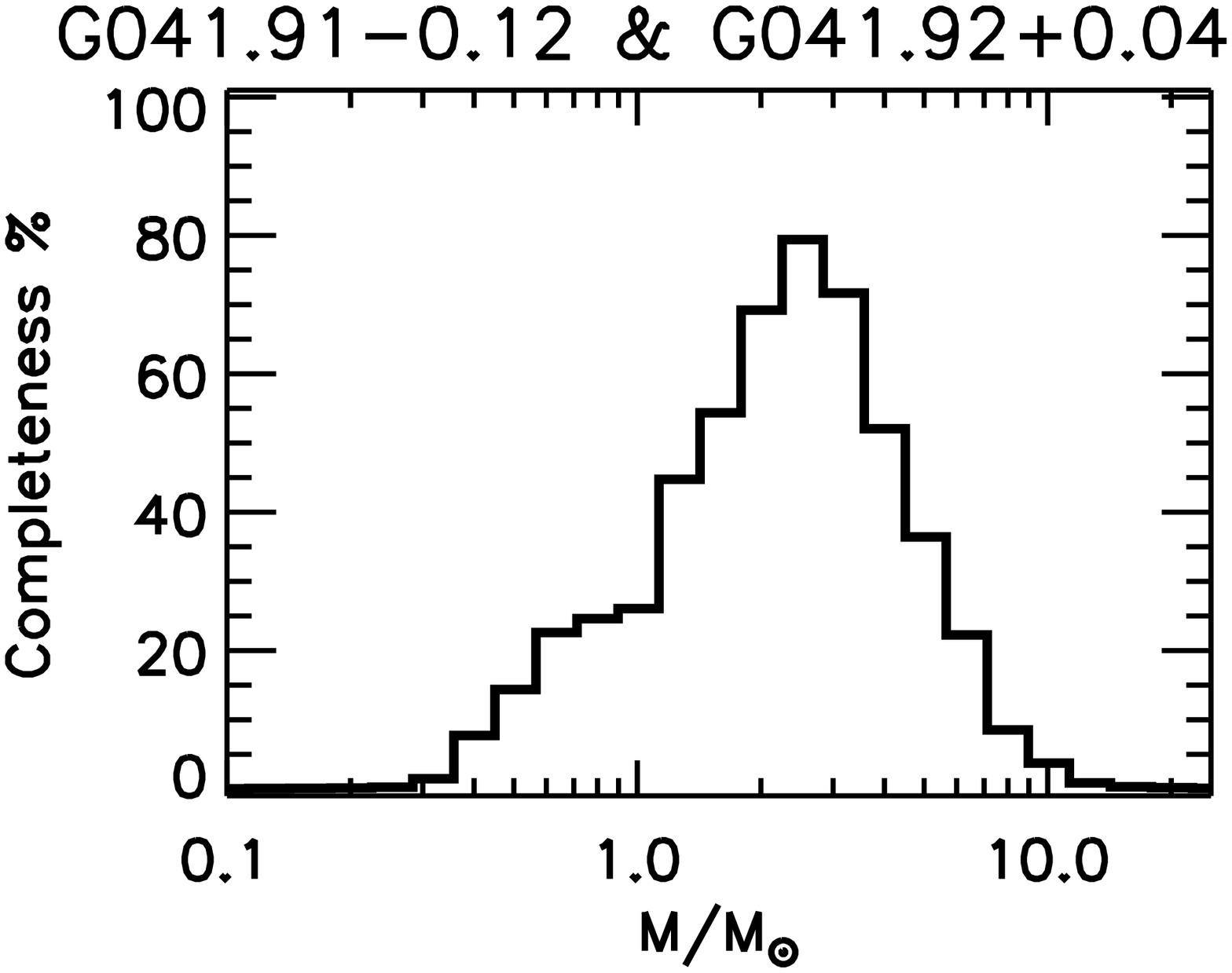}
\includegraphics[width=0.3\textwidth]{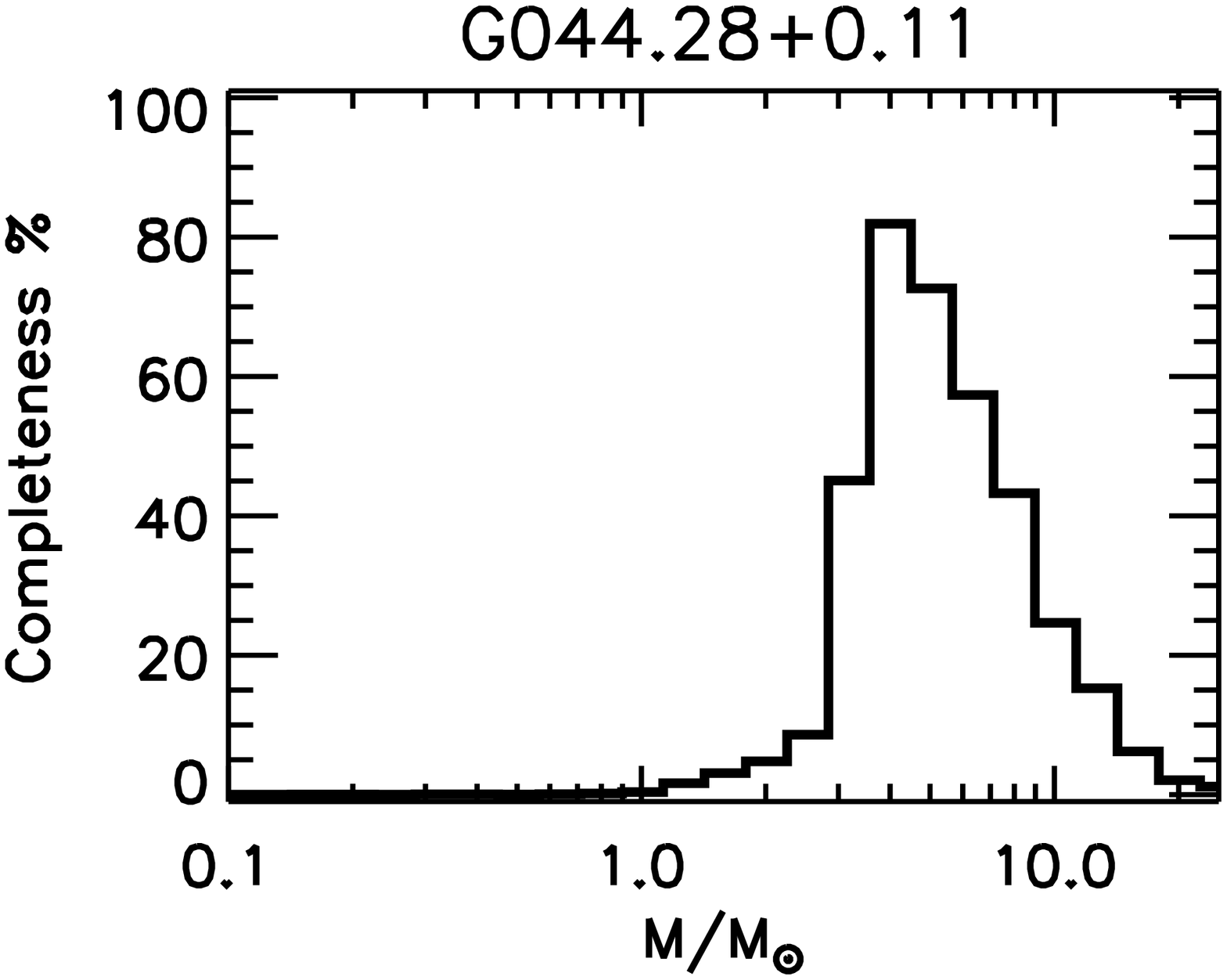}
\includegraphics[width=0.3\textwidth]{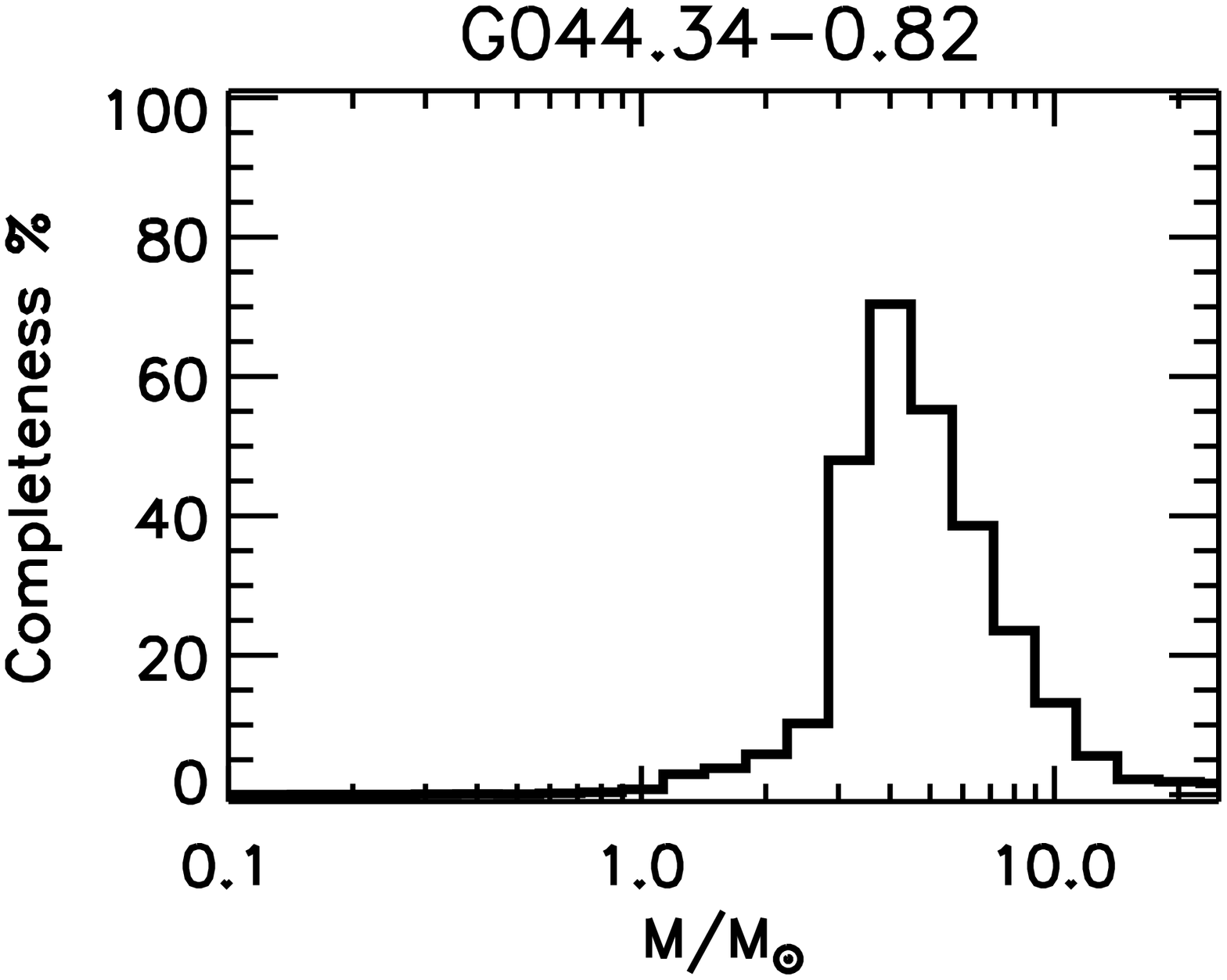}
\caption{Ratio of the two simulated distributions from Figure \ref{f5} (with and without applying observational and methodological effects) as an estimate of completeness as a function of mass. The incompleteness at the low-mass end is dominated by photometric sensitivity, while the incompleteness at the high mass end is dominated by color-magnitude cuts removing AGB contaminants. \label{f6}}
\end{center}
\end{figure}

\begin{figure}
\begin{center}
\includegraphics[width=0.3\textwidth]{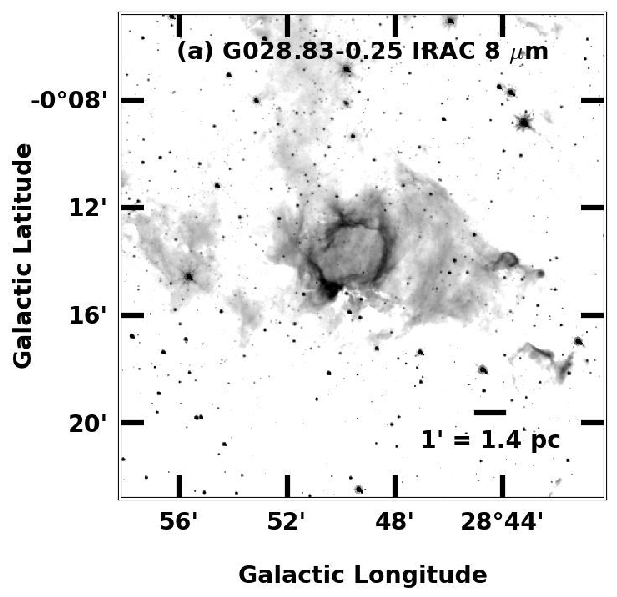}
\includegraphics[width=0.3\textwidth]{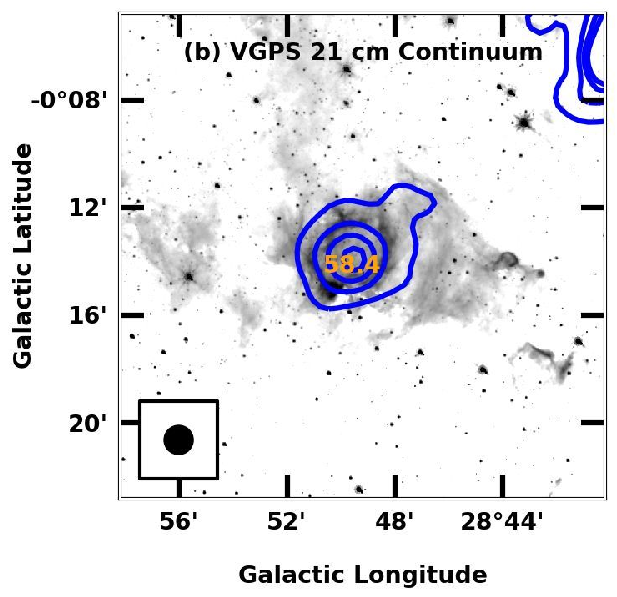}
\includegraphics[width=0.3\textwidth]{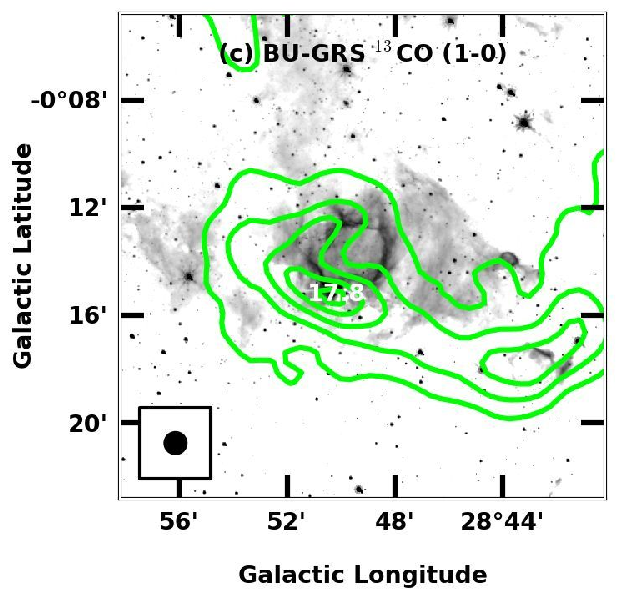}
\includegraphics[width=0.3\textwidth]{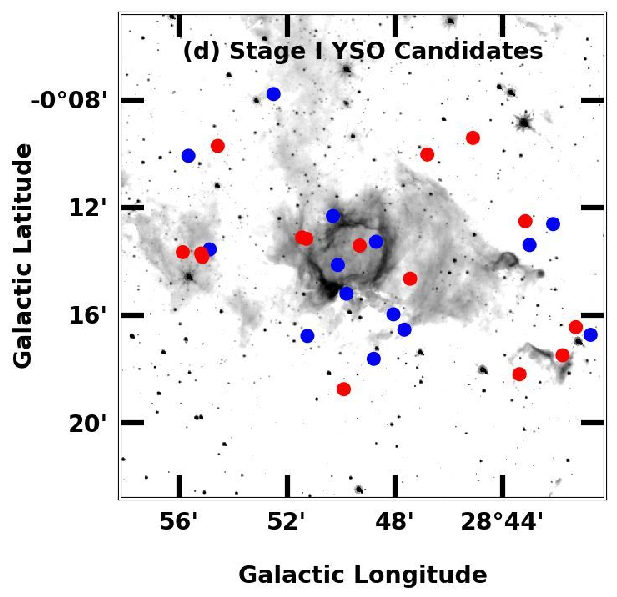}
\includegraphics[width=0.3\textwidth]{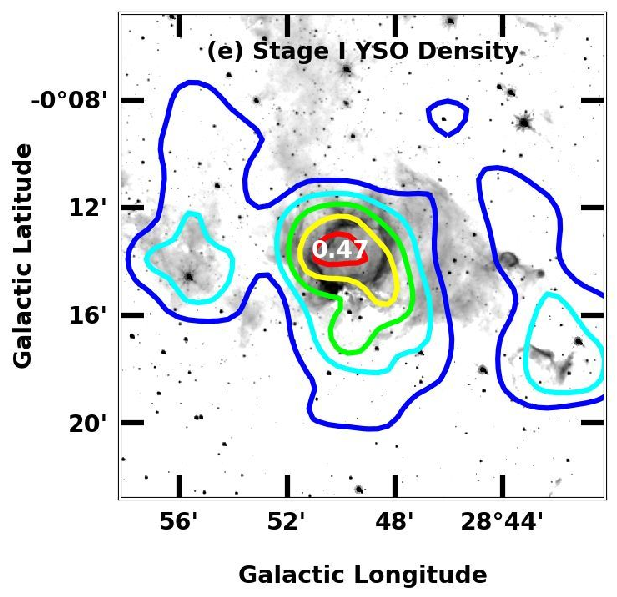}
\caption{Results for G028.83-0.25. (a) A \emph{Spitzer} 8 $\mu$m image, also used as the background in the other panels. The dense rim is visible. A 1' scale bar and physical scale at the near kinematic distance is in the lower right. (b) Radio continuum emission from the 21 cm VGPS. Contours are 95\%, 80\%, 60\%, 40\%, and 20\% of the peak brightness temperature, 58.4 K in this region, labeled in the panel. The H {\footnotesize II} region emission is visible and is coincident with the infrared bubble. The images have approximately 1' resolution, indicated by the beam in the lower left. (c) ${}^{13}$CO (1-0) emission from the BU-GRS over the velocity range 83.8 -- 90.0 km s${}^{-1}$. Contours are 95\%, 80\%, 60\%, 40\%, and 20\% of the peak integrated antenna temperature, 17.8 K km s${}^{-1}$ in this region, labeled in the panel. The 46'' beam is in the lower left. (d) Stage I (least evolved) YSOs identified by SED fitting, plotted with circles. (In the online version, red circles indicate sources with 24 $\mu$m photometry, while blue circles indicate sources with upper limits at 24 $\mu$m). Likely AGB contaminants have been removed. (e) Density of Stage I YSOs from SED fitting, sampled at 1.5 arcminutes. Contours are 95\%, 80\%, 60\%, 40\%, and 20\% of the peak density, 0.47 YSOs per square arcminute in this region, labeled in the panel. (In the online version, contour colors range from blue (low density) to red (high density)). The maxima indicate areas of enhanced clustering of relatively unevolved YSOs. There is an enhanced YSO population at the center of the bubble as compared to the surrounding field. \label{f7}}
\end{center}
\end{figure}

\begin{figure}
\begin{center}
\includegraphics{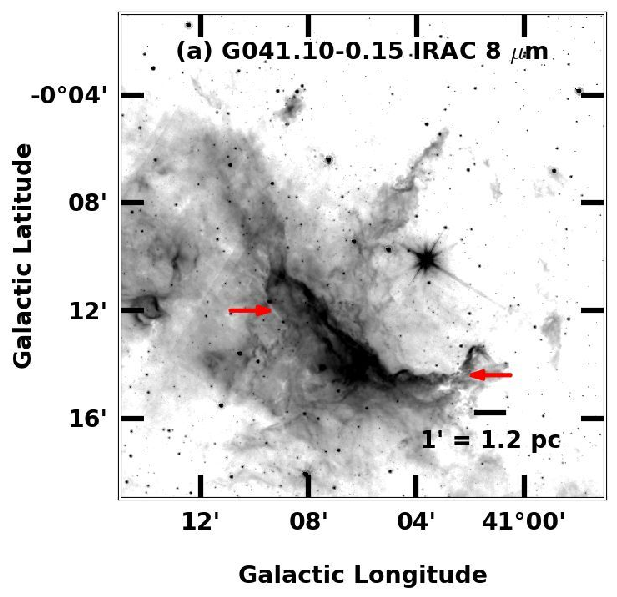}
\includegraphics{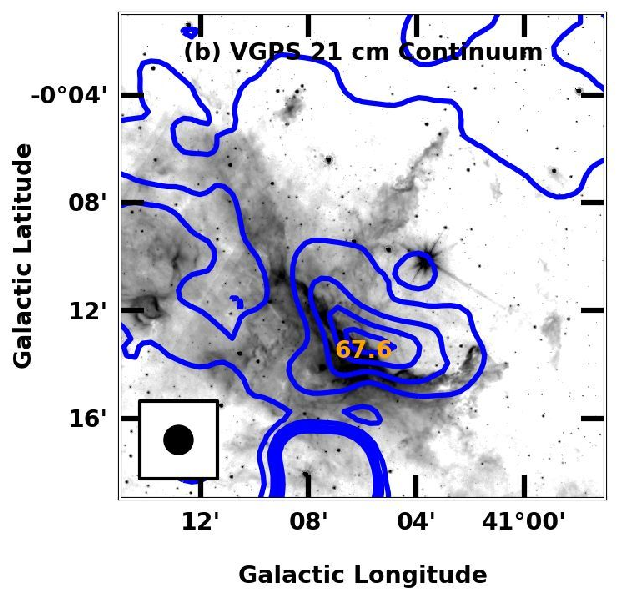}
\includegraphics{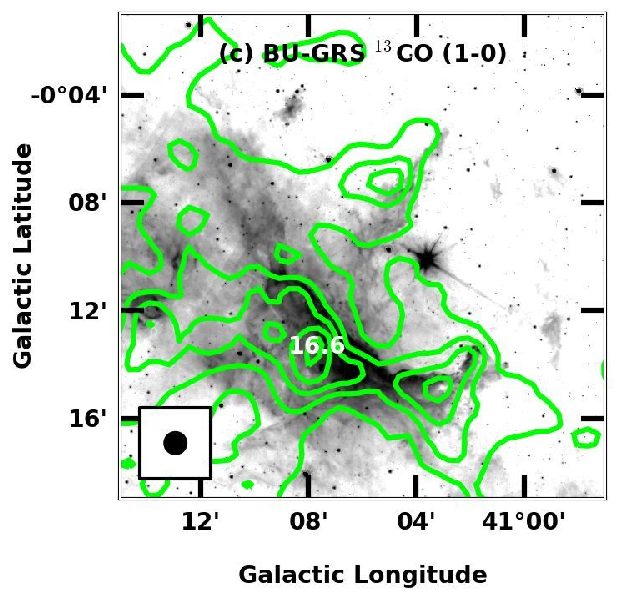}
\includegraphics{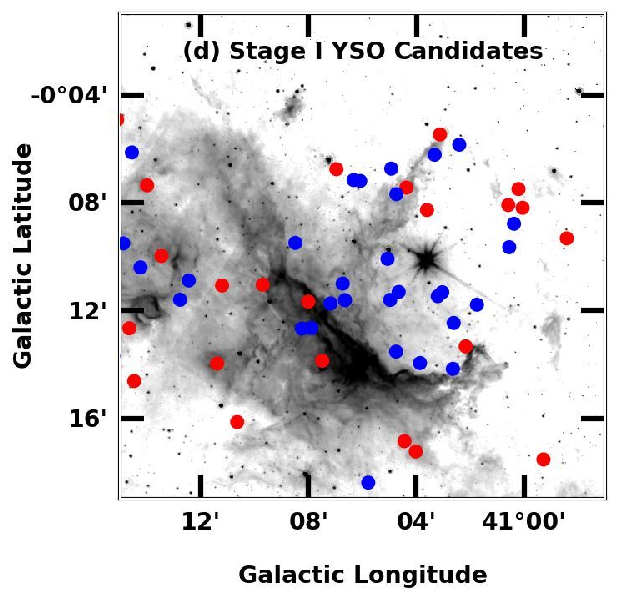}
\includegraphics{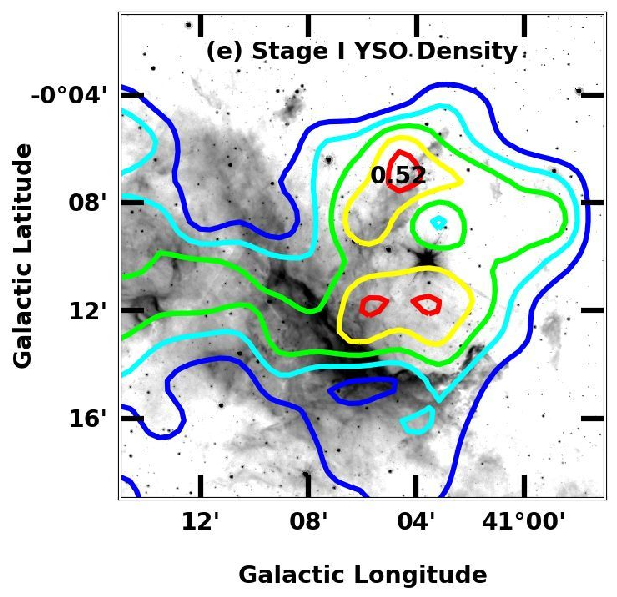}
\caption{Results for G041.10-0.15. Panels are as in Figure \ref{f7}. (In the online version, IRDCs are marked by red arrows in panel (a)). A portion of the unrelated supernovae remnant 3C 397 \citep{2010ApJ...712.1147J} is visible at the bottom of panel (b). The ${}^{13}$CO (1-0) emission in panel (c) is integrated over the velocity range 54.7 -- 68.2 km s${}^{-1}$. There is an enhanced YSO population located within and around the bubble as compared to the surrounding field, seen in panels (d) and (e). \label{f8}}
\end{center}
\end{figure}

\begin{figure}
\begin{center}
\includegraphics{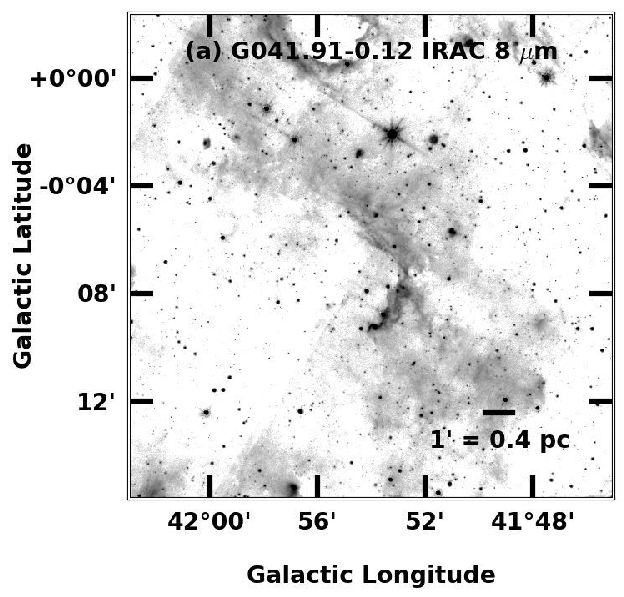}
\includegraphics{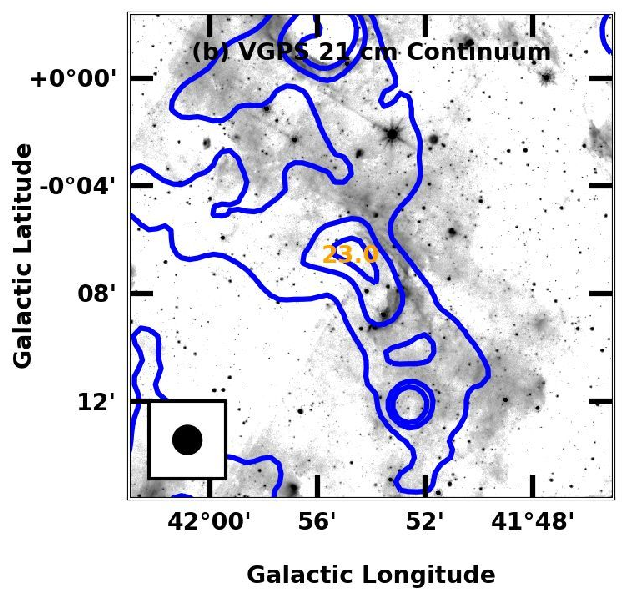}
\includegraphics{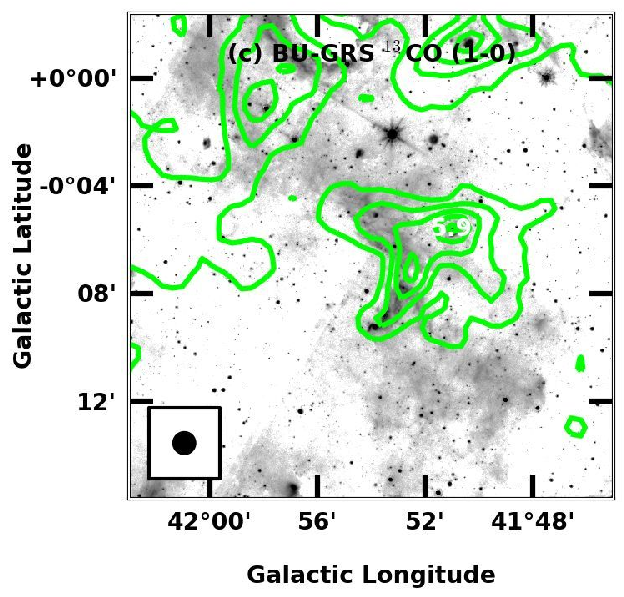}
\includegraphics{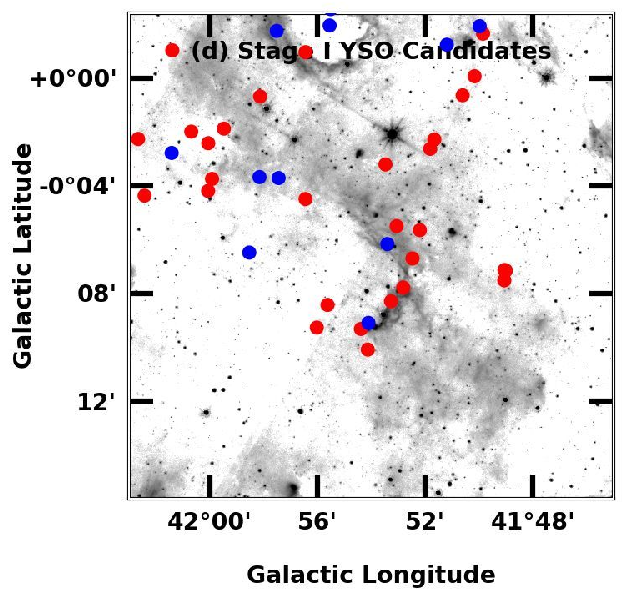}
\includegraphics{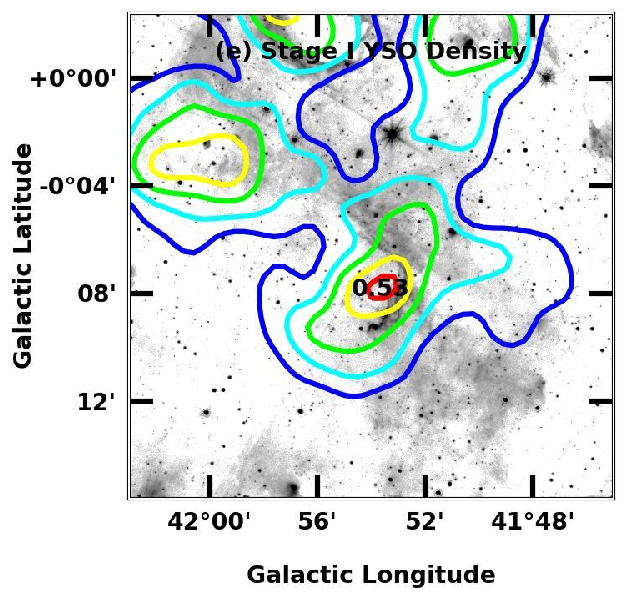}
\caption{Results for G041.91-0.12. Panels are as in Figure \ref{f7}. The ${}^{13}$CO (1-0) emission in panel (c) is integrated over the velocity range 12.0 -- 20.8 km s${}^{-1}$. There is an enhanced YSO population located around the infrared rim as compared to the surrounding field, seen in panels (d) and (e). \label{f9}}
\end{center}
\end{figure}

\begin{figure}
\begin{center}
\includegraphics{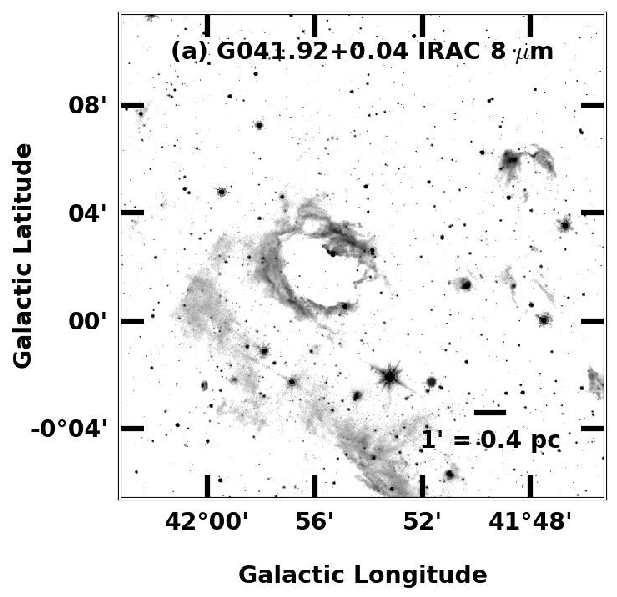}
\includegraphics{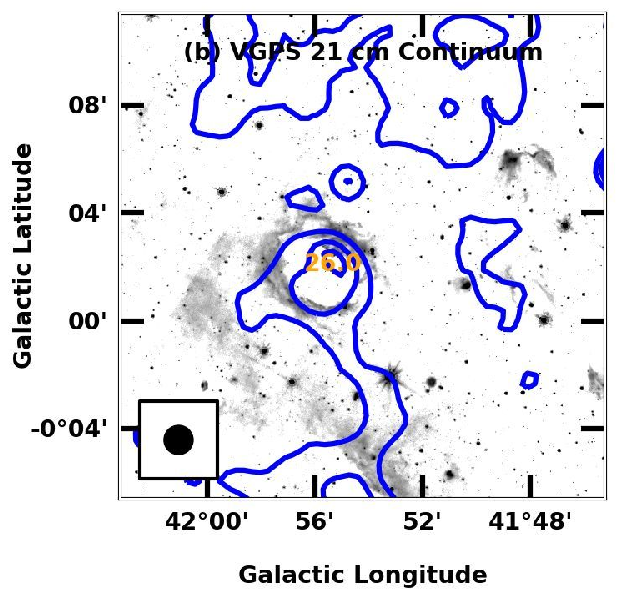}
\includegraphics{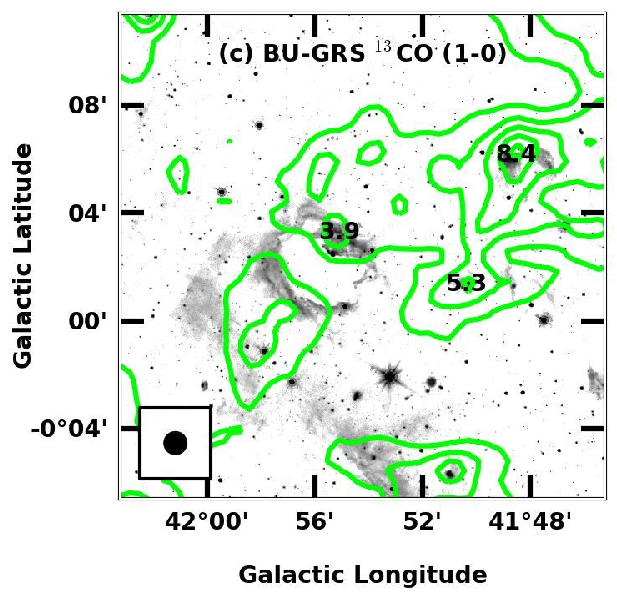}
\includegraphics{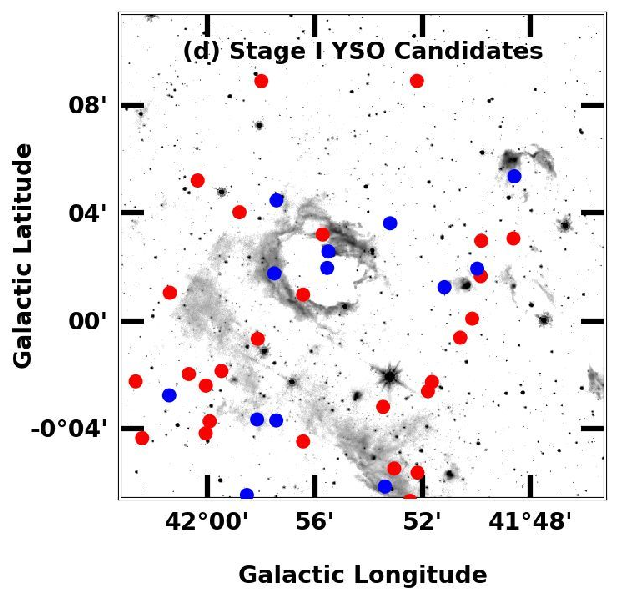}
\includegraphics{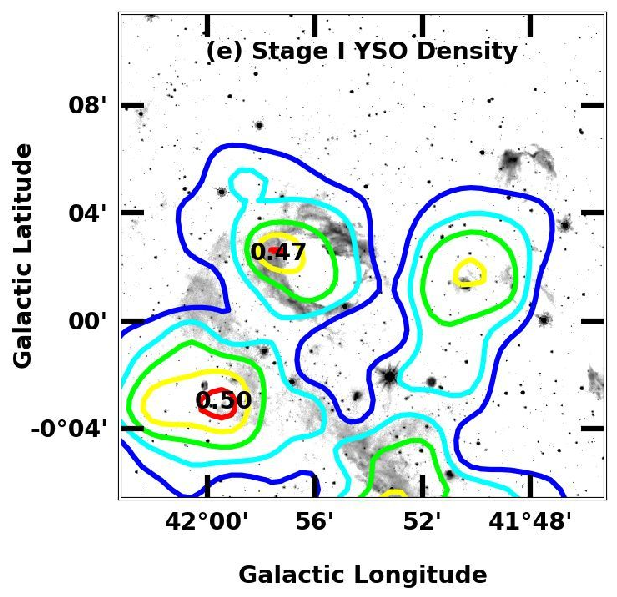}
\caption{Results for G041.92+0.04. Panels are as in Figure \ref{f7}. The ${}^{13}$CO (1-0) emission in panel (c) is integrated over the velocity range 12.0 -- 20.8 km s${}^{-1}$. There is a slightly enhanced YSO population located in the infrared bubble as compared to much of the surrounding field, seen in panels (d) and (e). \label{f10}}
\end{center}
\end{figure}

\begin{figure}
\begin{center}
\includegraphics{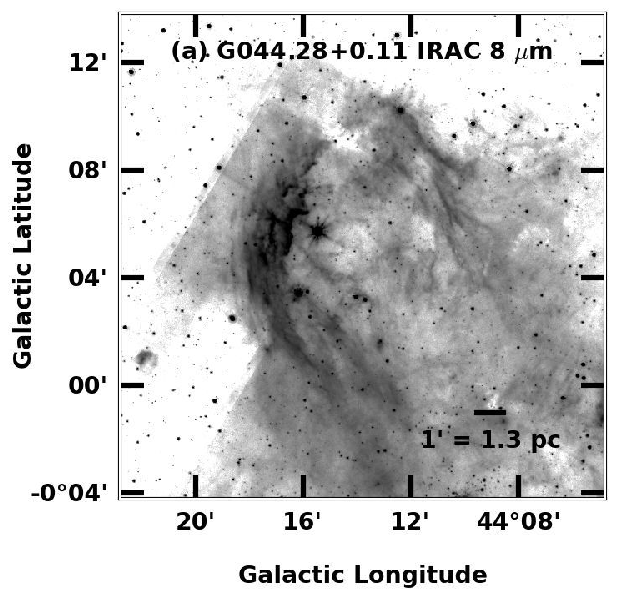}
\includegraphics{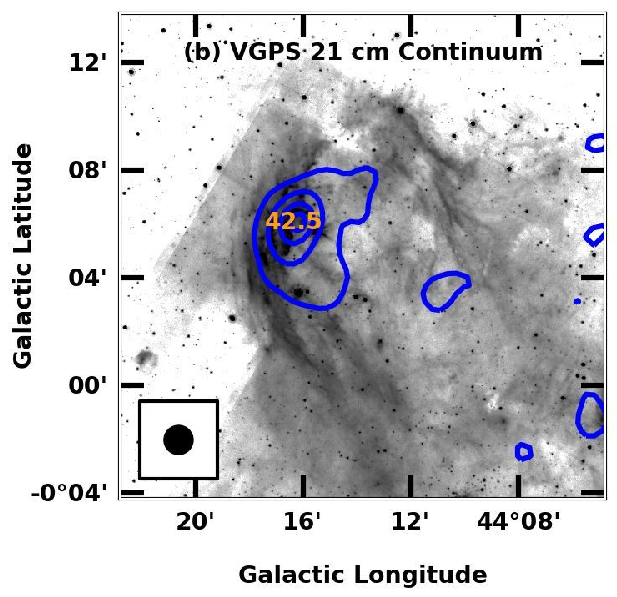}
\includegraphics{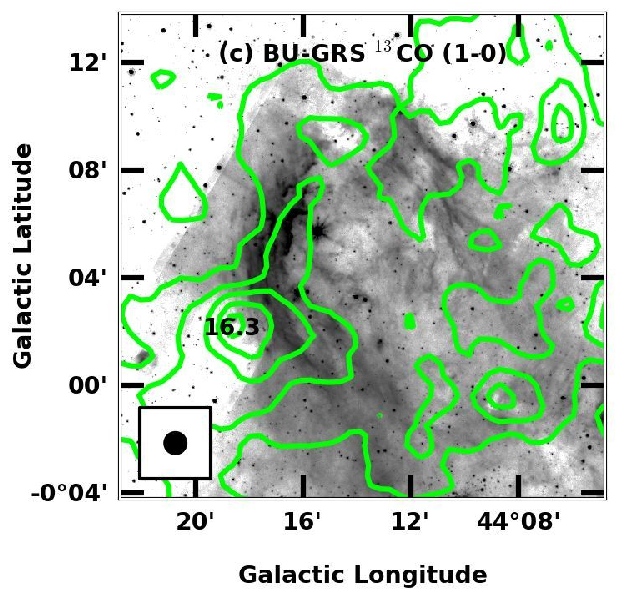}
\includegraphics{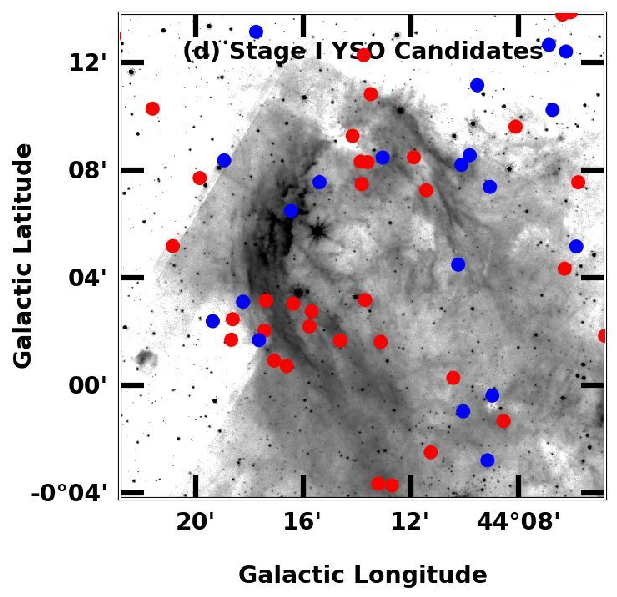}
\includegraphics{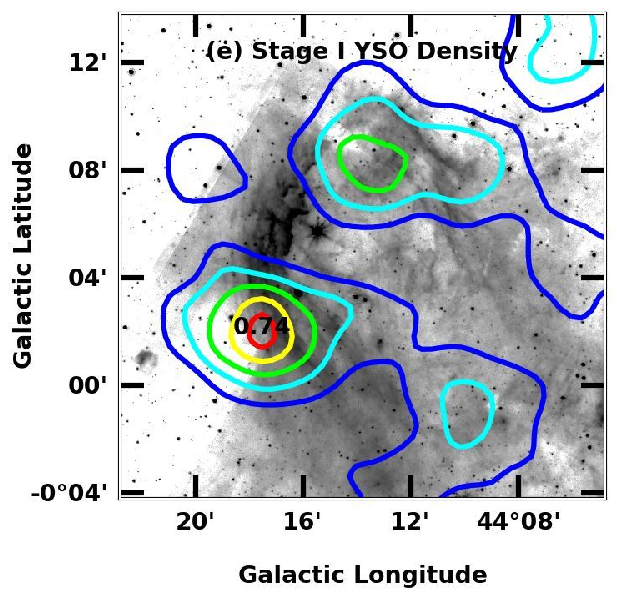}
\caption{Results for G044.28+0.11. Panels are as in Figure \ref{f7}. The ${}^{13}$CO (1-0) emission in panel (c) is integrated over the velocity range 52.2 -- 68.4 km s${}^{-1}$. There are enhanced YSO populations located on the infrared rim as compared to much of the surrounding field, seen in panels (d) and (e). \label{f11}}
\end{center}
\end{figure}

\begin{figure}
\begin{center}
\includegraphics{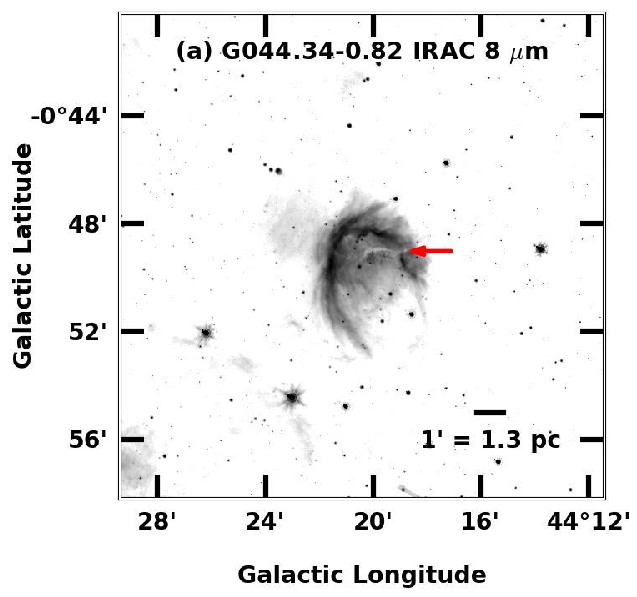}
\includegraphics{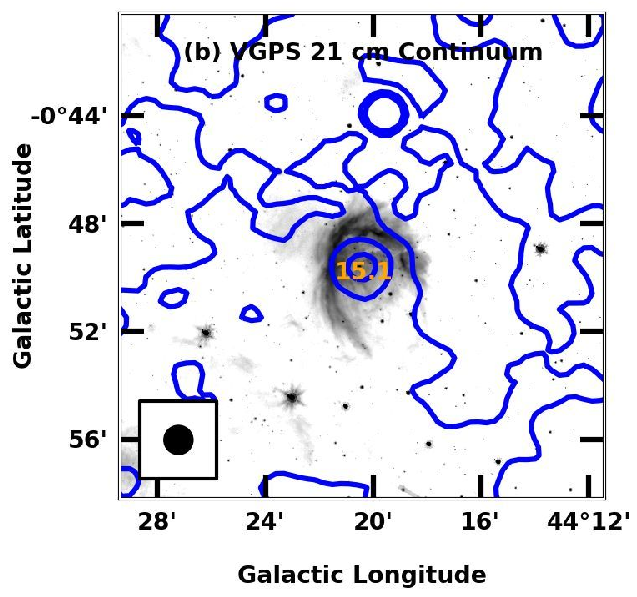}
\includegraphics{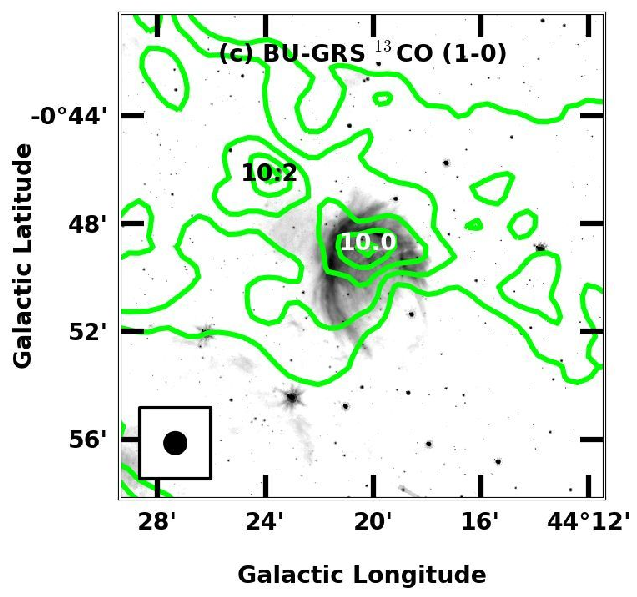}
\includegraphics{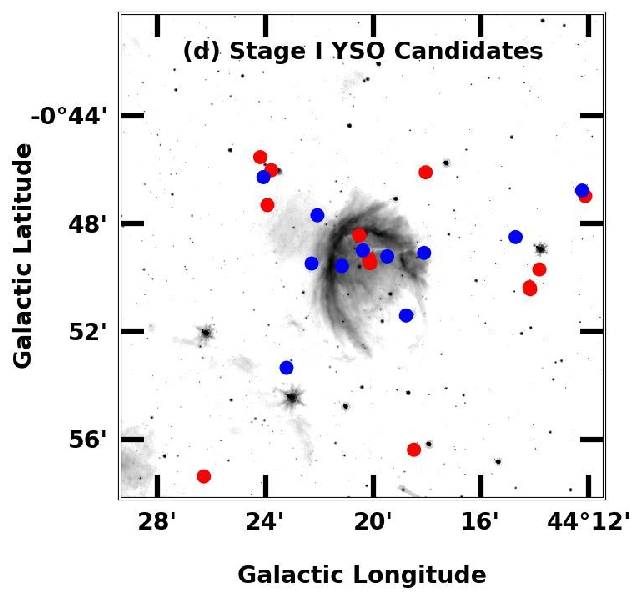}
\includegraphics{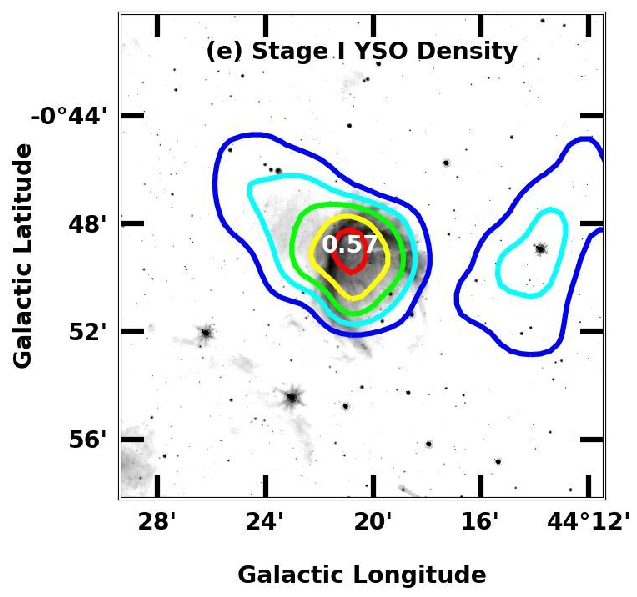}
\caption{Results for G044.34-0.82. Panels are as in Figure \ref{f7}. (In the online version, the IRDC is marked by a red arrow in panel (a)). The ${}^{13}$CO (1-0) emission in panel (c) is integrated over the velocity range 56.4 -- 67.1 km s${}^{-1}$. There is an enhanced YSO population located in the infrared bubble as compared to the surrounding field, seen in panels (d) and (e). \label{f12}}
\end{center}
\end{figure}

\begin{figure}
\begin{center}
\includegraphics{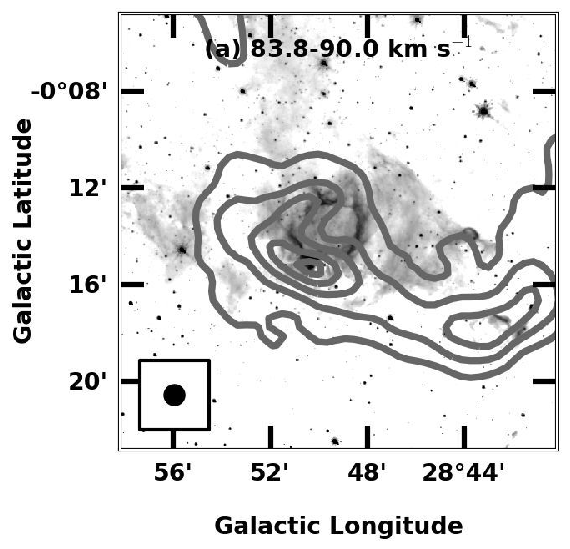}
\includegraphics{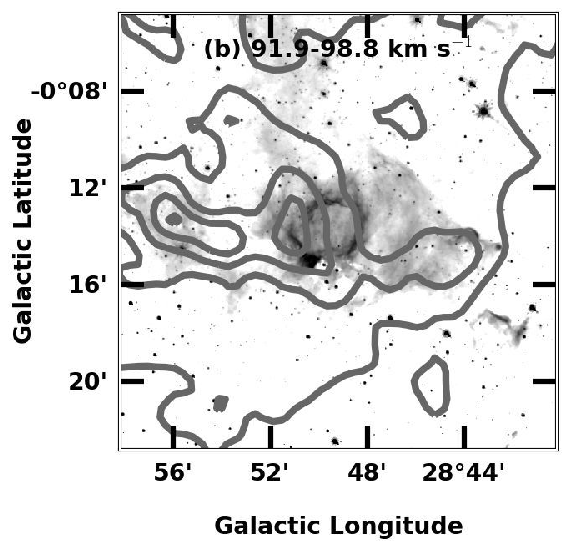}
\caption{Contours of BU-GRS ${}^{13}$CO (1-0) emission integrated over two velocity components possibly associated with the H {\footnotesize II} region G028.83-0.25, plotted over the 8 $\mu$m IRAC image. The first component (\emph{left}) is integrated over the velocity range 83.8 -- 90.0 km s${}^{-1}$, and the second component (\emph{right}) is integrated over the velocity range 91.9 -- 98.8 km s${}^{-1}$. The 46'' beam is shown in the lower left corner of panel (a). Contours are plotted as 95, 80, 60, 40, and 20\% of the peak value in panel (a), 17.8 K km s${}^{-2}$ antenna temperature.\label{f13}}
\end{center}
\end{figure}

\begin{figure}
\begin{center}
\includegraphics[width=0.6\textwidth]{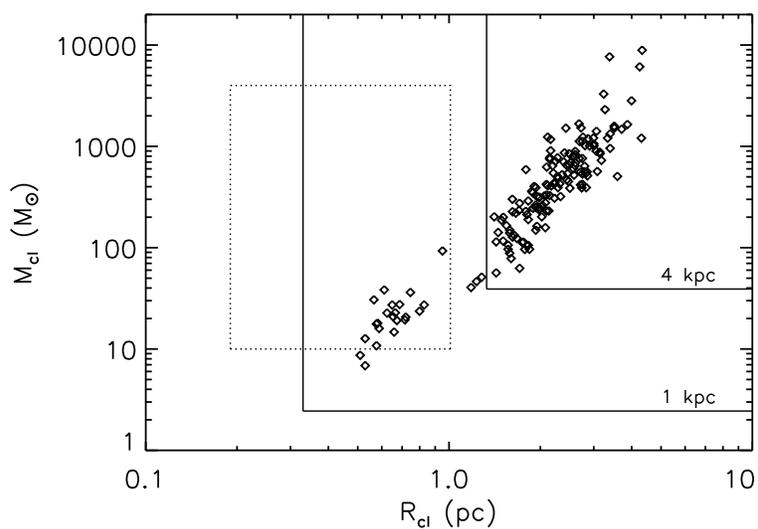}
\caption{Effective radius, $R_{\textrm{cl}}$, versus calculated mass of molecular gas, $M_{\textrm{cl}}$, for every clump identified by Clumpfind remaining in our sample after our quality control cuts. The solid lines mark approximate sensitivity and resolution limits using Clumpfind and our quality control cuts on the BU-GRS data, assuming 1 kpc and 4 kpc as labeled. The dotted box shows the range of masses and radii of molecular clumps associated with young stellar clusters within 1 kpc as observed by \cite{2003AJ....126..286R}. \label{f14}}
\end{center}
\end{figure}

\clearpage

\begin{center}

\begin{table}
\begin{center}
\begin{footnotesize}
\caption{H {\tiny II} Region Sample \label{t1}}
\begin{tabular}{lcccccr@{.}lr@{.}l}
\tableline\tableline
 \multirow{2}{*}{Regions} & \multirow{2}{*}{N\#\tablenotemark{a}} & RA (J2000) & Dec (J2000) & Principle & $v_{rad}$ & \multicolumn{4}{c}{\underline{Kinematic Distance}\tablenotemark{b}} \\
 &  & hh:mm:ss.s & dd:mm:ss & Morphology & $(\textrm{km s}^{-1})$ & \multicolumn{2}{c}{Near (kpc)} &  \multicolumn{2}{c}{Far (kpc)} \\
\tableline
G028.83-0.25 & N49 & 18:44:44.3 & -03:45:34 & Bubble & 90.6\tablenotemark{c} & 5&07 & 9&65 \\
G041.10-0.15 & ... & 19:06:48.9 & 07:10:55 & Cometary & 59.4\tablenotemark{c} & 3&99 & 8&67 \\
G041.91-0.12 & ... & 19:08:21.1 & 07:55:20 & Cometary & 18.1\tablenotemark{d} & 1&4 & 11&1 \\
G041.92+0.04 & N80 & 19:07:51.2 & 08:00:33 & Bubble & 17.7\tablenotemark{e} & 1&32 & 11&18 \\
G044.28+0.11 & N91 & 19:11:57.7 & 10:07:05 & Cometary & 59.6\tablenotemark{c} & 4&33 & 7&7 \\
G044.34-0.82 & N92 & 19:15:28.1 & 09:44:24 & Cometary & 62.0\tablenotemark{e} & 4&59 & 7&43 \\
\end{tabular}
\tablenotetext{a}{Identifier in \cite{2006ApJ...649..759C}.}
\tablenotetext{b}{Using the galactic rotation curve of \cite{2009ApJ...700..137R}}
\tablenotetext{c}{Radio recombination line velocity from \cite{1989ApJS...71..469L}}
\tablenotetext{d}{Radio recombination line velocity from \cite{1996ApJ...472..173L}}
\tablenotetext{e}{JCMT CO (3-2) velocity from \cite{2010ApJ...709..791B}}
\end{footnotesize}
\end{center}
\end{table}

\begin{table}
\begin{center}
\begin{footnotesize}
\caption{YSO Fitting Parameters \label{t2}}
\begin{tabular}{lcccc}
\tableline\tableline
 \multirow{2}{*}{Regions} & $d_{\textrm{min}}$\tablenotemark{a} & $d_{\textrm{max}}$\tablenotemark{a} & Range of Sample Coverage\tablenotemark{b} & Coverage Area \\
 & (kpc) & (kpc) & $(\ell,b:\ell,b)$ & (arcmin${}^{2}$) \\
\tableline
\multirow{2}{*}{G028.83-0.25} & \multirow{2}{*}{3.5} & \multirow{2}{*}{5.5} & (28${}^{\circ}$.65, -0${}^{\circ}$.36 : 28${}^{\circ}$.95, -0${}^{\circ}$.22) & \multirow{2}{*}{250} \\
 &  &  & (28${}^{\circ}$.7, -0${}^{\circ}$.22 : 28${}^{\circ}$.95, -0${}^{\circ}$.11) &  \\
 &  &  &  &  \\
G041.10-0.15 & 3.5 & 5.5 & (40${}^{\circ}$.95, -0${}^{\circ}$.36 : 41${}^{\circ}$.38, -0${}^{\circ}$.07) & 448 \\
 &  &  &  &  \\
G041.91-0.12 & \multirow{2}{*}{0.5} & \multirow{2}{*}{2.5} & (41${}^{\circ}$.80, -0${}^{\circ}$.18 : 42${}^{\circ}$.00, 0${}^{\circ}$.16) & \multirow{2}{*}{290} \\
and G041.92+0.04 &  &  & (42${}^{\circ}$.00, -0${}^{\circ}$.09 : 42${}^{\circ}$.05, 0${}^{\circ}$.16) &  \\
 &  &  &  &  \\
\multirow{2}{*}{G044.28+0.11} & \multirow{2}{*}{3.5} & \multirow{2}{*}{5.5} & (43${}^{\circ}$.95, -0${}^{\circ}$.15 : 44${}^{\circ}$.50, 0${}^{\circ}$.35) & \multirow{2}{*}{1303} \\
 &  &  & (44${}^{\circ}$.00, -0${}^{\circ}$.27 : 44${}^{\circ}$.60, 0${}^{\circ}$.00) &  \\
 &  &  &  &  \\
G044.34-0.82 & 3.5 & 5.5 & (44${}^{\circ}$.20, -0${}^{\circ}$.97 : 44${}^{\circ}$.46, -0${}^{\circ}$.72) & 234 \\
\end{tabular}
\tablenotetext{a}{Distance ranges for SED fitting are chosen to be consistent with the near kinematic distances from \S\ref{sec-dist} while being as homogenous as possible across regions.}
\tablenotetext{b}{Bounds of the area on the sky over which we searched for YSOs. These encompass significant area outside of the ``bubble'' or ``cometary'' regions to get a significant field sample as a control.}
\end{footnotesize}
\end{center}
\end{table}

\begin{table}
\begin{center}
\begin{footnotesize}
\caption{A summary of the point source sample sizes and results \label{t3}}
\begin{tabular}{lr@{ + }lr@{ + }lr@{ + }lr@{ + }l}
\tableline\tableline
\multirow{2}{*}{Region} & \multicolumn{2}{c}{Sample\tablenotemark{a}} & \multicolumn{2}{c}{Stellar\tablenotemark{b}} & \multicolumn{2}{c}{AGB\tablenotemark{c}} & \multicolumn{2}{c}{YSO\tablenotemark{d}} \\
 & (GPSC&MI) & (GPSC&MI) & (GPSC&MI) & (GPSC&MI) \\
\tableline
G028.83-0.25 & 1932&16 & 1614&2 & 32&0 &  48&4 \\
G041.10-0.15 & 2250&13 & 1845&1 & 28&0 & 93&11 \\
G041.91-0.12 and G041.92+0.04 & 1749&11 & 1526&0 & 23&0 & 78&11 \\
G044.28+0.11 & 8673&25 & 7911&1 & 46&1 & 155&23 \\
G044.34-0.82 & 1081&13 & 995&1 & 9&1 & 26&9 \\
\tablenotetext{a}{Includes sources from the GLIMPSE Point Source Catalog (GPSC) and additional sources identified visually in the 8 and 24 $\mu$m \emph{Spitzer} images (MI). Only sources detected in at least four photometric bands are included in this count.}
\tablenotetext{b}{Sources consistent with stellar atmosphere models from \cite{2005ESASP.576..565B}.}
\tablenotetext{c}{Sources initially identified as YSO candidates, but likely to be AGB stars (based on color-magnitude cuts) and thus removed from the final list of YSOs.}
\tablenotetext{d}{Sources identified as YSOs by the SED fitter of \cite{2007ApJS..169..328R} and not likely to be AGB stars.}
\end{tabular}
\end{footnotesize}
\end{center}
\end{table}

\begin{table}[!h]
\begin{center}
\begin{tiny}
\caption{YSOs identified by SED fitting \label{t4}}
\begin{tabular}{lccccr@{$\pm$}lr@{$\pm$}lr@{$\pm$}lr@{$\pm$}l}
\tableline\tableline
 \multirow{2}{*}{Region\tablenotemark{a}} & \multirow{2}{*}{YSO} & \multirow{2}{*}{Stage\tablenotemark{b}} & \multirow{2}{*}{$\left( \chi_{\textrm{best}}^{2}/n_{\textrm{data}} \right)$} & \multirow{2}{*}{24 $\mu$m} & \multicolumn{2}{c}{$M_{*}$\tablenotemark{c}} & \multicolumn{2}{c}{$L_{*}$} & \multicolumn{2}{c}{$\dot{M_{\textrm{env}}}$\tablenotemark{d}} & \multicolumn{2}{c}{$M_{\textrm{disk}}$} \\
  &  &  &  &  & \multicolumn{2}{c}{$\left( M_{\odot} \right) $} & \multicolumn{2}{c}{$\left( L_{\odot} \right) $} & \multicolumn{2}{c}{$\left( M_{\odot}~\textrm{yr}^{-1} \right) $} & \multicolumn{2}{c}{$\left( M_{\odot} \right) $} \\
\tableline
G28.83 & G028.6534-00.2539 & III & 2.89 & N & $14.6 $ & $ 0.0$ & $ 10^{4.3} $ & $ 0.0 $ & $ 0.0 $ & $ 0.0 $ & $ 10^{-8.0} $ & $ 0.0 $ \\
G28.83 & G028.6608-00.2305 & II & 3.00 & N & $5.5 $ & $ 1.4$ & $ 10^{3.0} $ & $ 10^{3.1} $ & $ 0.0 $ & $ 0.0 $ & $ 10^{-2.1} $ & $ 10^{-1.7} $ \\
G28.83 & G028.6788-00.2786 & I & 0.05 & N & $3.8 $ & $ 1.1$ & $ 10^{2.3} $ & $ 10^{2.6} $ & $ 10^{-4.7} $ & $ 10^{-4.1} $ & $ 10^{-1.7} $ & $ 10^{-1.4} $ \\
G28.83 & G028.6879-00.2739 & I & 0.31 &  Y  & $3.6 $ & $ 1.3$ & $ 10^{2.0} $ & $ 10^{2.1} $ & $ 10^{-4.1} $ & $ 10^{-3.8} $ & $ 10^{-1.5} $ & $ 10^{-1.2} $ \\
G28.83 & G028.6962-00.2913 & I & 0.47 &  Y  & $5.0 $ & $ 1.6$ & $ 10^{2.9} $ & $ 10^{3.1} $ & $ 10^{-4.8} $ & $ 10^{-3.9} $ & $ 10^{-1.9} $ & $ 10^{-1.5} $ \\
G28.83 & G028.7020-00.2101 & I & 0.71 & N & $2.7 $ & $ 0.9$ & $ 10^{1.7} $ & $ 10^{1.7} $ & $ 10^{-5.5} $ & $ 10^{-4.9} $ & $ 10^{-2.1} $ & $ 10^{-1.7} $ \\
G28.83 & G028.7166-00.2231 & I & 0.01 & N & $2.0 $ & $ 1.1$ & $ 10^{1.5} $ & $ 10^{1.9} $ & $ 10^{-5.4} $ & $ 10^{-4.8} $ & $ 10^{-2.2} $ & $ 10^{-1.8} $ \\
G28.83 & G028.7190-00.1813 & II & 0.67 &  Y  & $4.3 $ & $ 1.1$ & $ 10^{2.6} $ & $ 10^{2.9} $ & $ 10^{-6.0} $ & $ 10^{-4.8} $ & $ 10^{-2.1} $ & $ 10^{-1.7} $ \\
G28.83 & G028.7191-00.2083 & I & 2.06 &  Y  & $1.1 $ & $ 1.2$ & $ 10^{1.7} $ & $ 10^{2.1} $ & $ 10^{-4.5} $ & $ 10^{-4.0} $ & $ 10^{-1.8} $ & $ 10^{-1.6} $ \\
G28.83 & G028.7347-00.1769 & II & 0.01 &  Y  & $4.5 $ & $ 1.0$ & $ 10^{2.6} $ & $ 10^{2.6} $ & $ 10^{-6.2} $ & $ 10^{-4.8} $ & $ 10^{-2.0} $ & $ 10^{-1.6} $ \\
\end{tabular}
\tablenotetext{a}{Nearest H {\footnotesize II} region in this sample: G28.83 = G028.83-0.25, G41.10 = G041.10-0.15, G41.9X = G041.91-0.12 \& G041.92+0.04, G44.28 = G044.28+0.11, G44.34 = G044.34-0.82}
\tablenotetext{b}{See \S\ref{sec-classification} for explanation of evolutionary stages.}
\tablenotetext{c}{Values for all quantities are determined by the parameters of model SEDs that fit the source such that $\left( \chi^{2} - \chi^{2}_{\textrm{best}} \right) /n_{\textrm{data}} < 6$. Averages and uncertainties are the mean and standard deviation values of the fit parameters weighted by the probability of the corresponding model, $\exp \left( -\chi^{2}/2 \right)$ (See \S\ref{sec-irdata}). Uncertainties of 0.0 indicate no spread in the models that fit the data.}
\tablenotetext{d}{The data are sometimes fit by disk-only models with no accreting envelope, represented by a value of 0.0.}
\tablenotetext{}{(This table is available in its entirety in machine-readable and Virtual Observatory (VO) forms in the online journal. A portion is shown here for guidance regarding its form and content.)}
\end{tiny}
\end{center}
\end{table}

\begin{table}
\begin{center}
\begin{footnotesize}
\caption{H {\tiny II} Region Properties \label{t5}}
\begin{tabular}{lccccccccc}
\tableline\tableline
 \multirow{3}{*}{Region} &  &  & \multicolumn{3}{c}{\underline{Radio Continuum}} & \multirow{3}{*}{$\alpha_{\textrm{cm}}$\tablenotemark{b}} & \multirow{3}{*}{$\log_{10}\left( \dfrac{Q_{\textrm{Ly}}}{\textrm{s}^{-1}} \right)$} &  &  \\
  & \multicolumn{2}{c}{\underline{Diameter}} & $S_{21\textrm{~cm}}$ & $S_{21\textrm{~cm}}$\tablenotemark{a} & $S_{11\textrm{~cm}}$ &  &  & \multicolumn{2}{c}{\underline{Single Ionizing Source}} \\
  & (') & (pc) & (Jy) & (Jy) & (Jy) &  &  & Sp. Type\tablenotemark{c} & Sp. Type\tablenotemark{d} \\
\tableline
G028.83-0.25 &   2.8 & 2.0 & 1.01 & 0.95 & 0.81 & -0.24 & 48.3 & O9.5-B0 & O8-O9 \\
G041.10-0.15 & 10.3 & 6.0 & 5.45 & 6.01 & 6.61 &   0.15 & 48.8 & O8-O8.5 & O7-O7.5 \\
G041.91-0.12 &   4.0 & 0.8 &   0.5 & 0.57 & 0.45 & -0.36 & 46.9 & $<$B0.5 & B0.5-B1 \\
G041.92+0.04 &  3.2 &  0.6 & 0.45 &   0.4 & 0.23 & -0.84 & 46.8 & $<$B0.5 & B0.5-B1 \\ 
G044.28+0.11 &  8.9 &  5.6 & 1.32 &   1.3 & 1.22 &   -0.1 & 48.3 & O9.5-B0 & O8-O9 \\ 
G044.34-0.82 &   5.1 & 3.4 & 0.12 &   0.1 & 0.09 & -0.26 & 47.3 & $<$B0.5 & B0-B0.5 \\
\end{tabular}
\tablenotetext{a}{Smoothed to the resolution of the 11 cm images for calculating the spectral index.}
\tablenotetext{b}{Assuming 30\% errors in the radio continuum measurements, the uncertainties in the spectral indices are approximately 0.7-0.8.}
\tablenotetext{c}{Determined from $\log_{10}\left( Q_{\textrm{Ly}} \right)$ and \cite{1996ApJ...460..914V}, assuming a dwarf (luminosity class V) star.}
\tablenotetext{d}{Determined from $\log_{10}\left( Q_{\textrm{Ly}} \right)$ and \cite{2002MNRAS.337.1309S}, assuming solar metallicity and a dwarf (luminosity class V) star.}
\end{footnotesize}
\end{center}
\end{table}

\begin{table}[!h]
\begin{center}
\begin{scriptsize}
\caption{Molecular Gas Clump Parameters} \label{t6}
\begin{tabular}{lccccccccccc}
\tableline\tableline
 Region\tablenotemark{a} & $\ell_{peak}$ & $b_{peak}$ & $v_{peak}$             & $R_{cl}$\tablenotemark{b} & $\sigma_{cl}$          &  $d_{cl}$\tablenotemark{c} & $N_{cl}(\textrm{H}_{2})$\tablenotemark{d} & $n_{cl}(\textrm{H}_{2})$ & $M_{cl}(\textrm{H}_{2})$ & $\alpha_{vir}$  \\
 & (deg)         & (deg)      & $(\textrm{km s}^{-1})$ & (pc)                      & $(\textrm{km s}^{-1})$ & (pc)                      & $(10^{21} \textrm{~cm}^{-2})$             & $(\textrm{cm}^{-3})$     & ($M_{\odot}$)   & $(M_{vir}/M_{cl})$   \\
\tableline
G28.83a & 28.850 &   -0.24 &  88.26 &    4.3 &    1.9 &    1.7 &    18.7 &    274 &      6103 &    0.5 \\
G28.83a & 28.844 &   -0.21 &  85.93 &    3.3 &    1.6 &    3.2 &    11.0 &    228 &      2311 &    0.8 \\
G28.83a & 28.887 &   -0.20 &  85.93 &    3.1 &    1.2 &    1.6 &     5.4 &    108 &       899 &    1.1 \\
G28.83a & 28.887 &   -0.22 &  87.41 &    3.5 &    1.0 &    1.1 &     7.1 &    121 &      1518 &    0.5 \\
G28.83a & 28.868 &   -0.24 &  86.14 &    2.7 &    1.5 &    1.7 &     7.4 &    298 &      1671 &    0.9 \\
G28.83a & 28.887 &   -0.23 &  85.93 &    2.1 &    1.2 &    1.1 &     5.2 &    233 &       625 &    1.0 \\
G28.83a & 28.887 &   -0.26 &  85.93 &    2.9 &    1.6 &    2.1 &     6.5 &    141 &      1010 &    1.8 \\
G28.83a & 28.795 &   -0.23 &  86.14 &    3.7 &    2.0 &    4.5 &     4.8 &     99 &      1478 &    2.1 \\
G28.83a & 28.868 &   -0.30 &  87.20 &    3.9 &    1.8 &    4.0 &     3.9 &     97 &      1645 &    2.0 \\
G28.83b & 28.930 &   -0.22 &  95.63 &    3.4 &    3.0 &    4.3 &    21.8 &    684 &      7668 &    1.0 \\
\end{tabular}
\tablenotetext{a}{Nearest H {\footnotesize II} region in this sample: G28.83a = G028.83-0.25 (83.8-90.0 km/s), G28.83b = G028.83-0.25 (91.9-98.8 km/s), G41.10 = G041.10-0.15, G41.91 = G041.91-0.12, G41.92 = G041.92+0.04, G44.28 = G044.28+0.11, G44.34 = G044.34-0.82}
\tablenotetext{b}{Clump effective radius}
\tablenotetext{c}{Nearest neighbor (peak-to-peak) separation}
\tablenotetext{d}{Peak column density}
\tablenotetext{}{(This table is available in its entirety in a machine-readable form in the online journal. A portion is shown here for guidance regarding its form and content.)}
\end{scriptsize}
\end{center}
\end{table}

\begin{table}
\begin{center}
\begin{footnotesize}
\caption{Summary of Molecular Gas Properties \label{t7}}
\begin{tabular}{lcccccccc}
\tableline\tableline
\multirow{3}{*}{Region} & \multicolumn{4}{c}{Median Clump Values} & \multicolumn{4}{c}{Region-wide Values} \\
  & $\left< R_{\textrm{cl}} \right>$\tablenotemark{c} & $\left< d_{\textrm{cl}} \right>$\tablenotemark{d} & $\left< M_{\textrm{cl}} \right>$ & $\left< N_{\textrm{cl}} \right>$ & $M_{tot}$\tablenotemark{e} & $n_{i}$ & $t_{\textrm{HII}}$\tablenotemark{f} & $\Phi_{\textrm{Ly}}$\tablenotemark{g} \\
  & (pc) & (pc)& $(M_{\odot})$ & $(10^{21} \textrm{~cm}^{-2})$  & $(M_{\odot})$ & ($10^{3}$ cm${}^{-3}$) & ($10^{6}$ yr) & ($10^{9}$ cm${}^{-2}$ s${}^{-1}$) \\
\tableline
G028.83-0.25\tablenotemark{a} & 3.3 & 1.7 & 1518 & 6.5 & 17264 & 4.79 & 0.79 & $>$16.5 \\
G028.83-0.25\tablenotemark{b} & 3.1 & 1.8 & 1194 & 2.8 & 31133 & 4.46 & 0.83 & $>$16.5 \\
G041.10-0.15 & 1.8 & 1.0 & 202 & 2.6 & 19913 & 1.35 & 2.1 & $>$6.4 \\
G041.91-0.12 & 0.58 & 0.7 & 18 & 3.4 & 177 & 3.13 & 0.31 & $>$3.8 \\
G041.92+0.04 & 0.67 & 0.3 & 23 & 2.3 & 357 & 3.27 & 0.19 & $>$5.6 \\
G044.28+0.11 & 2.5 & 1.7 & 534 & 3.7 & 29180 & 1.19 & 2.41 & $>$2.1 \\
G044.34-0.82 & 2.5 & 1.6 & 656 & 4.0 & 9534 & 1.73 & 2.12 & $>$0.6 \\
\end{tabular}
\tablenotetext{a}{83.8 -- 90.0 km s${}^{-1}$}
\tablenotetext{b}{91.9 -- 98.8 km s${}^{-1}$}
\tablenotetext{c}{Median molecular clump radius.}
\tablenotetext{d}{Median clump peak-to-peak separation.}
\tablenotetext{e}{Sum of molecular clump masses.}
\tablenotetext{f}{Dynamical age of the H {\tiny II} region.}
\tablenotetext{g}{Ionizing flux.}
\end{footnotesize}
\end{center}
\end{table}

\begin{table}
\begin{center}
\begin{footnotesize}
\caption{Predicted Molecular Fragments Properties for Triggered Star Formation \label{t8}}
\begin{tabular}{lcccccccc}
\tableline\tableline
 \multirow{2}{*}{Region} & $t_{\star}$ & \multirow{2}{*}{$t_{\textrm{HII}}/t_{\star}$\tablenotemark{c}} & $t_{\textrm{frag}}$ & $R_{\textrm{frag}}$ & $N_{\textrm{frag}}$ & $M_{\textrm{frag}}$ & $d_{\textrm{frag}}$ & \multirow{2}{*}{$t_{\textrm{HII}}/t_{\textrm{frag}}$\tablenotemark{d}} \\
  & ($10^{6}$ yr) &  & ($10^{6}$ yr) & (pc) & ($10^{21}$ cm${}^{-2}$) & ($M_{\odot}$) & (pc) &  \\
\tableline
G028.83-0.25\tablenotemark{a} & $<$0.08 & $>$9.9 & 0.9 & 2.1 & 10.5 & 13.1 & 0.5 & 0.9 \\
G028.83-0.25\tablenotemark{b} & $<$0.08 & $>$10.4 & 0.9 & 2.2 & 10.3 & 13.4 & 0.5 & 0.9 \\
G041.10-0.15 & $<$0.10 & $>$21.0 & 1.4 & 4.8 & 6.6 & 20.8 & 0.7 & 1.5 \\
G041.91-0.12 & $<$0.12 & $>$2.6 & 1.4 & 2.0 & 6.5 & 21.3 & 0.8 & 0.2 \\
G041.92+0.04 & $<$0.11 & $>$1.7 & 1.4 & 1.9 & 6.5 & 21.3 & 0.8 & 0.1 \\
G044.28+0.11 & $<$0.15 & $>$16.1 & 1.7 & 4.5 & 5.6 & 24.7 & 0.9 & 1.4 \\
G044.34-0.82 & $<$0.23 & $>$9.2 & 1.7 & 3.0 & 5.4 & 25.5 & 0.9 & 1.2 \\
\end{tabular}
\tablenotetext{a}{83.8 -- 90.0 km s${}^{-1}$}
\tablenotetext{b}{91.9 -- 98.8 km s${}^{-1}$}
\tablenotetext{c}{The ratio of the dynamical age of the region to the timescale for radiatively driven implosion to begin. A value greater than 1 indicates that this is a plausible triggering scenario.}
\tablenotetext{c}{The ratio of the dynamical age of the region to the timescale for collect and collapse to begin. A value greater than 1 indicates that this is a plausible triggering scenario. The ratios for G028.83-0.25 are within uncertainty of a value of 1.}
\end{footnotesize}
\end{center}
\end{table}

\begin{table}
\begin{center}
\begin{footnotesize}
\caption{Correspondence between collect and collapse predictions and observation \label{t9}}
\begin{tabular}{lccccc}
\tableline\tableline
 Region & $t_{\textrm{frag}}$ & $R_{\textrm{frag}}$ & $N_{\textrm{frag}}$ & $M_{\textrm{frag}}$ & $d_{\textrm{frag}}$ \\
\tableline
G028.83-0.25\tablenotemark{a} & Y\tablenotemark{c} & Y & Y & N & Y \\
G028.83-0.25\tablenotemark{b} & Y & Y & N & N & N \\
G041.10-0.15 & Y & N & N & N & Y \\
G041.91-0.12 & N & N & Y & Y & Y \\
G041.92+0.04 & N & N & Y & Y & Y \\
G044.28+0.11 & Y & Y & Y & N & Y \\
G044.34-0.82 & Y & Y & Y & N & Y \\
\end{tabular}
\tablenotetext{a}{83.8 -- 90.0 km s${}^{-1}$}
\tablenotetext{b}{91.9 -- 98.8 km s${}^{-1}$}
\tablenotetext{c}{Cell values indicate whether the median clump values are (Y) or are not (N) consistent with the predicted values for collect and collapse. In the case of $t_{\textrm{frag}}$, the value is consistent if the dynamical age of the region is at least as the lower bound on the uncertainty range.}
\end{footnotesize}
\end{center}
\end{table}

\end{center}

\end{document}